\documentclass[journal,draftcls,onecolumn,12pt,twoside]{IEEEtran}
\IEEEoverridecommandlockouts

\usepackage{graphicx}
\usepackage{tikz}
\usepackage{pgfplots}
\usepackage{xcolor}
\usepackage{color}
\usepackage{colortbl}
\usepackage{amsthm}
\usepackage{amsmath}
\usepackage{bbm}
\usepackage{amsfonts}
\usepackage{amssymb}
\usepackage{mathtools}
\usepackage{cite}
\usepackage[nolist]{acronym}
\usepackage{booktabs}
\usepackage{diagbox}		
\usepackage{upgreek}
\usepackage{bbm}
\usepackage{Files/bm}
\usepackage{Files/winsnotation}
\usepackage{Files/Symbols_OPL}
\usepackage{Files/LVcolours}
\usepackage[ruled,vlined,noresetcount]{algorithm2e}
\usepackage{setspace}
\usepackage{comment}
\usepackage{array}
\usepackage{bbold}
\usepackage{lipsum}

\newcommand{\popsize}{K}
\newcommand{\nslot}{N_{\mathrm{s}}}
\newcommand{\T}{T}

\newcommand{\load}{\mathsf{G}}

\newcommand{\pactive}{\pi} 

\newcommand{\csaefficiency}{\eta}

\newcommand{\TMPR}{T}
\newcommand{\avgG}{\mathsf{G}}

\newcommand{\Ka}{K_{\mathrm{a}}}

\definecolor{emerald}{rgb}{0.31, 0.78, 0.47}

\makeatletter
\def\endthebibliography{%
  \def\@noitemerr{\@latex@warning{Empty `thebibliography' environment}}%
  \endlist
}
\makeatother
\IEEEoverridecommandlockouts

\theoremstyle{definition}
\newtheorem{remark}{Remark}
\newtheorem{theorem}{Theorem}
\newtheorem{lemma}{Lemma}
\newtheorem{corollary}{Corollary}
\newtheorem{example}{Example}
\newtheorem{definition}{Definition}

\pgfplotsset{compat=1.14}

\begin{document}

\title{IRSA-based Random Access\\ over the Gaussian Channel}

\author{%
  Velio Tralli,~\IEEEmembership{Senior Member,~IEEE,} and Enrico Paolini,~\IEEEmembership{Senior Member,~IEEE}
\thanks{V. Tralli is with CNIT/WiLab, DE, University of Ferrara, Italy. Email: velio.tralli@unife.it. E. Paolini is with CNIT/WiLab, DEI, University of Bologna, Italy. Email: e.paolini@unibo.it. 
An earlier version of this paper was presented in part at the 12th International Symposium on Topics in Coding (ISTC), Brest, France, Sept. 2023, IEEE [DOI: 10.1109/ISTC57237.2023.10273462].
}
}

\maketitle 

\begin{acronym}
\small
\acro{ACK}{acknowledgement}
\acro{AWGN}{additive white Gaussian noise}
\acro{BAC}{binary adder channel}
\acro{BCH}{Bose Chaudhuri Hocquenghem}
\acro{BPR}{Bar-David, Plotnik, Rom}
\acro{BS}{base station}
\acro{CDF}{cumulative distribution function}
\acro{CRA}{coded random access}
\acro{CRC}{cyclic redundancy check}
\acro{CRDSA}{contention resolution diversity slotted ALOHA}
\acro{CSA}{coded slotted ALOHA}
\acro{eMBB}{enhanced mobile broad-band}
\acro{FER}{frame error rate}
\acro{GMAC}{Gaussian multiple access channel}
\acro{IFSC}{intra-frame spatial coupling}
\acro{i.i.d.}{independent and identically distributed}
\acro{IoT}{Internet of Things}
\acro{IRSA}{irregular repetition slotted ALOHA}
\acro{LDPC}{low-density parity-check}
\acro{LOS}{line of sight}
\acro{MAC}{medium access control}
\acro{MIMO}{multiple input multiple output}
\acro{ML}{maximum likelihood}
\acro{MMA}{massive multiple access}
\acro{mMTC}{massive machine-type communication}
\acro{MPR}{multi-packet reception}
\acro{MRC}{maximal ratio combining}
\acro{PAB}{payload aided based}
\acro{PDF}{probability density function}
\acro{PGF}{probability generating function}
\acro{PHY}{physical}
\acro{PLP}{packet loss probability}
\acro{PLR}{packet loss rate}
\acro{PMF}{probability mass function}
\acro{PRCE}{perfect replica channel estimation}
\acro{QPSK}{quadrature phase-shift keying}
\acro{RF}{radio-frequency}
\acro{SA}{slotted ALOHA}
\acro{SC}{spatial coupling}
\acro{SIC}{successive interference cancellation}
\acro{SIS}{successive interference subtraction}
\acro{SNB}{squared norm based}
\acro{SNR}{signal-to-noise ratio}
\acro{TAC}{ternary adder channel}
\acro{URLLC}{ultra-reliable and low-latency communication}
\acro{IMDS}{irregular MDS}
\end{acronym}
\setcounter{page}{1}

\begin{abstract}
A framework for the analysis of synchronous grant-free massive multiple access schemes based on the irregular repetition slotted ALOHA (IRSA) protocol and operating over the Gaussian multiple access channel is presented. IRSA-based schemes are considered here as an instance of the class of unsourced slotted random access codes, operating over a frame partitioned in time slots, and are obtained by concatenation of a medium access control layer code over the entire frame and a physical layer code over each slot. In this framework, an asymptotic analysis is carried out in presence of both collisions and slot decoding errors due to channel noise, which allows the derivation of density-evolution equations, asymptotic limits for minimum packet loss probability and average load threshold, and a converse bound for threshold values. This analysis is exploited as a tool for the evaluation of performance limits in terms of minimum signal-to-noise ratio required to achieve a given packet loss probability, and also provides convergence boundary limits that hold for any IRSA scheme with given physical layer coding scheme. The tradeoff between energy efficiency and spectrum efficiency is numerically evaluated comparing some known coding options, including those achieving random coding bounds at slot level. It is shown that IRSA-based schemes have a convergence boundary limit within few dB from the random coding bound 
when the number of active transmitters is sufficiently large.
\end{abstract}

\begin{IEEEkeywords}
Grant-free multiple access, iterative decoding, irregular repetition slotted ALOHA, massive machine-type communications, successive interference cancellation, unsourced random access.
\end{IEEEkeywords}

\section{Introduction}



Wireless \ac{IoT} communications in cellular networks have shown an impressive growth in the recent years  and will become pervasive in the future 6G systems \cite{Nguyen2021:IoT}.
\ac{IoT} networks connect a large set of battery-powered 
devices autonomously transmitting short packets to a common \ac{BS}. The short activity periods are separated by random idle periods. We usually refer to \ac{MMA} 
\cite{Chen2021:Massive} when the number of devices is very large with respect to the number of channel uses available for the transmission of one data packet, and the number of active devices is an unknown subset of the whole set of devices.

Future massive \ac{IoT} networks represent a natural venue for \emph{grant-free} and \emph{uncoordinated} communication protocols 
\cite{Paolini2015:magazine,Liu2018:sparse,Chisci2019:Uncoordinated}. In this scenario, the coordination of a massive number of  devices 
that become active at unpredictable instants and for very short periods
would be significantly inefficient, as the overhead   required  may become even larger than data, with a negative impact on  scalability, latency, and energy efficiency.  
Accessing the channel in a grant-free fashion, i.e., without any prior agreement with the \ac{BS} and without any coordination with the other devices that are active at the same time, is much more efficient and simplifies the operations at the device side, though increasing the computational effort at the \ac{BS}.

To understand the fundamental
limitations of uncoordinated massive random access, Y. Polyanskiy \cite{Polyanskiy2017:Perspective} proposed a framework where i) each active user transmits a packet with a fixed amount of information bits within a finite frame length; ii) the users are required to share the same codebook, i.e., they are unidentifiable, unless they put their identity in the payload, and the decoder is only required to provide an unordered list of user messages;  iii) the error probability is defined per-user, as the probability that the user message is not decoded. 
This framework enabled the evaluation of  finite-blocklength performance bounds as, for example, the bound for the minimal energy per bit required to support communication at given error probability. 
Further extensions of this analysis have appeared more recently. The work in \cite{Ngo2021:Random_user_activity} extended the random-coding achievability bound for the \ac{GMAC} presented  in \cite{Polyanskiy2017:Perspective} to the case where the number of active users is random and a-priori unknown: both mis-detection and false-alarm events, that together replace  the generic error event in this scenario, have been considered. 
The work in \cite{Kowshik2020:Energy_Efficient} extended the previous framework to consider quasi-static fading providing performance bounds and  a practical coding scheme. 

The comparison with fundamental bounds has shown that the well known existing schemes such as \ac{SA} \cite{Abramson1970:ALOHA,Roberts72:ALOHA} are very far from the analytical random coding bounds.
To improve the performance, a $T$-fold \ac{SA} approach has been proposed in \cite{Ordentlich2017:low_complexity}, where the frame is divided into slots and, if the number of users transmitting in each slot is no more than $\TMPR$, then the decoder tries to decode all corresponding messages; otherwise, nothing is decoded. This approach relies on the \ac{MPR} capability of the decoder that is able to estimate the number of transmitted packets in a slot and is able to decode them if they are no more than $\TMPR$. A practical coding scheme is proposed to allow this possibility, which is based on the concatenation of a code for zero error probability detection on a \ac{BAC} and a binary linear code. The code for \ac{BAC} was originally proposed in \cite{Bar-David1993:Forward} and is hereafter referred to as \ac{BPR} code.

The main limitations of the basic \ac{SA} scheme without retransmission is the large packet loss rate  since each slot works as a collision channel in which a single packet is successfully received if no collision occurs. Effective techniques have been developed in \cite{casini2007:contention,liva2011:irsa} to improve \ac{SA} by transmitting multiple packet replicas over a frame and then applying \ac{SIC} to remove the signals of decoded packets from slots with collisions. A fixed number of repetitions is adopted in the \ac{CRDSA} protocol \cite{casini2007:contention}, while a random number of them is used in the \ac{IRSA} protocol \cite{liva2011:irsa}. An interesting view of the \ac{SIC} process has been provided in \cite{liva2011:irsa,paolini2015:csa } where the analogy with decoding codes on sparse graph is exploited for the analysis  and  design of these techniques.  In particular, density evolution analysis has been exploited to derive, for each scheme, the average load threshold that allows operations without packet losses over the collision channel in asymptotic conditions. This analysis is also useful for the design of protocol parameters and for driving the search for good schemes with large load threshold. An upper bound for the load threshold acting as a converse bound has been also derived for the collision channel.
As shown in \cite{paolini2015:csa}, by using packet segmentation and encoding, \ac{CSA} schemes can be obtained, offering different tradeoffs between throughput and spectrum efficiency, and being able to approach 1 successful transmission per slot with negligible packet-loss rate over a collision channel when the efficiency of the protocol gets close to zero.
IRSA/CSA schemes can be also enhanced to exploit MPR in each slot. Even if the MPR capability is obtained at the expense of bandwidth expansion, a net throughput improvement is still achievable for CSA over collision channel, as shown in \cite{stefanovic2018:multipacket}.

Differently from the previous work on the analysis of \ac{IRSA} and \ac{CSA} random access schemes, here we investigate the behavior of \ac{IRSA} schemes when used for the unsourced random access over a Gaussian channel. The objective is two-fold: (i) analyze the performance of \ac{IRSA}-based schemes over a noisy channel that introduces a floor on packet-loss probability and limitations on load threshold; (ii) evaluate the performance of \ac{IRSA}-based unsourced random access schemes seeking for achievable limits on packet loss probability and energy efficiency.

Recently, two studies have also investigated the performance achievable by \ac{IRSA}-based schemes with \ac{MPR} for unsourced random access over a noisy channel. In \cite{vem2019:unsourced}  a coding scheme is proposed where the  packets in each slot  are encoded with an interleaved LDPC code and transmitted with side information related to the interleaving configuration, that is first decoded using sparse detection techniques. This scheme achieves good energy efficiency, improving the results obtained with the scheme proposed in \cite{Ordentlich2017:low_complexity}, but performance evaluation is carried out 
under the simplified assumption that \ac{SIC} is operating with errorless slot decoding. 
In \cite{glebov2019} a density evolution technique is exploited to derive packet loss probability over \ac{GMAC} considering 
a decoder able to provide an output list of up to $T$ messages with unknown number of transmissions. Here, the main objective was to find good \ac{IRSA} protocols able to minimize signal-to-noise ratio for a given target packet loss probability, without investigating the asymptotic behavior  of the random access scheme in terms of packet error probability floor or average load threshold. 
Moreover, in the scheme investigated in \cite{glebov2019}  a slot is considered as resolved even if not all the messages are correctly decoded after an iteration of the \ac{SIC} process, and the correct part of the outcome of the slot decoder is used to  cancel interference in other slots, while in the basic \ac{IRSA} processing investigated in our paper a slot is resolved and decoded messages are used to cancel interference in other slots only when all the messages in the slot are correctly decoded.

More recently, other two works 
have investigated \ac{IRSA}-based random access schemes addressing the issue of imperfect interference cancellation in the \ac{SIC} process, beyond the preliminary investigation given in \cite{liva2011:irsa,paolini2015:csa}. In \cite{dumas2021:canc_err} a density evolution technique with suitable modeling  of the cancellation efficiency at each iteration is proposed to evaluate packet loss probability in a noiseless channel, and to search for good codes that maximize average load for a given packet loss probability. In \cite{haghighat2023:canc_err} 
the effects of imperfect packet recovery  are investigated in a noisy channel with a different approach based on the disjoint treatment of MAC and PHY layers. Here, the \ac{SIC} process includes the cancellation of all the signals that can be decoded, even if they are decoded with errors, and density evolution technique is exploited to track equivalent noise plus interference power along the SIC process.

The practical solutions discussed so far are based on conventional channel coding and decoding. Alternative solutions based on compressed sensing have been proposed in \cite{Amalladinne2020,Fengler2021:SPARCs,Calderbank2019}.
In the scheme in \cite{Amalladinne2020}, referred to as ``coded compressed sensing'', each packet is partitioned in small sub-blocks transmitted over  consecutive slots. Within a slot, all sub-blocks are recovered with compressed sensing algorithms. The sub-blocks are further encoded with systematic linear codes that allow piecing together the recovered blocks into the original messages.
Other solutions have been recently explored in \cite{pradhan2021:icc} considering the combined use of spreading sequences, to control multiple access interference, and  powerful channel coding with iterative belief propagation decoding.  This type of scheme represents today the state-of-the-art solution that provides the best energy efficiency among the schemes proposed so far for unsourced random access. It also allows to approach the achievability bound in \cite{Polyanskiy2017:Perspective} when the number of transmitting devices is small, while staying  within 2 dB from this bound for larger numbers.


This paper investigates \ac{IRSA}-based synchronous grant-free random access  schemes operating over a Gaussian channel. Such schemes can be viewed as an instance of slotted random access codes that are built for a frame partitioned in time slots and are obtained with the concatenation of \ac{MAC} layer code extended over the entire frame and a \ac{PHY} layer code for the transmission over each slot. 
More specifically, the paper provides a framework for performance analysis in asymptotic conditions of \ac{IRSA}-based MAC layer codes in presence of both collisions and slot decoding errors due to channel noise. The framework includes density-evolution equations,  asymptotic limits for minimum packet loss probability and average load threshold,  and a converse bound  for threshold values, which defines a set of  average loads for which packet loss probability can never converge to values below a suitable minimum.
Moreover, it provides performance evaluation and comparison of \ac{IRSA}-based random access schemes with two basic options for PHY layer coding: i) optimum coding for random access with up to $T$ active transmitters, ii) \ac{BPR} coding for $T$-user binary memoryless adder channel, concatenated with an ideal linear inner code for binary-input output-symmetric channel, as in \cite{Ordentlich2017:low_complexity}. The analysis developed for the asymptotic condition is used as a tool to evaluate actual performance. The results show the performance limits  of \ac{IRSA}-based random-access schemes in terms of achievable $E_b/N_0$ and tradeoff between energy and spectrum efficiency.

The main contributions of the paper may be summarized as follows:
\begin{itemize}
    \item a framework for performance analysis in asymptotic conditions of IRSA-based schemes in presence of both collisions and decoding errors due to channel noise is provided;
    \item  density-evolution equations, asymptotic limits for the minimum packet loss probability, and average load threshold are derived, by considering a decoder based on perfect estimation of the number of packets in a slot and on ideal detection of errors in the list of the decoded packets;
    \item conditions on system parameters  to obtain  a well-defined average load threshold are derived;
    \item a converse bound for threshold values is derived, which defines a set of average loads for which packet loss probability can never converge to values below a given target;
    \item  performance limits of IRSA-based random-access schemes are numerically evaluated in terms of achievable energy efficiency and spectrum efficiency, by exploiting the tools of asymptotic analysis.
\end{itemize}
Our work is different from the main literature on IRSA protocol analysis in the following aspects:
in \cite{liva2011:irsa,paolini2015:csa,stefanovic2018:multipacket} IRSA protocol is analyzed over a collision channel, while we investigate IRSA schemes for the unsourced random access over a Gaussian channel; 
in \cite{vem2019:unsourced} the performance of the proposed scheme is evaluated  with a simplified SIC decoder that operates with zero errors in slots where no more that $T$ packets occupy it, while we consider decoding errors due to noise; 
in \cite{glebov2019} the SIC process considers a slot as resolved even if not all the messages are correctly decoded at each iteration, and uses  the outcome of the slot decoder to cancel interference in other slots, while in this paper a slot is resolved and decoded messages are used to cancel interference in other slots only when all the messages in the slot are correctly decoded.

The rest of the paper is organized as follows. Section \ref{Sec:model} presents the system model, the class of the slotted random access codes and the \ac{IRSA}-based codes. The performance of MAC layer component of \ac{IRSA}-based codes over \ac{GMAC} is analyzed in Section~\ref{Sec:MAC_performance} considering asymptotic threshold, error floor and convergence boundary, while PHY layer component options are discussed in Section~\ref{Sec:PHY_performance}. The framework for evaluating achievable $E_b/N_0$ and tradeoff between energy and spectrum efficiency is presented in Section~\ref{Sec:achievable_EbNo}, 
and the related numerical results are discussed in Section~\ref{Sec:results}. Finally,  conclusions are drawn in Section~\ref{Sec:conclusions}.


\section{System model}\label{Sec:model}

We consider an \ac{MMA}  scenario with a large number $K$ of transmitters (representing IoT devices, wireless sensors, smart meters, etc.) and one receiver.  
The common receiver is a \ac{BS} of the radio access network.
The time is organized in frames and all transmitters are frame-synchronous.
Each transmitter becomes active when it has a new data packet to transmit and remains in idle state otherwise. 
The activation process is random, independent frame by frame, and 
characterized by the activation probability, denoted by $\pi$. 
As a consequence, the  number of transmitters that are active in a frame and have one data packet to transmit, denoted by $\Ka$,  is a random variable which is binomially distributed with mean value $\mathbb{E}[\Ka]=\pi K$.
In the most general case, this random variable is unknown to the receiver.
All data packets are assumed to have the same length of $k$ information bits, and each active transmitter  may transmit up to $N$ symbols per frame. 
A null symbol is assumed for those channel uses where no transmission is performed. 
All symbols are assumed to be real in the simplest  setting.

We consider access protocols that are both grant-free and uncoordinated. 
According to the unsourced framework proposed in \cite{Polyanskiy2017:Perspective},
we also assume that all transmitters use a common codebook and the receiver only needs to decode the list of the messages transmitted by the active devices. The active transmitter identities 
need not be recovered.
However, they can be embedded in the payload and can be recovered at the higher layers of the protocol stack. 

All active nodes have the same encoder that maps a message $m \in [M]=\{ 1,\ldots, M\}$ onto a sequence of $N$ symbols, as
$$
f_{RA}: \;\; m\in [M] \longmapsto f_{RA}(m)=(x_1,\ldots, x_N) \in \mathbb{R}^N
$$
where the codebook size $M$  is less than or equal to $2^k$. 
Each user $j$ with a $k$-bit packet picks the corresponding message $W_j\in [M] $ and encodes it into $f_{RA}(W_j)$. The encoding rate of the active transmitter is $R=(\log_2 M)/N$.
Each symbol of the encoded sequence is transmitted over a real \ac{GMAC} whose output, at the generic time instant of the frame, is given by
\begin{equation}
Y= \sum_{j=1}^{\Ka} X_j +Z
\end{equation}
where $X_j$, $j=1,\ldots,\Ka$, are the input symbols of each active transmitter and $Z\sim \mathcal{N}(0,\sigma^2)$ is the additive noise. 

In the particular case  of perfect knowledge of $\Ka$, the receiver tries to build the list of $\Ka$ transmitted messages by using a suitably defined decoding function
$$
g: \;\; Y^N=(y_1,\ldots,y_N)  \longmapsto g(Y^N) = \{\hat{m}_1,\ldots, \hat{m}_{\Ka}\}
$$
with $m_j \in [M]$.
The pair of encoding and decoding functions $f_{RA}$ and $g$ defines the random access code for
the $\Ka$-user GMAC channel.
In \cite{Polyanskiy2017:Perspective} a random coding bound on the per-user error probability has been derived. It is an achievability result that gives a fundamental limit for random access codes.
%
%
%
%
Accordingly, under a finite energy constraint $\|f_{RA}(W_j)\|^2_2 \leq N\bar{P}$ and for given $N$, $M$, $\Ka$, and $\bar{P}/\sigma^2$, there exists a random access code for a $\Ka$-user GMAC such that $\epsilon_\text{pu} \leq F_1(N,M,\bar{P}/\sigma^2,\Ka)$, where $\epsilon_\text{pu}$ is the per-user error probability, defined as $\epsilon_\text{pu} =\frac{1}{\Ka} \sum_{j=1}^{\Ka}\mathbb{P}[E_j]$, and $E_j = \{W_j \notin g(Y^N)\} \cup \{W_j = W_i$ for some $i \neq j\} $ is the $j$-th user error event, for  $W_1 , \ldots, W_{\Ka}$  independent and uniform on $[M ]$.
The bounding expression $F_1(\cdot)$ is specified in \cite{Polyanskiy2017:Perspective}. It
is not in simple form and includes implicit maximization/minimization with respect to some parameters which can be evaluated numerically. 
It can be also related to the signal-to-noise ratio
\begin{align*}
\frac{E_b}{N_0} = 
\frac{N}{2\log_2 M} \frac{\bar{P}}{\sigma^2}
\end{align*}
where the energy per bit is $N\bar{P}/\log_2M$ and the noise variance is  $\sigma^2=N_0/2$.

The result summarized above has been extended in \cite{Ngo2021:Random_user_activity} to include the cases where the number of active transmitters $\Ka$ is random and unknown to the decoder,  which has to estimate it. In \cite{Ngo2021:Random_user_activity}, the achievability result is also formulated for a complex-valued GMAC with phase-synchronous transmitters. 

\subsection{Slotted Random Access Codes}\label{subsec:random_access_codes}

In this paper we consider the class of multiple access schemes where the available frame of $N$ channel uses is partitioned into $\nslot$ slots of length $n=N/\nslot$. 
The number of slots is an additional parameter of the access scheme which can be designed to optimize the performance metrics. 
In a slotted multiple access scheme each active transmitter, without any coordination  with the others, randomly selects  a set of one or more slots for transmission, while sending null symbols in the other slots. 
This 
is a generalized view of frame-based \ac{SA}, where each transmitter simply picks one slot of the frame and uses it to send its packet. 
In 
a general scheme each transmitter may send 
its encoded packet over  a random number $D$ of slots, drawing the $D$ slots according to some probability distribution, e.g., uniform, over all possible slot $D$-tuples.

We define a slotted random access code as a random access code where the codewords of the codebook are in the form
\begin{align}\label{eq:slotcode}
 f_{RA}(m)=(\mathbf{x}_1,\ldots, \mathbf{x}_{\nslot}) ,  \; 
\;  \mathbf{x}_i=(x_{i1},\ldots,x_{in})=f_{s,i}(m)
\end{align}
with the energy constraint 
$||\mathbf{x}_{i}||^2_2 \leq nP$ if slot $i$ is selected for transmission and $||\mathbf{x}_{i}||^2_2 =0$ otherwise.
In the model, the length-$N$ codeword generated by an active transmitter is a deterministic function of the message $m$ only: the same message is always mapped onto the same codeword.
As an example, in \ac{SA} random access the slot index may be obtained as $m\,\, \mathrm{mod}\,\,\nslot$, which maps message $m\in [M]$ onto a slot.
If $\bar{d}=\mathbb{E}[D]$ is the expected number of slots selected for transmission,  the signal to noise ratio for slotted random access codes can be defined as 
\begin{align}\label{eq:ebno}
\frac{E_b}{N_0} = \frac{n \bar{d}}{2\log_2M}\frac{P}{\sigma^2} .
\end{align}

The schemes based on \ac{IRSA}  \cite{liva2011:irsa} and \ac{CSA} \cite{paolini2015:csa}  belong to the class of slotted random access codes. 
In this paper we analyze the performance of IRSA-based random access codes, which can be also viewed as codes on sparse graphs, taking frame-based \ac{SA} as a reference for comparison. 
In the IRSA protocol, at the beginning of the frame, each active device  samples, independently of the other devices, a discrete random variable $D \in \{1,\dots,d_\text{max}\}$ with \ac{PGF} 
$\Lambda(x) = \sum_{d} \Lambda_d x^d$,  where $\Lambda_d = \mathbb{P}(D=d)$. 
The distribution $\mathbf{\Lambda}=\{\Lambda_1,\Lambda_1,\dots ,\Lambda_{d_\text{max}}\}$ is the same for all transmitters and is hereafter referred to as the \ac{IRSA} distribution.
The transmitter then draws, uniformly at random and without replacement, $D$ integers between $0$ and $\nslot-1$ and transmits $D$ replicas of its encoded packet in the corresponding slots of the frame.

\begin{figure}[t]
\centering
         \resizebox{0.5\columnwidth}{!}{%
     	\tikzset{every picture/.style={line width=0.75pt}} 

\begin{tikzpicture}[x=0.75pt,y=0.75pt,yscale=-1,xscale=1]

\draw   (100,98) .. controls (100,93.58) and (103.58,90) .. (108,90) -- (199,90) .. controls (203.42,90) and (207,93.58) .. (207,98) -- (207,122) .. controls (207,126.42) and (203.42,130) .. (199,130) -- (108,130) .. controls (103.58,130) and (100,126.42) .. (100,122) -- cycle ;
\draw   (100,178) .. controls (100,173.58) and (103.58,170) .. (108,170) -- (199,170) .. controls (203.42,170) and (207,173.58) .. (207,178) -- (207,202) .. controls (207,206.42) and (203.42,210) .. (199,210) -- (108,210) .. controls (103.58,210) and (100,206.42) .. (100,202) -- cycle ;
\draw    (155,130) -- (155,166.5) ;
\draw [shift={(155,169.5)}, rotate = 270] [fill={rgb, 255:red, 0; green, 0; blue, 0 }  ][line width=0.08]  [draw opacity=0] (7.14,-3.43) -- (0,0) -- (7.14,3.43) -- cycle    ;
\draw    (155,50.5) -- (155,87) ;
\draw [shift={(155,90)}, rotate = 270] [fill={rgb, 255:red, 0; green, 0; blue, 0 }  ][line width=0.08]  [draw opacity=0] (7.14,-3.43) -- (0,0) -- (7.14,3.43) -- cycle    ;
\draw    (155,209.5) -- (155,246) ;
\draw [shift={(155,249)}, rotate = 270] [fill={rgb, 255:red, 0; green, 0; blue, 0 }  ][line width=0.08]  [draw opacity=0] (7.14,-3.43) -- (0,0) -- (7.14,3.43) -- cycle    ;

\draw (115,103) node [anchor=north west][inner sep=0.75pt]   [align=left] {{\small $\!\!$frame encoding}};
\draw (120,183) node [anchor=north west][inner sep=0.75pt]   [align=left] {{\small $\!\!$slot encoding}};
\draw (169.5,49) node [anchor=north west][inner sep=0.75pt]  [font=\footnotesize] [align=left] {$\displaystyle m$};
\draw (169.5,141) node [anchor=north west][inner sep=0.75pt]   [align=left] {{\footnotesize $\displaystyle ( 0,m,0,0,m,0,0,0,0,m,0,0,0)$}};
\draw (169.5,227) node [anchor=north west][inner sep=0.75pt]   [align=left] {{\footnotesize $\displaystyle (\mathbf{0} ,\mathbf{x} ,\mathbf{0} ,\mathbf{0} ,\mathbf{x} ,\mathbf{0} ,\mathbf{0} ,\mathbf{0} ,\mathbf{0} ,\mathbf{x} ,\mathbf{0} ,\mathbf{0} ,\mathbf{0})$}};
\draw (221.5,248.5) node [anchor=north west][inner sep=0.75pt]  [font=\footnotesize] [align=left] {$\displaystyle \mathbf{x} =f_{P}( m)$};

\end{tikzpicture}
     	}%
       \caption{An example of encoding function for IRSA.  A nonzero message $m\in [M]$ is mapped onto a sequence $f_M(m)$ of length $\nslot$, where the number $D$ of copies of message $m$ and the positions of them in the sequence are all functions of the message $m$ only. Each message in  $f_M(m)$ is encoded into a slot codeword  which can be $\mathbf{x}=f_P(m)$ or $\mathbf{0}=f_P(0)$.}
    \label{fig:encoder_irsa}
\end{figure}
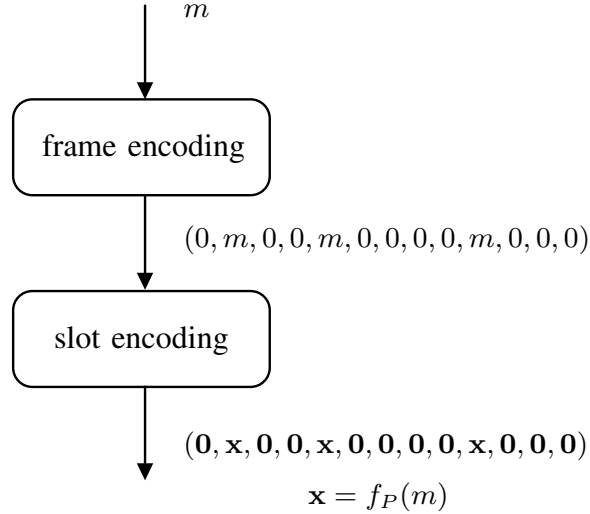

In \ac{IRSA}-based random access codes  the encoding function  can be formally described as illustrated in Fig.~\ref{fig:encoder_irsa}. 
The message $m=0$ is used by the idle transmitters and is encoded into the all-zero codeword,
whereas the message $m \in [M]$ 
is encoded as follows. 
As a first encoding step,  a sequence of $\nslot$ messages is formed: $D$ positions out of $\nslot$ are filled with a replica of $m$, whereas the remaining $\nslot-D$ positions are filled with the $0$ message.  Both the repetition degree $D$ and the positions of the $D$ activity words are random, but functions of the message $m$. 
In \cite{paolini2022:isit} a method to map message $m$ to $D$ and to the set of $D$ positions in the frame, according to the statistics of random variables, has been discussed.
This first encoding step is denoted as
\begin{align}
f_M: \;\; m\in [M] \longrightarrow &f_M(m)=(m_{s,0},\ldots, m_{s,\nslot-1}) 
\end{align}
where $m_{s,i}\in\{ 0,m\}$.
As a second encoding step, each message of the sequence is encoded into a codeword of length $n$ symbols; the $0$ message is encoded into the all zero codeword. 
This encoding step is denoted as
\begin{align} \label{eq:fP_enc}
f_P: \;\; m_{s,i}\in [M]\cup \{ 0\} \longrightarrow f_P(m_{s,i})=(x_{i1},\ldots, x_{in}) \in \mathbb{R}^n \, .
\end{align}
Note that this two-step encoding is in agreement with \eqref{eq:slotcode}.
%
%
%
It is interesting to note that the encoder of a slotted random access code is obtained with the concatenation of two encoding functions, i.e., a low-layer encoding operating with channel symbols, and a high-layer encoding operating with slots. We will denote them as PHY layer encoding (or slot encoding) and MAC layer encoding (or frame encoding), respectively. 

\begin{figure}[t]
\centering
         \resizebox{0.45\textwidth}{!}{%
     		\tikzset{every picture/.style={line width=0.75pt}} 

\begin{tikzpicture}[x=0.75pt,y=0.75pt,yscale=-1,xscale=1]

\draw [line width=0.25pt]  (18,66) -- (78,130);
\draw [line width=0.25pt]  (18,66) -- (178,130);

\draw [line width=0.25pt]  (38,66) -- (38,130);
\draw [line width=0.25pt]  (38,66) -- (78,130);
\draw [line width=0.25pt]  (38,69) -- (238,130);

\draw [line width=0.25pt]  (78,66) -- (78,130);
\draw [line width=0.25pt]  (78,66) -- (178,130);

\draw [line width=0.25pt]  (118,66) -- (78,130);
\draw [line width=0.25pt]  (118,66) -- (98,130);
\draw [line width=0.25pt]  (118,66) -- (138,130);
\draw [line width=0.25pt]  (118,66) -- (178,130);

\draw [line width=0.25pt]  (138,66) -- (98,130);
\draw [line width=0.25pt]  (138,66) -- (138,130);

\draw [line width=0.25pt]  (178,66) -- (178,130);
\draw [line width=0.25pt]  (178,66) -- (238,130);

\draw [line width=0.25pt]  (218,69) -- (-1,130);
\draw [line width=0.25pt]  (218,69) -- (238,130);

\draw (18,56) node  [font=\scriptsize] [align=left] {$\displaystyle u_{0}$};

\draw  [color={rgb, 255:red, 92; green, 138; blue, 168 }  ,draw opacity=1 ][fill={rgb, 255:red, 240; green, 240; blue, 255 }  ,fill opacity=1 ] (13,66) .. controls (13,63.24) and (15.24,61) .. (18,61) .. controls (20.76,61) and (23,63.24) .. (23,66) .. controls (23,68.76) and (20.76,71) .. (18,71) .. controls (15.24,71) and (13,68.76) .. (13,66) -- cycle ;

\draw (38,56) node  [font=\scriptsize] [align=left] {$\displaystyle u_{1}$};

\draw  [color={rgb, 255:red, 92; green, 138; blue, 168 }  ,draw opacity=1 ][fill={rgb, 255:red, 240; green, 240; blue, 255 }  ,fill opacity=1 ] (33,66) .. controls (33,63.24) and (35.24,61) .. (38,61) .. controls (40.76,61) and (43,63.24) .. (43,66) .. controls (43,68.76) and (40.76,71) .. (38,71) .. controls (35.24,71) and (33,68.76) .. (33,66) -- cycle ;

\draw (58,56) node  [font=\scriptsize] [align=left] {$\displaystyle u_{2}$};

\draw  (53,66) .. controls (53,63.24) and (55.24,61) .. (58,61) .. controls (60.76,61) and (63,63.24) .. (63,66) .. controls (63,68.76) and (60.76,71) .. (58,71) .. controls (55.24,71) and (53,68.76) .. (53,66) -- cycle ;

\draw (78,56) node  [font=\scriptsize] [align=left] {$\displaystyle u_{3}$};

\draw  [color={rgb, 255:red, 92; green, 138; blue, 168 }  ,draw opacity=1 ][fill={rgb, 255:red, 240; green, 240; blue, 255 }  ,fill opacity=1 ] (73,66) .. controls (73,63.24) and (75.24,61) .. (78,61) .. controls (80.76,61) and (83,63.24) .. (83,66) .. controls (83,68.76) and (80.76,71) .. (78,71) .. controls (75.24,71) and (73,68.76) .. (73,66) -- cycle ;

\draw (98,56) node  [font=\scriptsize] [align=left] {$\displaystyle u_{4}$};

\draw  (93,66) .. controls (93,63.24) and (95.24,61) .. (98,61) .. controls (100.76,61) and (103,63.24) .. (103,66) .. controls (103,68.76) and (100.76,71) .. (98,71) .. controls (95.24,71) and (93,68.76) .. (93,66) -- cycle ;

\draw (118,56) node  [font=\scriptsize] [align=left] {$\displaystyle u_{5}$};

\draw  [color={rgb, 255:red, 92; green, 138; blue, 168 }  ,draw opacity=1 ][fill={rgb, 255:red, 240; green, 240; blue, 255 }  ,fill opacity=1 ] (113,66) .. controls (113,63.24) and (115.24,61) .. (118,61) .. controls (120.76,61) and (123,63.24) .. (123,66) .. controls (123,68.76) and (120.76,71) .. (118,71) .. controls (115.24,71) and (113,68.76) .. (113,66) -- cycle ;

\draw (138,56) node  [font=\scriptsize] [align=left] {$\displaystyle u_{6}$};

\draw  [color={rgb, 255:red, 92; green, 138; blue, 168 }  ,draw opacity=1 ][fill={rgb, 255:red, 240; green, 240; blue, 255 }  ,fill opacity=1 ] (133,66) .. controls (133,63.24) and (135.24,61) .. (138,61) .. controls (140.76,61) and (143,63.24) .. (143,66) .. controls (143,68.76) and (140.76,71) .. (138,71) .. controls (135.24,71) and (133,68.76) .. (133,66) -- cycle ;

\draw (158,56) node  [font=\scriptsize] [align=left] {$\displaystyle u_{7}$};

\draw  (153,66) .. controls (153,63.24) and (155.24,61) .. (158,61) .. controls (160.76,61) and (163,63.24) .. (163,66) .. controls (163,68.76) and (160.76,71) .. (158,71) .. controls (155.24,71) and (153,68.76) .. (153,66) -- cycle ;

\draw (178,56) node  [font=\scriptsize] [align=left] {$\displaystyle u_{8}$};

\draw  [color={rgb, 255:red, 92; green, 138; blue, 168 }  ,draw opacity=1 ][fill={rgb, 255:red, 240; green, 240; blue, 255 }  ,fill opacity=1 ] (173,66) .. controls (173,63.24) and (175.24,61) .. (178,61) .. controls (180.76,61) and (183,63.24) .. (183,66) .. controls (183,68.76) and (180.76,71) .. (178,71) .. controls (175.24,71) and (173,68.76) .. (173,66) -- cycle ;

\draw (198,56) node  [font=\scriptsize] [align=left] {$\displaystyle u_{9}$};

\draw  (193,66) .. controls (193,63.24) and (195.24,61) .. (198,61) .. controls (200.76,61) and (203,63.24) .. (203,66) .. controls (203,68.76) and (200.76,71) .. (198,71) .. controls (195.24,71) and (193,68.76) .. (193,66) -- cycle ;

\draw (220,56) node  [font=\scriptsize] [align=left] {$\displaystyle u_{10}$};

\draw  [color={rgb, 255:red, 92; green, 138; blue, 168 }  ,draw opacity=1 ][fill={rgb, 255:red, 240; green, 240; blue, 255 }  ,fill opacity=1 ] (213,66) .. controls (213,63.24) and (215.24,61) .. (218,61) .. controls (220.76,61) and (223,63.24) .. (223,66) .. controls (223,68.76) and (220.76,71) .. (218,71) .. controls (215.24,71) and (213,68.76) .. (213,66) -- cycle ;

\draw [color={rgb, 255:red, 92; green, 138; blue, 168 }  ,draw opacity=1 ][fill={rgb, 255:red, 240; green, 240; blue, 255 }  ,fill opacity=1 ]  (-7,130) -- (3,130) -- (3,140) -- (-7,140) -- cycle ;
\draw (-1,148) node  [font=\scriptsize] [align=left] {$\displaystyle s_{0}$};

\draw   (13,130) -- (23,130) -- (23,140) -- (13,140) -- cycle ;
\draw (19,148) node  [font=\scriptsize] [align=left] {$\displaystyle s_{1}$};

\draw [color={rgb, 255:red, 92; green, 138; blue, 168 }  ,draw opacity=1 ][fill={rgb, 255:red, 240; green, 240; blue, 255 }  ,fill opacity=1 ]  (33,130) -- (43,130) -- (43,140) -- (33,140) -- cycle ;
\draw (39,148) node  [font=\scriptsize] [align=left] {$\displaystyle s_{2}$};

\draw   (53,130) -- (63,130) -- (63,140) -- (53,140) -- cycle ;
\draw (59,148) node  [font=\scriptsize] [align=left] {$\displaystyle s_{3}$};

\draw [color={rgb, 255:red, 237; green, 28; blue, 36 }  ,draw opacity=1 ][fill={rgb, 255:red, 255; green, 227; blue, 224 }  ,fill opacity=1 ]  (73,130) -- (83,130) -- (83,140) -- (73,140) -- cycle ;
\draw (79,148) node  [font=\scriptsize] [align=left] {$\displaystyle s_{4}$};

\draw  [color={rgb, 255:red, 92; green, 138; blue, 168 }  ,draw opacity=1 ][fill={rgb, 255:red, 240; green, 240; blue, 255 }  ,fill opacity=1 ] (93,130) -- (103,130) -- (103,140) -- (93,140) -- cycle ;
\draw (99,148) node  [font=\scriptsize] [align=left] {$\displaystyle s_{5}$};

\draw   (113,130) -- (123,130) -- (123,140) -- (113,140) -- cycle ;
\draw (119,148) node  [font=\scriptsize] [align=left] {$\displaystyle s_{6}$};

\draw  [color={rgb, 255:red, 92; green, 138; blue, 168 }  ,draw opacity=1 ][fill={rgb, 255:red, 240; green, 240; blue, 255 }  ,fill opacity=1 ] (133,130) -- (143,130) -- (143,140) -- (133,140) -- cycle ;
\draw (139,148) node  [font=\scriptsize] [align=left] {$\displaystyle s_{7}$};

\draw   (153,130) -- (163,130) -- (163,140) -- (153,140) -- cycle ;
\draw (159,148) node  [font=\scriptsize] [align=left] {$\displaystyle s_{8}$};

\draw [color={rgb, 255:red, 237; green, 28; blue, 36 }  ,draw opacity=1 ][fill={rgb, 255:red, 255; green, 227; blue, 224 }  ,fill opacity=1 ]  (173,130) -- (183,130) -- (183,140) -- (173,140) -- cycle ;
\draw (179,148) node  [font=\scriptsize] [align=left] {$\displaystyle s_{9}$};

\draw   (193,130) -- (203,130) -- (203,140) -- (193,140) -- cycle ;
\draw (201,148) node  [font=\scriptsize] [align=left] {$\displaystyle s_{10}$};

\draw   (213,130) -- (223,130) -- (223,140) -- (213,140) -- cycle ;
\draw (221,148) node  [font=\scriptsize] [align=left] {$\displaystyle s_{11}$};

\draw [color={rgb, 255:red, 237; green, 28; blue, 36 }  ,draw opacity=1 ][fill={rgb, 255:red, 255; green, 227; blue, 224 }  ,fill opacity=1 ]  (233,130) -- (243,130) -- (243,140) -- (233,140) -- cycle ;
\draw (241,148) node  [font=\scriptsize] [align=left] {$\displaystyle s_{12}$};

\end{tikzpicture}
     	}%
       \caption{Graphical representation of \ac{IRSA} access with $\popsize=11$ users ($\Ka=7$ active) and $\nslot=13$ slots. Light-blue circles are active users and blank circles idle users. Assuming $\T=2$, light-blue squares are resolvable slots, light-red squares unresolvable slots, and blank squares empty slots.}
    \label{fig:frame}
\end{figure}
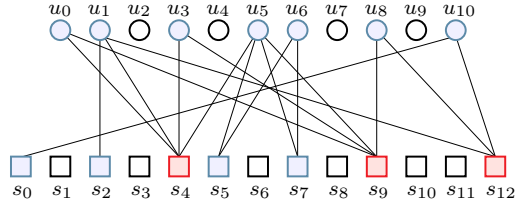

The status of the frame after transmissions can be described by means of a bipartite graph with $\popsize$ user nodes, one per transmitter, and $\nslot$ slot nodes, one per slot.
An edge connects user node $u_i$, $i\in\{0,\dots,\popsize-1\}$, to slot node $s_j$, $j\in\{0,\dots,\nslot-1\}$, if transmitter $i$ sent a packet replica in slot $j$.
An example with $\popsize = 11$ users ($u_0$, $\dots$, $u_{10}$) and $\nslot = 13$ slots ($s_0$, $\dots$, $s_{12}$) is depicted in Fig.~\ref{fig:frame}.
Users are represented by circles and slots by squares.
There are $\Ka= 7$ active transmitters, represented by colored circles; the number of transmitted packet replicas may change from user to user.
Blank circles represent transmitters that are idle in the current frame.

The efficiency of \ac{IRSA}, denoted by $\eta$, is defined as
\begin{align}\label{eq:csa_efficiency}
    \eta =\frac{1}{\bar{d}} = \frac{1}{\sum_{d=1}^{{d}_{\text{max}}} d \Lambda_d}= [\Lambda'(1)]^{-1} .
\end{align}
%
We also define the average load as 
$\load = \mathbb{E}[\Ka]/\nslot\,\mathrm{[packets/slot]}$.
Note that the average number of packet replicas per slot is thus given by $\load / \eta$.

\subsection{Decoding of Slotted Random Access Codes}

The receiver for slotted random access codes has no knowledge of the number $\Ka$ of active users and is based on per-slot processing of the \ac{GMAC} output symbols, which reduces implementation complexity as for the simple  \ac{SA} scheme. This means that there is a decoder that attempts  to recover PHY layer codewords, and related encoded messages, slot-by-slot. In an equivalent view, the receiver may be thought as equipped with $\nslot$ PHY layer (or slot) decoding processors, one for each slot. These processors can cooperate with each other in  an iterative \ac{SIC}-based decoding procedure that performs MAC layer decoding. 

The PHY layer  decoding processor (or slot decoder) may be designed to decode up to $T$  codewords transmitted in the same slot, with $T$ fixed. The simplest decoder has $T=1$, as in SA schemes. When $T>1$ we say that the receiver has \ac{MPR} capability. 
We model the slot decoder as follows. It includes an estimator of the number $t_i$ of transmissions in  slot $i$ and a decoder for $t_i$ PHY layer codewords, with $t_i\leq T$.   We  assume to have a decoder with perfect $t_i$ estimation and ideal error detection capability.
It provides at its output a list of correctly detected messages and a binary flag $f_i$ indicating if a slot is resolved, i.e., completely detected, or if it is unresolved, i.e. there is at least one undetected PHY layer codeword.
More specifically, a slot is resolved if $t_i=0$ or if $0<t_i\leq T$ and the list of correctly detected messages has $t_i$ elements; a slot is unresolved if $t_i>T$ or if $0<t_i\leq T$ and the list of correctly detected messages has less than $t_i$ elements.
In practical decoder implementations the estimator of $t_i$ is not perfect and has to be carefully designed to keep the probability of estimation error negligible. {A discussion on practical estimation methods and an evaluation of the effects of estimation errors is provided in Appendix \ref{appendix:t_estimation}.

In the \ac{IRSA}-based schemes the MAC layer decoder attempts the recovery of all messages using an iterative \ac{SIC}-based procedure.
At the beginning, the receiver tries to decode the signals received in all  slots with the slot decoder. It marks each slot as resolved, or unresolved, according to slot decoder outcomes. 
Then, an iterative process starts where each iteration may be described as follows, having in mind that each decoded replica carries the information about the number and the positions of the other replicas.
 For each PHY layer codeword correctly decoded in a resolved non-empty slot, the decoded messages are re-encoded and the corresponding signals of every replica of the decoded messages are subtracted, as interference, from the signals received in the slots where the replicas are located. 
 Again, the slot decoder tries to decode the signals received in all the still unresolved slots. The slots where decoding is successful are marked as resolved. If at least one new non-empty resolved slot is found, then a new iteration is started. Otherwise, 
 the iterative procedure stops. 
 If all  slots are resolved, the decoding procedure is successful, under the assumption of ideal error detection and perfect estimation of $t_i$, and the output list of the decoder includes all the transmitted messages.
 If one or more slots remain unresolved, the output list of the decoder collects all the messages decoded in the resolved slots and all the messages, if any, correctly detected by slot-decoders in unresolved slots. In this case the decoding procedure is in general unsuccessful, as one or more transmitted messages may not be present in the output list.
 
The iterative \ac{SIC}-based decoding can be described as a message-passing algorithm on the bipartite graph, as in \cite{liva2011:irsa}.
With reference again to Fig.~\ref{fig:frame}, assume $\T=2$ and that in all slots with $t_i\leq2$ all transmitted packets are correctly decoded.
Since slots $s_0$, $s_2$, $s_5$, and $s_7$, are resolvable, packet replicas transmitted by users $u_1$, $u_5$, $u_6$, and $u_{10}$ in these slots are correctly received at the first iteration.
Interference cancellation in the first iteration makes slots $s_4$ and $s_{12}$ resolvable.
In the second iteration, at least one packet replica transmitted by users $u_0$, $u_3$, and $u_8$ is correctly received in these slots.
Interference cancellation in the second iteration makes all slots resolved and the procedure terminates successfully.

The IRSA protocols have been investigated  in \cite{liva2011:irsa}, \cite{paolini2015:csa}  with $T=1$, and in \cite{Ghanbarinejad2013:Irregular,stefanovic2018:multipacket} with $T>1$,
over the simple noiseless collision channel under the following additional assumptions:
       
\emph{Assumption~1}: If there are at most $\T$ arrivals in a slot, a condition detected by the receiver, then all of transmitted codewords are correctly received with zero error probability; if there are more than $\T$ arrivals in a slot, then none of these transmitted codewords can be decoded.

\emph{Assumption~2}: Interference subtraction across slots is ideal.

In this work Assumption~1 is removed, as the noiseless collision channel is replaced by the \ac{GMAC}, and zero error probability detection of PHY layer codewords is not guaranteed. We still
preserve Assumption~2. A discussion on the reliability of this assumption can be found in \cite{liva2011:irsa}. Further evaluations and discussion on the effect of imperfect interference cancellation are included in Appendix \ref{appendix:imperfect_sic}.

\section{MAC Layer Code Performance}\label{Sec:MAC_performance}

In this section we analyze the performance of MAC layer code. We denote with $E_\text{M}$ the error event where a given message $m$ encoded with PHY layer code and transmitted in a slot of the frame is not decoded in the slot, i.e., it does not appear in the output list of the slot decoder.  We also denote with $E_\text{D}$ the error event where one or more  messages encoded with the PHY layer code and transmitted in a slot of the frame are not decoded in the slot, i.e., the output list of the slot decoder is different in some way from the list of transmitted messages. 
With other words, $E_\text{M}$ is the event that a single message $m$ transmitted in a given slot is not correctly decoded by the PHY layer decoder in the considered slot. In contrast, the event $E_\text{D}$ is the event that the PHY layer decoder sets the flag $f_i$ to zero in the considered slot, i.e., it was not able to decode all the messages in the slot. Clearly, if $E_\text{M}$ holds, then $E_\text{D}$ must hold as well, but the vice versa is not necessarily true.
The occurrence of these error events is strictly dependent on the number of transmissions, i.e., colliding codewords, in the slot. Such events occur if the number of colliding codewords exceeds $T$, or if noise induces a decoding error.
The probabilities of message and decoding error, conditioned to $t$ transmissions  in the slot, are denoted as $P_{\text{M}| t} = \mathbb{P}(E_\text{M} | t)$ and $P_{\text{E}| t} = \mathbb{P}(E_\text{D} | t)$, respectively. 
They depend on PHY layer coding  and will be derived in the next section for each specific PHY layer coding option.
Note that the two error events are the same when $t=1$, i.e., $P_{\text{E}| 1} = P_{\text{M}| 1}$, while in general we have $P_{\text{E}| t} \geq P_{\text{M}| t}$. It is also expected that $P_{\text{E}| t+1} \geq {P_{\text{E}| t}}$, which will be assumed always true throughout the paper.

In the next subsections we will derive the performance parameters of the MAC layer coding schemes considered in this paper as a function of error probabilities $P_{\text{E}| t}$ and $P_{\text{M}| t}$.
The first subsection summarizes the well-known information for frame-based \ac{SA} schemes. The second subsection provides the background information available from the literature for the analysis of \ac{IRSA} schemes over noiseless channels. The third and fourth subsections provide the new results for  \ac{IRSA} schemes over \ac{GMAC}.

\subsection{Frame-Based SA with and without \ac{MPR} Capability}

In the SA scheme the encoded message of each active transmitter is transmitted only once, in one slot of the frame. The message transmitted in a slot is lost if the event $E_\text{M}$ occurs. Let us first consider a fixed number $\Ka$ of active transmitters in the frame. The probability of message loss, i.e. error in the transmission, commonly denoted as packet loss probability, is evaluated as
\begin{equation}
P_L=\sum_{t=1}^{\Ka} {P_c}(t-1) P_{\text{M}| t}
\end{equation}
being ${P_c}(t-1)$ the probability of other $t-1$ codewords in the slot colliding with the reference codeword. Since in the frame there are $\Ka$ active devices and each one selects one slot with uniform probability $1/\nslot$, the number of transmissions in a slot is a binomial random variable. Thus, we have
\begin{align}
{P_c}(t)  & =  \mathbb{P}\big( \text{Bin}(\Ka-1, 1/\nslot)=t \big) \nonumber \\
& 
=  \binom{\Ka-1}{t} \left(\frac{1}{\nslot}\right)^{t} \left( 1-\frac{1}{\nslot}\right)^{\Ka-1-t} 
\end{align}
for $0\leq t<\Ka$, and $P_c(t)=0$ for $t\geq \Ka$.

Due to  $P_{\text{M}| t} = 1$ for $t>T$, with simple manipulations the packet loss probability becomes
\begin{equation}
P_L=1- \sum_{t=1}^T {P_c}(t-1)(1-P_{\text{M}| t}) \, .
\end{equation}
In the special case where $P_{\text{M}| t}=P_{\text{M}}$, $\forall t\in [1,T]$, i.e. it does not depend on $t$,  the packet loss probability simply becomes $P_L=1- (1-P_{\text{M}})\sum_{t=1}^T {P_c}(t-1)$. 
When the SA scheme does not include MPR capabilty, i.e. $T=1$, it further simplifies in $P_L=1- (1- P_{\text{M}}) (1-1/\nslot)^{\Ka-1}$.

Note that in the asymptotic scenario where $\Ka,\nslot \rightarrow \infty$ with constant $\Ka/\nslot=\beta$, the binomial distribution converges to a Poisson distribution with parameter $\beta$, leading to ${P_c}(t) = e^{-\beta} \beta^t /t!$. The Poisson approximation becomes tight for moderately large values of $\Ka$ and $\nslot$. The packet loss probability thus  becomes
\begin{equation}
\label{eq:slotted_aloha}
P_{L\infty}=1- \sum_{t=1}^T e^{-\beta} \frac{\beta^{t-1}}{(t-1)!}(1-P_{\text{M}| t}) \, .
\end{equation}

We consider now $\Ka$ as the result of a random activation process with probability $\pi$. In the asymptotic scenario where $K,\nslot \rightarrow \infty$ with constant density of devices per slot, $K/\nslot=\alpha$, the probability distribution of $\Ka$ tends to concentrate around $\mathbb{E}[\Ka]=\pi K$. Also in this case the Poisson approximation can be used to describe the number of transmissions per slot conditioned to a given number $\Ka$ of active devices, but in the asymptotic scenario the parameter of Poisson distribution becomes  $\beta=\mathbb{E}[\Ka]/\nslot=\pi \alpha$, when $K,\nslot \rightarrow \infty$, which is independent of $\Ka$. Therefore, the expression of the packet loss probability in \eqref{eq:slotted_aloha} also holds in the case of random activation of the users, and the parameter $\beta$ becomes the average load of the system, i.e., $\beta=\avgG$, according to the definition given at the end of Section~\ref{Sec:model}-A.
We finally note that this analysis provides the same results in both settings, namely with $\Ka$ fixed variable and $\Ka$ random variable.

\subsection{IRSA-based Random Access Codes over a Noiseless Channel}

In this subsection, we review the performance framework of IRSA-based random access schemes  in the basic scenario where the multiple-access channel is a simple noiseless collision channel.
In this case, as already pointed out at the end of Section~II, the error events $E_\text{M}$ and $E_\text{D}$ at the PHY layer occur only if the number of colliding codewords exceeds $T$, which is the MPR capability enabled by the PHY layer coding, and these error events are fully detected. The content of the section provides  background notations and methods that will be 
further developed and exploited in the remaining part of the paper. 

The \ac{SIC}-based procedure in the IRSA scheme has been analyzed in the asymptotic setting $\nslot \rightarrow \infty$, $\mathbb{E}[\Ka] \rightarrow \infty$, with constant user density $\alpha$, in \cite{liva2011:irsa}, \cite{paolini2015:csa}  for $T=1$, and in \cite{Ghanbarinejad2013:Irregular,stefanovic2018:multipacket} for $T>1$, under the additional assumption of ideal interference subtraction across slots.
The asymptotic analysis has been carried out by considering independent and memoryless device activation processes with  activation probability $\pactive$. This analysis also holds in the setting where $\Ka$ is a fixed variable and $\mathbb{E}[\Ka] =\Ka$, as for frame-based \ac{SA}.

According to the description of the \ac{SIC} process as a message passing algorithm on the bipartite graph, let $\ell$ be the \ac{SIC} iteration index, $p_{\ell}$ be  the probability that the generic edge is connected to a slot node not yet resolved at the end of iteration $\ell$, and $q_{\ell}$ be the probability that the generic edge is connected to a user node not yet detected at the end of iteration $\ell$.
In the asymptotic setting, the evolution of $p_{\ell}$ over \ac{SIC} iterations can be described through a recursion $p_{\ell} = f(p_{\ell-1})$, where the structure of the function $f(\cdot)$ depends on the considered channel. This recursion is analogous to density evolution in the framework of codes on sparse graphs \cite{Richardson2001:capacity,Richardson2001:design}.

The main result for the asymptotic analysis of  \ac{IRSA} schemes with  $\TMPR$-\ac{MPR} 
is found in \cite{paolini2015:csa,stefanovic2018:multipacket} and is summarized by the following lemma.
\begin{lemma}
\label{th:density_evolution_mpr_ch}
Let $\nslot \rightarrow \infty$ and $\mathbb{E}[\Ka] \rightarrow \infty$ for constant $\avgG=\mathbb{E}[\Ka]/\nslot=\pi \alpha$. 
Let $\csaefficiency=1/\bar{d}$ be the efficiency of the \ac{IRSA} protocol as defined in \eqref{eq:csa_efficiency}.
Then, at the $\ell$-th iteration of the \ac{SIC} process we have
\begingroup
\allowdisplaybreaks
\begin{align}\label{eq:density_evolution_mpr_ch}
p_{\ell} &= 1 - \exp\left( -{ \avgG} \bar{d}q_{\ell} \right) \sum_{t = 0}^{\TMPR - 1} \frac{ 1 }{t!} \left( { \avgG} \bar{d}q_{\ell} \right)^{t} = f_s(q_\ell)\\
q_{\ell} &=(1/\bar{d}) \sum_{d} \Lambda_d \, d \, p_{\ell-1}^{d-1} = \Lambda'(p_{\ell-1})/\bar{d}=f_b(p_{\ell-1})
\label{eq:density_evolution_mpr_ch_2}
\end{align}
\endgroup
where the starting point of the recursion is $p_0=1-\exp ( - \avgG \bar{d} ) \sum_{t = 0}^{\TMPR - 1} \left(  \avgG \bar{d} \right)^{t} / t!$.
\end{lemma}

Note that \eqref{eq:density_evolution_mpr_ch} reduces to $p_{\ell} = 1 - \exp \big ( -{ \avgG} \bar{d} q_{\ell} \big ) = f_s(q_\ell)$ for the basic case of $\TMPR=1$ \cite{paolini2015:csa}. 
Note also that the quantity $p_{\ell}^d$ represents the probability that the packet of a device which  transmitted $d$ packet replicas, is not yet correctly decoded after $\ell$ \ac{SIC} iterations.

Next, with reference to recursion \eqref{eq:density_evolution_mpr_ch}-\eqref{eq:density_evolution_mpr_ch_2}, it is possible to prove that, for \ac{IRSA} protocols with $\Lambda_1=0$, there exists a value $\avgG^{\star}$ of the average load such that: (i) If $\avgG < \avgG^{\star}$ then $p_{\ell}$ tends to zero as $\ell$ tends to infinity; (ii) If $\avgG > \avgG^{\star}$ then $p_{\ell}$ converges to a value that is bounded away from zero as $\ell$ tends to infinity. 
The condition $p_{\ell} \rightarrow 0$ corresponds to a vanishing packet loss probability in the asymptotic setting.
The value of $\avgG^{\star}$ depends on the \ac{PGF} 
$\Lambda(x)$, or equivalently on the IRSA distribution $\bm{\Lambda}$, and on the value of $T$, and is called the \emph{asymptotic threshold} of the \ac{IRSA} protocol over the slot-synchronous collision channel. 
Formally, we can write
\begin{align}\label{eq:coll_ch_asym_th_G}
\avgG^{\star}( \bm{\Lambda},T) = \sup \{ \avgG  > 0 : p_{\ell} \rightarrow 0 \text{ as } \ell \rightarrow \infty \} .
\end{align}
We can interpret  $\avgG^{\star}(\bm{\Lambda},T)$ as the largest traffic (in data packets per slot) that an \ac{IRSA} scheme with distribution $\bm{\Lambda}$, and $\Lambda_1=0$, can support reliably, i.e., with a vanishing packet loss probability, over the considered channel, when both $\mathbb{E}[\Ka]$ and $\nslot$ are large.  
The value of $\avgG^{\star}$ can be found by testing the convergence of the recursion \eqref{eq:density_evolution_mpr_ch}-\eqref{eq:density_evolution_mpr_ch_2} for different values of $\avgG$.

According to the asymptotic analysis, when the load threshold is known for an \ac{IRSA} protocol, a simple approximated model for determining the packet loss probability $ P_L$ can be formulated as $ P_L\approx P_{L\infty}$, where 
\begin{align}
P_{L\infty} =0 \; \; \text{if } \avgG < \avgG^{\star}
\end{align}
being $P_{L\infty}\gg0$ otherwise. 
This approximation becomes asymptotically tight for $\avgG$ below the threshold. According to this approximation, the generic requirement $P_L<\epsilon$ translates into the requirement $\avgG < \avgG^{\star}$, if $\epsilon$ is sufficiently small, which is an upper bound of the actual requirement on $\avgG$ in non-asymptotic conditions.

A fundamental limit on the asymptotic load threshold of \ac{IRSA}-type protocols can be also found as 
an upper bound that applies to \emph{any} \ac{IRSA} scheme of some efficiency $\eta$ ($\leq 0.5$, when $\Lambda_1=0$), regardless of the specific configuration employed.
It is expressed by the following lemma \cite{stefanovic2018:multipacket}.
\begin{lemma} 
\label{th:threshold_bound}
For any \ac{IRSA} scheme with $\TMPR$-\ac{MPR} capability, distribution $\bm{\Lambda}$ and $\Lambda_1=0$, the load threshold $\avgG^{\star}$ and the efficiency $\eta$ of the protocol must fulfill the inequality
\begin{align}\label{eq:G_threshold_bound}
    \frac{\avgG^{\star}}{\TMPR} \leq 1 - \frac{1}{\TMPR} \, \exp \left( - \frac{\avgG^{\star}}{\eta} \right) \sum_{k=0}^{\TMPR-1} \frac{\TMPR - k}{k!} \left( \frac{\avgG^{\star}}{\eta} \right)^k \, . 
\end{align}
\end{lemma}

Lemma~\ref{th:threshold_bound} also applies to more general \ac{CSA} schemes with $\eta\leq 1$, and in the special case of $\TMPR=1$ the result was originally developed in \cite{paolini2015:csa} with the simple form: $\avgG^{\star} \leq 1 - \exp \left( -{ \avgG^{\star}}/{\eta} \right) $.
It is relatively easy to show that, for any integer $\TMPR \geq 1$ and any $0 < \eta \leq 1$, inequality \eqref{eq:G_threshold_bound} sets an implicit upper bound on the load threshold $\avgG^{\star}$, which can be expressed in the form
\begin{align}\label{eq:G_threshold_bound2}
    \avgG^{\star} \leq \mathbb{G}(\eta,\TMPR)
\end{align}
where $\mathbb{G}(\eta,\TMPR)$ is the unique positive real root of
\begin{align}
     \frac{G}{\TMPR} = 1 - \frac{1}{\TMPR} \, \exp \left( - \frac{G}{\eta} \right) \sum_{k=0}^{\TMPR-1} \frac{\TMPR - k}{k!} \left( \frac{G}{\eta} \right)^k \, .
\end{align}
Notably, \eqref{eq:G_threshold_bound2} is a \emph{converse bound} which defines a region for the values of $\avgG$ where packet loss probability does not converge to 0 as $\ell \rightarrow \infty$.
The boundary of this region is $\mathbb{G}(\eta,\TMPR)$ which is referred here to as the \emph{convergence boundary} of \ac{IRSA} protocol over a collision channel.
Achievability of $\mathbb{G}(\eta,\TMPR)$ for any rate $\eta$ has never been proved to date, although some optimized actual schemes can approach the boundary quite closely at specific efficiencies.

\subsection{IRSA-based Random Access over \ac{GMAC}: Asymptotic Threshold}

Here, we extend the results of the previous subsection and obtain novel results for the \ac{GMAC}, where the error probability in the detection of PYH-layer codewords is not zero when  their number in a slot is less than or equal to $T$. 
As for the noiseless channel, the results will be also valid in the setting where $\Ka$ is a fixed variable and $\mathbb{E}[\Ka] =\Ka$.
The first result for the asymptotic analysis of  \ac{IRSA}-based  schemes with  $\TMPR$-\ac{MPR} capability  is the following.

\begin{theorem}\label{th:density_evolution_mpr}
Let $\nslot \rightarrow \infty$ and $\mathbb{E}[\Ka] \rightarrow \infty$ for constant $\avgG=\mathbb{E}[\Ka]/\nslot=\pi \alpha$. 
Let $\csaefficiency=1/\bar{d}$ be the efficiency of the \ac{IRSA} protocol as defined in \eqref{eq:csa_efficiency}. 
Then, at the $\ell$-th iteration of the \ac{SIC} process we have
\begingroup
\allowdisplaybreaks
\begin{align}\label{eq:density_evolution_mpr}
p_{\ell} &=1-\exp (-{ \avgG} \bar{d}q_{\ell}) \sum_{t = 1}^{\TMPR } (1 - P_{\text{E}| t}) \frac{\left({\avgG} \bar{d}q_{\ell}\right)^{t-1}}{(t-1)!} =f_s(q_\ell) \;\;\; \\
q_{\ell} &=(1/\bar{d}) \sum_{d} \Lambda_d \, d \, p_{\ell-1}^{d-1} = \Lambda'(p_{\ell-1})/\bar{d}=f_b(p_{\ell-1}) \label{eq:density_evolution_mprq}
\end{align}
\endgroup
 where the starting point of the recursion is $p_0=1-\exp ( - \avgG \bar{d} ) \sum_{t = 1}^{\TMPR } (1 - P_{\text{E}| t})\left(  \avgG \bar{d} \right)^{t-1} / (t-1)!$.\\
In the special case with $T=1$, i.e., no \ac{MPR}, equation \eqref{eq:density_evolution_mpr} at the $\ell$-th iteration becomes
\begin{align}\label{eq:density_evolution}
p_{\ell} &= 1- (1 - P_{\text{E}| 1}) \exp \big ( -{\avgG} \bar{d} q_{\ell} \big )
\end{align}
and the starting point of the recursion becomes $p_0=1- (1-P_{\text{E}| 1})\exp ( -\avgG \bar{d} )$. 
\end{theorem}
\begin{IEEEproof} See Appendix A
\end{IEEEproof}

It is important now to highlight the differences in the asymptotic behavior of IRSA in presence of non zero probability of erroneous detection of PHY layer codewords, with respect to the case with zero error probability. 
Let us first consider function $f_s(q)$ in \eqref{eq:density_evolution_mpr}. We can easily note that $f_s(0)=P_{\text{E}|1}$. Moreover,  by looking at its derivative $f'_s(q)$, given by
\begin{align}\label{eq:density_evolution_deriv}
f'_s(q)&= { \avgG} \bar{d}\exp (-{ \avgG} \bar{d}q) \sum_{t = 1}^{\TMPR } (P_{\text{E}| t+1} -P_{\text{E}| t}) \frac{\left({\avgG} \bar{d}q_{\ell}\right)^{t-1}}{(t-1)!}
\end{align}
with $P_{\text{E}| T+1}=1$, we also note that $f'_s(q)\geq 0$  in the normal setting $P_{\text{E}|t+1}\geq P_{\text{E}|t}$. The following result is then also obtained, which illustrates the effects of non-zero error probabilities.
\begin{corollary}\label{cor:pl_limit}
In the iterative \ac{SIC} process, $p_{\ell}$ can only converge to a value  $p_{\infty}\geq P_{\text{E}|1} \text{ as } \ell \rightarrow \infty$, i.e., there must be a nonzero probability, lower-bounded by $P_{\text{E}|1}$, that slot nodes remain unresolved at the end of the process.
When $\avgG \rightarrow 0$, then $p_{\infty} \rightarrow P_{\text{E}|1}$.
A residual nonzero value for $p_\infty$ translates into a non zero value for packet loss probability given by
\begin{align}\label{eq:PL}
P_{L\infty}=\sum_d \Lambda_d p_{\infty}^d \geq \sum_d \Lambda_d P_{\text{E}|1}^d =P_{L\min} \, .
\end{align}
\end{corollary}

\begin{figure}[t]
\centering
\includegraphics[width=\columnwidth]{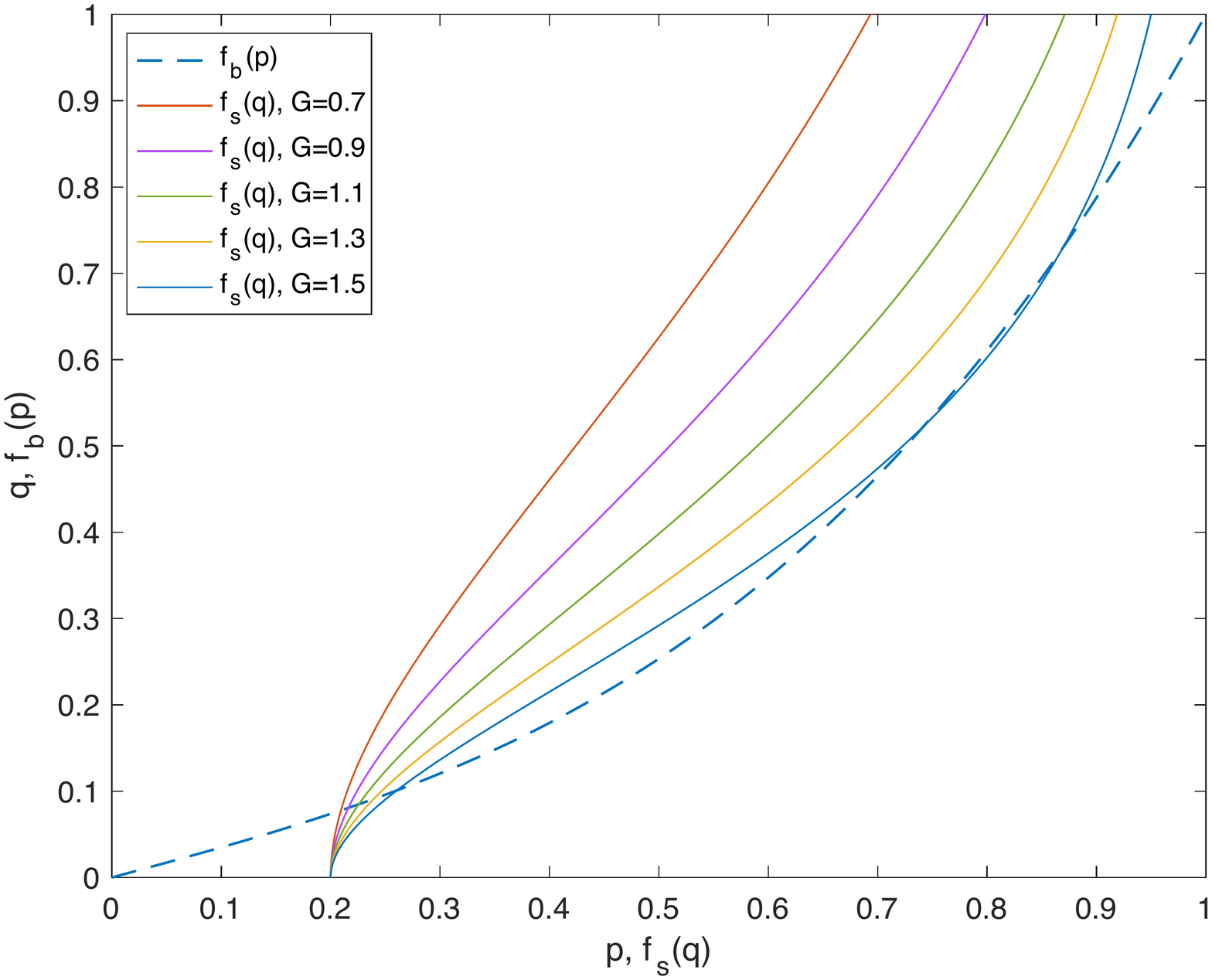}
      \caption{EXIT chart for \ac{IRSA} protocol obtained with $d_{\max}=4$, $\mathbf{\Lambda}=\{0, 0.5102, 0, 0.4898\}$, $\TMPR=2$, $P_{\text{E}|1}=P_{\text{E}|2}=0.2$ and different values of $\avgG$.}
    \label{fig:exit}
\end{figure}

The next example is proposed to better understand the asymptotic solution of the  \ac{SIC} process.

\begin{figure}[t]
\centering
\includegraphics[width=\columnwidth]{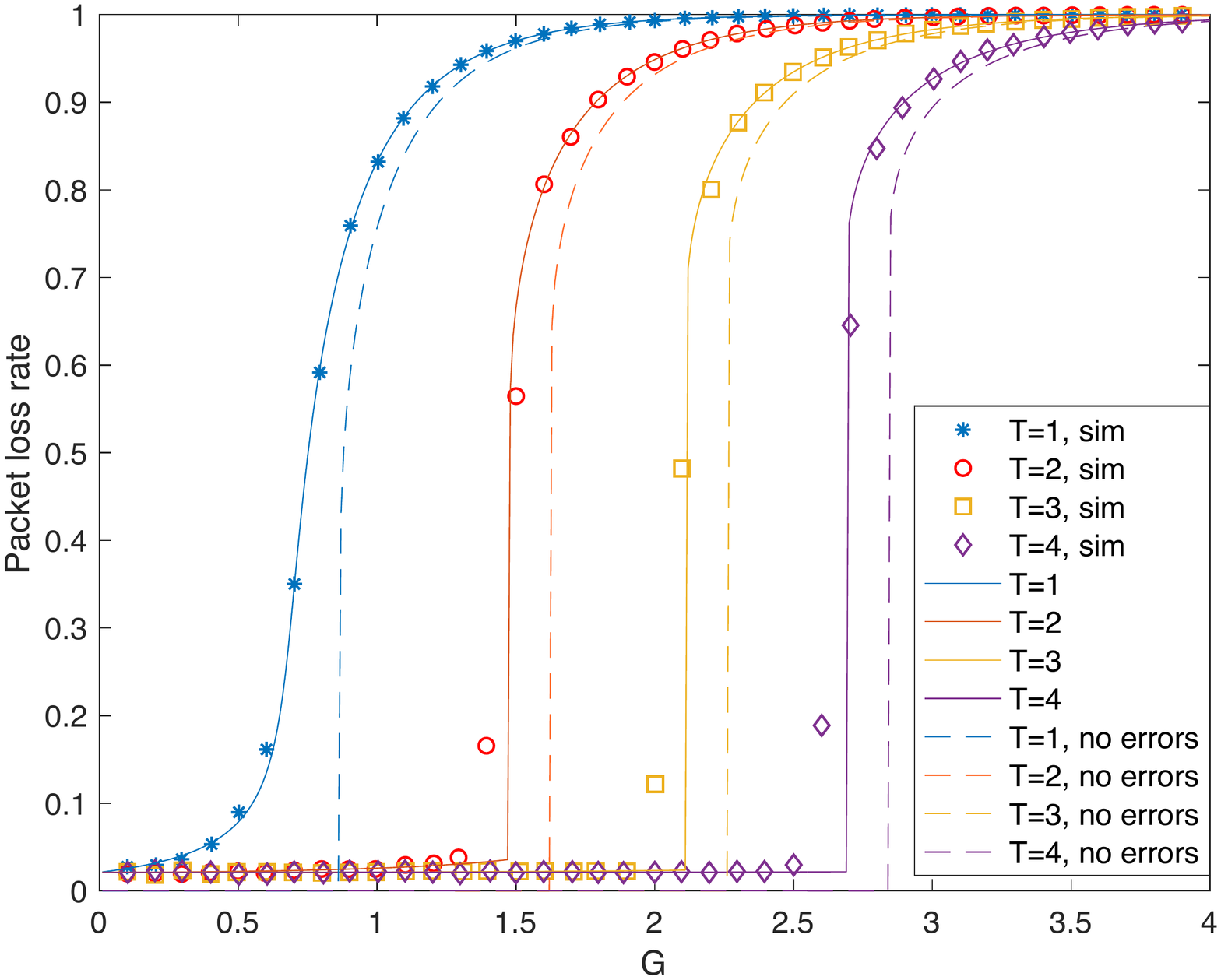}
      \caption{Packet loss probability versus the average load $\avgG$ for an \ac{IRSA} protocol with $d_{\max}=4$, $\mathbf{\Lambda}=\{0, 0.5102, 0, 0.4898\}$. Solid: asymptotic packet loss probability for $P_{\text{E}|t}=0.2, \;\forall t \leq \TMPR$. Dashed: asymptotic packet loss probability for $P_{\text{E}|t}=0, \;\forall t \leq \TMPR$. Markers: Finite frame size Monte Carlo simulation for $\nslot=400$ and $P_{\text{E}|t}=0.2, \;\forall t \leq \TMPR$.} 
    \label{fig:PLev}
\end{figure}

\begin{example}\label{example:EXIT}
Let us consider an IRSA protocol with $d_{\max}=4$, $\mathbf{\Lambda}=\{0, 0.5102, 0, 0.4898\}$, and $\eta=1/\bar{d}=0.3356$ (proposed in \cite{liva2011:irsa}, Table~I, for the noiseless channel). 
The density evolution recursion, given by \eqref{eq:density_evolution_mpr} and \eqref{eq:density_evolution_mprq}, can be graphically visualized through an EXIT chart that includes the functions $q=f_b(p)$ and $p=f_s(q)$ in the $(p,q)$ plane, as shown in Fig.~\ref{fig:exit}. 
The iteration between the two functions, starting from the  point $(p_0,1)$, aims at finding the first stable crossing point between them, $(p_{\infty},q_{\infty})$, wishing that this point results as close as possible to $(0,0)$.
The function $q=f_b(p)$ depends on the IRSA protocol only, i.e., the MAC code, whereas the function $p=f_s(q)$ depends on the error probability vector and on the average load $\avgG$. The chart is plotted for  $\TMPR=2$, $P_{\text{E}|1}=P_{\text{E}|2}=0.2$ and different values of $\avgG$. 
With reference to Fig.~\ref{fig:exit}, we notice the existence of a value of the load, $\avgG_0 \approx 1.48$ (determined numerically), such that: for $\avgG < \avgG_0$ there is only one crossing point with $p$ close to $P_{\text{E}|1}$; 
for $\avgG>\avgG_0$ another stable crossing point appears with $p$ deviating substantially from $P_{\text{E}|1}$.
The example shows a pronounced ``on-off'' threshold effect in terms of $\avgG$.
Below threshold the system achieves a value $p_{\infty}$ close to its lower bound $P_{\text{E}|1}$, while above threshold $p_\infty$ jumps away from $P_{\text{E}|1}$.
%
Fig.~\ref{fig:PLev} displays the asymptotic packet loss probability (solid curves) evaluated as a function of $\avgG$ for the same \ac{IRSA} protocol with $\TMPR \in \{1,2,3,4\}$ and $P_{\text{E}|t}=0.2$ for all $t \leq \TMPR$.
We can note that the theoretically minimum asymptotic packet loss probability, $P_{L\min} = 0.0212$, is closely approached for a wide range of $\avgG$ values when $\TMPR>1$, making the above-mentioned ``on-off'' evident, similar to that typical of a collision channel with $P_{\text{E}|t}=0$ for all $t \leq \TMPR$ (dashed curves). 
With respect to this latter case, however, the obtained threshold values are smaller as it can be expected.
We also note that the threshold effect is much less pronounced for  $\TMPR=1$ and $P_{\text{E}|t}>0$, in which case the packet loss probability increases slowly from its minimum, as $G$ increases.
As a sanity check, Fig.~\ref{fig:PLev} also shows the corresponding packet loss rates obtained by Monte Carlo simulation\footnote{In the simulation the values of $P_{\text{E}|t}$, $t>0$, are fixed as input parameters and the decoding does not take into account the inter-dependencies between error events in a slot occurring in different interference cancellation rounds.} for $\nslot=400$ slots (markers), highlighting a very good match with the asymptotic curves both in the error floor and load threshold regions.
\end{example}

From the previous example we can also conclude that, in presence of PHY layer slot decoding errors, a good IRSA configuration should determine the existence of no more than one crossing point in the EXIT chart for a wide range of  $\avgG$ values (as for  IRSA over the collision channel) and a value of $p$ in the crossing point as close as possible to $P_{\text{E}|1}$.
We now try to get insight on how to obtain a crossing point with $p$ very close to $P_{\text{E}|1}$. We assume to work in a regime with small values of slot decoding error probabilities and to focus our analysis in the EXIT chart region where $p$ and $q$ are small. In this region, the functions $f_b(p)$ and $f_s(q)$ can be approximated~as
\begingroup
\allowdisplaybreaks
\begin{align}
f_b(p)& \approx \tilde{f}_b(p)=\left\{ \begin{array}{ll}
  \eta \Lambda_{d_{\min}}d_{\min} p^{d_{\min}-1}  &\text{if} \;\; d_{\min}>1  \\
  \eta (\Lambda_1 +2\Lambda_2p)  & \text{if} \;\; d_{\min}=1   
\end{array} \right. \label{eq:ftilde_b}\\
f_s(q)& \approx  \tilde{f}_s(q)=  P_{\text{E}|1}+f'_s(0) q \label{eq:ftilde_s}
\end{align}
\endgroup
where $d_{\min}$ is the minimum number of transmitted replicas and $f'_s(0)$, obtained from \eqref{eq:density_evolution_deriv}, is $(1-P_{\text{E}|1})\avgG /\eta$ if $\TMPR=1$ or $(P_{\text{E}|2}-P_{\text{E}|1})\avgG /\eta$ if $\TMPR>1$.
Note that \eqref{eq:ftilde_b} and \eqref{eq:ftilde_s} are analogous to the well-know stability condition for iterative decoders.
The intersection of the two functions, $q=f_b(p)$ and $p=f_s(q)$, has coordinate $p$ close to $P_{\text{E}|1}$ if at least one of the two is flat close to zero, i.e., if at least one of the two conditions holds: 
\begingroup
\allowdisplaybreaks
\begin{align}
\text{a)} \quad & d_{\min} \geq 3 \\
\text{b)} \quad & d_{\min}=2, \; \TMPR \geq 2, \; P_{\text{E}|2}=P_{\text{E}|1}
\end{align}
\endgroup
where $P_{\text{E}|2}=P_{\text{E}|1}$ can be relaxed to $P_{\text{E}|2}$ close enough to $P_{\text{E}|1}$ as discussed in Theorem~\ref{cor:suf_conditions} below.
This can be easily understood by looking at Fig.~\ref{fig:exit} where \ac{IRSA} protocol settings are $P_{\text{E}|1}=P_{\text{E}|2}$ and $d_{\min} =2$. This also explains the  good behavior of packet loss probability shown in Fig.~\ref{fig:PLev} for \ac{IRSA} with $T>1$ with respect to $T=1$. The last result is formally presented as follows. 
\begin{theorem}\label{cor:suf_conditions}
Let $P_{\text{E}|1} \approx 0$. Then, there exists $\avgG_0>0$ such that for all $\avgG<\avgG_0$ any of the two conditions:
\begin{align}
\text{a)} \quad & d_{\min} \geq 3 \\
\text{b)} \quad & d_{\min} =2 , \; \TMPR \geq 2 , \; P_{\text{E}|2}-P_{\text{E}|1} \ll \frac{1}{2 \Lambda_2 \avgG_0}  
\label{eq:pmin_conditions}
\end{align}
is sufficient for the  iterative \ac{SIC} process to converge to a value $p_{\infty}= P_{\text{E}|1} + O(P_{\text{E}|1}^2)$.
\end{theorem}
\begin{IEEEproof} See Appendix~\ref{appendix:proof_th2}.
\end{IEEEproof}

\begin{remark}
The value of $\avgG_0$ mentioned in the statement of Theorem~\ref{cor:suf_conditions} depends on the parameters $\bm{\Lambda}$, $\TMPR$, and $\mathbf{P}_{\text{E}}$.
It numerically coincides with the value of $\avgG$ for which a second stable crossing point appears in the EXIT chart, mentioned in Example~\ref{example:EXIT}.
\end{remark}

\begin{figure}[t]
\centering
\includegraphics[width=\columnwidth]{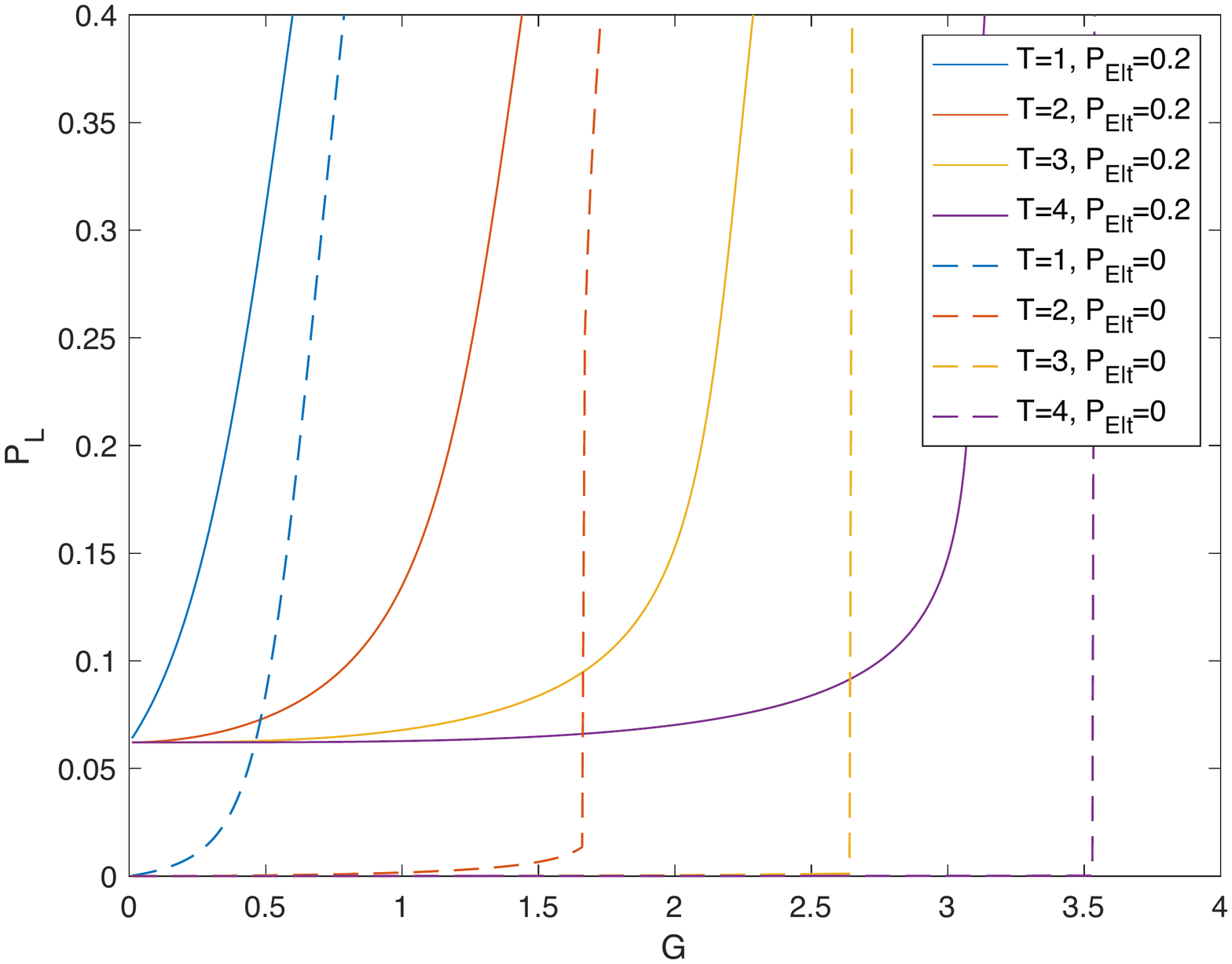}
      \caption{Asymptotic packet loss probability $P_L$ versus the average load $\avgG$ for an \ac{IRSA} protocol with $d_{\min}=1$, $d_{\max}=2$, $\mathbf{\Lambda}=\{0.1382, 0.8618\}$. Solid: noisy channel with  $P_{\text{E}|t}=0.2, \;\forall t \leq \TMPR$. Dashed: noiseless channel with $P_{\text{E}|t}=0, \;\forall t \leq \TMPR$.} 
    \label{fig:PLev_d1}
\end{figure}

We finally note that, in the case where $d_{\min}=1$, it is not always guaranteed for the  iterative \ac{SIC} process  to converge to a value $p_{\infty}= P_{\text{E}|1} + O(P_{\text{E}|1}^2)$. 
In fact, since $f_b(p)\geq \eta\Lambda_1$ for all $p\in[0,1]$, this condition precludes the possibility to have  $p_\infty$ close to $P_{\text{E}|1}$, unless $\avgG$ is tending to zero. 
The behavior  of an \ac{IRSA} protocol with $d_{\min}=1$, i.e., the one with  with $d_{\max}=2$, $\mathbf{\Lambda}=\{0.1382, 0.8618\}$, and $\eta=1/\bar{d}=0.5371$ (proposed in \cite{glebov2019}, Table~II, for $T=2$),   is illustrated in Fig.~\ref{fig:PLev_d1}. In the figure the asymptotic packet loss probability is plotted as a function of $\avgG$ and $\TMPR \in \{1,2,3,4\}$ for the two cases with $P_{\text{E}|t}=0.2$ for all $t \leq \TMPR$ (solid line) and $P_{\text{E}|t}=0$ for all $t \leq \TMPR$ (noiseless channel, dashed line).
We can note that the minimum asymptotic packet loss probability, $P_{L\min}$, which is 0.0621 and 0 in the two cases, respectively, is achieved only for $\avgG=0$. Even in the noiseless case packet losses can occur (with probability that decreases as $\TMPR$ increases). When $d_{\min}>1$, as shown in Fig.~\ref{fig:PLev}, there is no packet loss in the noiseless channel for  a wide range of $\avgG$ values.

As for the scenario with collision channel and zero slot decoding error probability, we can define an average load threshold under which the \ac{SIC} process converges to a  point in the $(p,q)$ plane with $p$ close to $P_{\text{E}|1}$. Since the solution $p=P_{\text{E}|1}$ can not be achieved exactly (unless for $\avgG\rightarrow 0$), we need to introduce a parameter that quantifies how much $p_\infty$ should be close to $P_{\text{E}|1}$. This is done by defining the interval $[P_{\text{E}|1}, P_{\text{E}|1}(1+e)]$ as the target range to be achieved by $p_\infty$.
\begin{definition}
Let $\nslot \rightarrow \infty$ and $\mathbb{E}[\Ka] \rightarrow \infty$ for constant $\avgG=\mathbb{E}[\Ka]/\nslot$. 
 The \emph{asymptotic threshold} of the \ac{IRSA}-based random access for a given parameter $e$ is defined as 
\begin{align}\label{eq:err_ch_asym_th_G}
& \avgG^{\star}( \bm{\Lambda},T,\mathbf{P}_{\text{E}};e) \notag \\
&= \sup \{ \avgG > 0 : 
p_{\ell} \rightarrow p_\infty \leq P_{\text{E}|1}(1+e) \text{ as } \ell \rightarrow \infty \} .
\end{align}
\end{definition}

The value of the threshold $\avgG^{\star}$ satisfies the properties: (i) If $\avgG < \avgG^{\star}$ then $p_{\ell}$ tends to $p_\infty \leq P_{\text{E}|1}(1+e)$ as $\ell$ tends to infinity; (ii) If $\avgG > \avgG^{\star}$ then $p_{\ell}$ converges to a value that is bounded away from  $P_{\text{E}|1}$ as $\ell$ tends to infinity. 
Given the target parameter $e$, the value of $\avgG^{\star}$ depends on $\bm{\Lambda}$, the value of $T$ and the set of error probabilities $\mathbf{P}_{\text{E}}=\{P_{\text{E}|t}, t=1,\ldots,T\}$. 
We can interpret  $\avgG^{\star}(\bm{\Lambda},T,\mathbf{P}_{\text{E}};e)$ as the largest traffic (in data packets per slot) that the \ac{IRSA}-based random access with distribution $\bm{\Lambda}$  can support reliably, i.e., with packet loss probability suitably close to its minimum, over the considered channel with error probabilities $\mathbf{P}_{\text{E}}$, when both $\mathbb{E}[\Ka]$ and $\nslot$ are large. 
Therefore, the \ac{IRSA} decoder with perfect interference subtraction, can achieve packet loss probability $P_{L\infty}$ close to $P_{L\min}$, defined in \eqref{eq:PL}, when $\avgG < \avgG^{\star}$. 

According to the asymptotic analysis, when the load threshold is known for a given \ac{IRSA} protocol and any of the the conditions of Theorem~\ref{cor:suf_conditions} holds, a simple approximated model for determining the packet loss probability is 
\begin{align}   
 P_L\approx P_{L\min} \;\; \text{if} \;\; \avgG < \avgG^{\star}(\bm{\Lambda},T,\mathbf{P}_{\text{E}};e)
\end{align}
being $P_L\gg P_{L\min}$ otherwise. On the other hand, the actual values of $P_L$, for a given \ac{IRSA} scheme and a given set of error probabilities $\mathbf{P}_{\text{E}}$, can be  obtained by simulation.

\begin{example}\label{ex:threshold}
The value of $\avgG^{\star}$ can be found by testing the convergence of the recursions \eqref{eq:density_evolution_mpr}-\eqref{eq:density_evolution_mprq} for different values of $\avgG$. Fig.~\ref{fig:Gstar_pe} shows the values of $\avgG^{\star}$ for the \ac{IRSA} protocol  as in the Example~\ref{example:EXIT}, evaluated as function of $P_{\text{E}|1}$. 
The schemes with \ac{MPR} capability have  $P_{\text{E}|1}=P_{\text{E}|2}$. 
Note that the asymptotic threshold for $\TMPR=1$ drops when $P_{\text{E}|1}>0$ with respect to the case when $P_{\text{E}|1}=0$, since $d_{\min}=2$ in this case (Theorem~\ref{cor:suf_conditions} is not satisfied). Differently, for $\TMPR>1$, just a slight decrease appears on the threshold when $P_{\text{E}|1}>0$. 
Note also that the threshold of the schemes with $\TMPR>1$ is quite insensitive to parameter $e$, whereas the threshold of the scheme with $\TMPR=1$ depends on both $e$ and $P_{\text{E}|1}$.
\end{example}

\begin{figure}[t]
\centering
\includegraphics[width=\columnwidth]{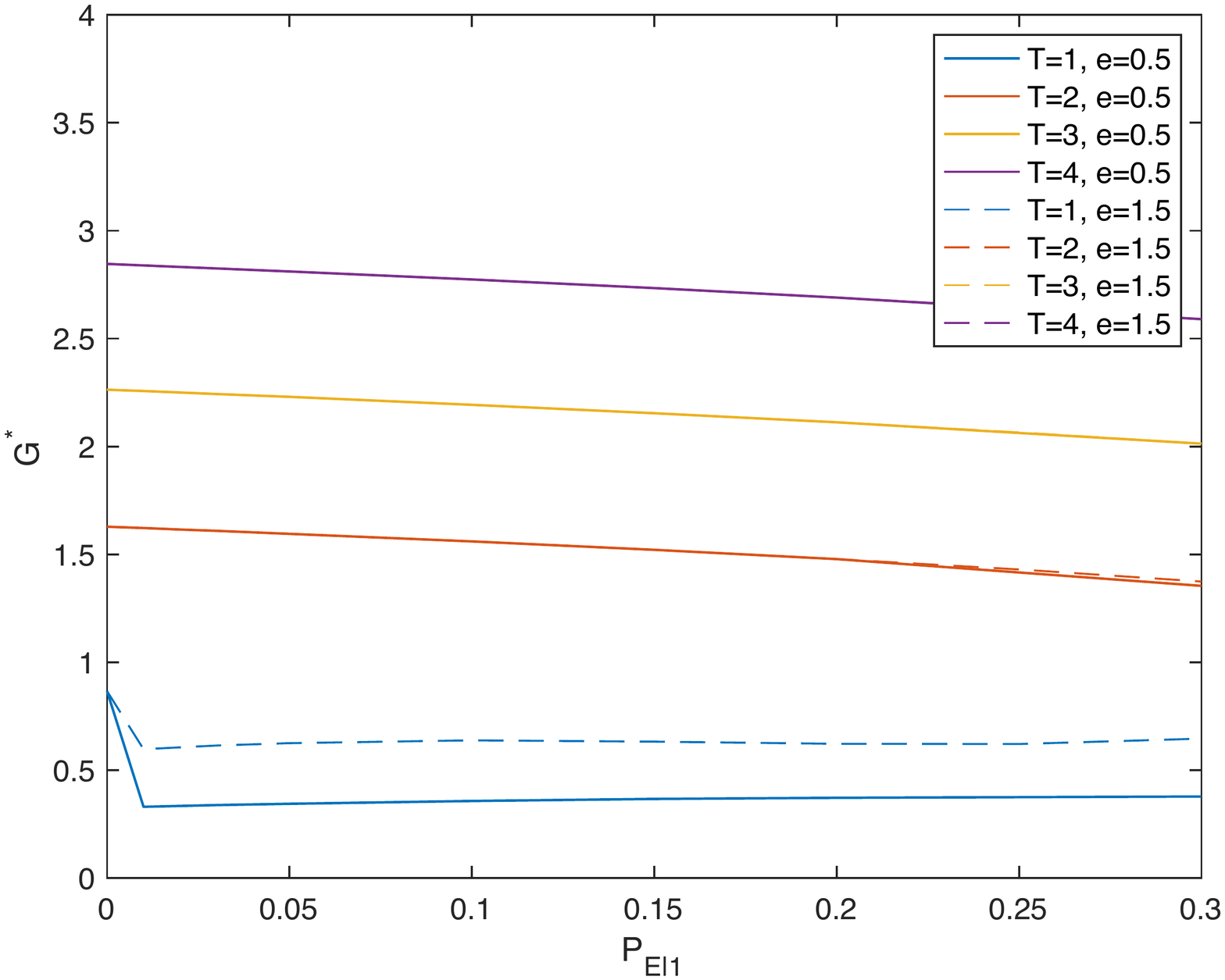}
      \caption{Asymptotic threshold as function of  $P_{\text{E}|1}$ ($=P_{\text{E}|2}$ if $\TMPR>1$) for $e=0.5$ and $e=1.5$. The \ac{IRSA} protocol is obtained with $d_{\max}=4$, $\mathbf{\Lambda}=\{0, 0.5102, 0, 0.4898\}$.}
    \label{fig:Gstar_pe}
\end{figure}

\begin{remark}
If any of the conditions of Theorem~\ref{cor:suf_conditions} holds, the probability $p_{\infty}$ falls to a value $P_{\text{E}|1} + O(P_{\text{E}|1}^2)$ as soon as $\avgG$ becomes smaller than $\avgG_0$. 
In this setting, unless $e$ is taken very small (even smaller than $P_{E|1}$), the threshold $\avgG^{\star}$ becomes numerically very close to $\avgG_0$ and exhibits a very low sensitivity to $e$.
This is not the case when the conditions of Theorem~\ref{cor:suf_conditions} do not hold, in which case we still define a threshold, that however is typically smaller than $\avgG_0$ and more sensitive to the parameter $e$.
We also point out that, if  any of the conditions of Theorem~\ref{cor:suf_conditions} holds and if both $\mathbb{E}[\Ka]$ and $\nslot$ are large, the threshold-based model becomes asymptotically tight for $\avgG$ values below threshold and the threshold becomes approximately independent of parameter~$e$.
\end{remark}
\begin{remark}
According to threshold-based model for packet loss probability, the generic requirement $P_L< \epsilon$ translates into the two requirements $\avgG < \avgG^{\star}(\bm{\Lambda},T,\mathbf{P}_{\text{E}};e)$ and $P_{L}<\sum_d \Lambda_d P_{\text{E}|1}^d(1+e)^d<\epsilon$. These two requirements might be used to drive a search for best \ac{IRSA}-based random access codes that maximize $\avgG$ subject to a maximum packet loss probability $\epsilon$. This would extend the results obtained in \cite{liva2011:irsa}  for a noiseless channel with $\epsilon=0$.
\end{remark}

\subsection{IRSA-based Random Access over \ac{GMAC}: Convergence Boundary}

In this subsection we derive a fundamental limit on the asymptotic load threshold of \ac{IRSA}-type schemes working over the \ac{GMAC}, where slot decoding has a finite error probability. This limit  applies to \emph{any} \ac{IRSA} scheme with efficiency $\eta$, regardless of the specific configuration employed, for a given set of error probabilities $\mathbf{P}_{\text{E}}$, when both $\mathbb{E}[\Ka]$ and $\nslot$ are large. 

Let us start with the following theorem.
\begin{theorem}\label{th:threshold_bound_new}
For any \ac{IRSA} scheme with $\TMPR$-\ac{MPR} capability and probability of 1 replica $\Lambda_1$, working over a \ac{GMAC} with a set of error probabilities $\mathbf{P}_{\text{E}}$, and with the \ac{SIC} process converging to $p_\infty \leq P_{\text{E}|1}(1+e)$ as $\ell \rightarrow \infty$, the load threshold $\avgG^{\star} = \avgG^{\star}(\bm{\Lambda},T,\mathbf{P}_{\text{E}};e)$ and the efficiency $\eta$ of the protocol must fulfill the two inequalities
\begin{align}\label{eq:G_threshold_bound_new}
  \avgG^{\star} [1 - & \Lambda_1 P_{\text{E}|1}(1+e)- (1-\Lambda_1)P_{\text{E}|1}^2(1+e)^2] \nonumber \\
   & 
\!\!\!\!\!\!\!\!\!\!\!\!\! \leq \sum_{t=1}^\TMPR (1- P_{\text{E}|t}) \left[ 1 -  \exp \left( - \frac{\avgG^{\star}}{\eta} \right) \sum_{k=0}^{t-1} \frac{1}{k!} \left( \frac{\avgG^{\star}}{\eta} \right)^k \right] \,  \\
\label{eq:G_threshold_bound_2}
 f_s(\Lambda_1) 
 &= 1-\exp \left(-\frac{ \avgG^* \Lambda_1}{\eta}\right) \sum_{t = 1}^{\TMPR } (1 - P_{\text{E}| t}) \frac{\left({\avgG^* \Lambda_1}/\eta \right)^{t-1}}{(t-1)!} 
 \leq P_{\text{E}|1}(1+e)
\end{align}
where the function $f_s(\cdot)$ is defined in \eqref{eq:density_evolution_mpr}. The first expression holds if $P_{\text{E}|1}(1+e)<0.5$.
\end{theorem}
\begin{IEEEproof} See Appendix C.
\end{IEEEproof}

Note that the theorem applies to the two classes of \ac{IRSA} protocols, i.e., the one with $\Lambda_1=0$ for which \eqref{eq:G_threshold_bound_2} is always satisfied, because $f_s(0)=P_{\text{E}|1}$, and the one with $\Lambda_1\neq 0$ for which \eqref{eq:G_threshold_bound_2} is a qualifying constraint. 
In the special case of $\TMPR=1$  the two inequalities of the theorem become 
\begin{align}
\avgG^{\star} &\leq \frac{1- P_{\text{E}|1}}{1 - \Lambda_1 P_{\text{E}|1}(1+e)- (1-\Lambda_1)P_{\text{E}|1}^2(1+e)^2} 
\left[ 1 - \exp \left( - \frac{\avgG^{\star}}{\eta} \right) \right]\, \\
 & 1-\exp \left(-\frac{ \avgG^* \Lambda_1}{\eta}\right) (1 - P_{\text{E}| 1}) 
 \leq P_{\text{E}|1}(1+e)\, .
\end{align}
Also note that in the case of slot decoding with zero error probability, i.e. $P_{\text{E}|t}=0$, the result of Lemma~\ref{th:threshold_bound} for \ac{IRSA} protocols with $\Lambda_1=0$ is obtained.

The following corollary can be also proved.
\begin{corollary}\label{cor:boundary}
For any integer $\TMPR \geq 1$, any probability $\Lambda_1$, and any $0 < \eta \leq (1- P_{\text{E}|1})/[1 -\Lambda_1 P_{\text{E}|1}(1+e)- (1-\Lambda_1)P_{\text{E}|1}^2(1+e)^2]$, let $\mathbb{G}_1(\eta,\Lambda_1,\TMPR,\mathbf{P}_{\text{E}};e)$ 
be the unique positive real root of the  equation
\begin{align}
\label{eq:G_threshold_bound2_new}
 & G [1 -\Lambda_1 P_{\text{E}|1}(1+e)- (1-\Lambda_1) P_{\text{E}|1}^2(1+e)^2] \nonumber \\
 &
 = \sum_{t=1}^\TMPR (1- P_{\text{E}|t}) \left[ 1 -  \exp \left( - \frac{G}{\eta} \right) \sum_{k=0}^{t-1} \frac{1}{k!} \left( \frac{G}{\eta} \right)^k \right]
 \end{align}
and $\mathbb{G}_2(\eta,\Lambda_1,\TMPR,\mathbf{P}_{\text{E}};e)$ be the unique positive real root of the  equation
\begin{align}
1-\exp \left(-\frac{ \avgG \Lambda_1}{\eta}\right) \sum_{t = 1}^{\TMPR } (1 - P_{\text{E}| t}) \frac{\left({\avgG \Lambda_1}/\eta \right)^{t-1}}{(t-1)!}
 = P_{\text{E}|1}(1+e)
\label{eq:G_threshold_bound2_new_2} 
\end{align}
that exists only if $\Lambda_1\neq 0$. 
Then, the load threshold $\avgG^{\star} = \avgG^{\star}(\bm{\Lambda},T,\mathbf{P}_{\text{E}};e)$ is upper bounded by 
%
%
%
\begin{align}\label{eq:G_threshold_bound3_new}
\avgG^{\star} &\leq \mathbb{G}(\eta,\Lambda_1,\TMPR,\mathbf{P}_{\text{E}};e) 
= \min \{ \mathbb{G}_1(\eta,\Lambda_1,\TMPR,\mathbf{P}_{\text{E}};e),\mathbb{G}_2(\eta,\Lambda_1,\TMPR,\mathbf{P}_{\text{E}};e)\} 
\end{align}
where  $\mathbb{G}_2(\eta,0,\TMPR,\mathbf{P}_{\text{E}};e)$ is conventionally defined as $+\infty$ if $\Lambda_1=0$, 
for any choice of $\bm{\Lambda}$ giving the efficiency $\eta$.
\end{corollary}
\begin{IEEEproof}
Let us first denote with $F_1(G)$ the right-hand side of \eqref{eq:G_threshold_bound_new} with $G$ replacing $\avgG^{\star}$. We can rewrite \eqref{eq:G_threshold_bound_new} as $\avgG [1 -\Lambda_1 P_{\text{E}|1}(1+e)- (1-\Lambda_1)P_{\text{E}|1}^2(1+e)^2] \leq F_1(\avgG)$. It can be easily noted that $F_1(0)=0$, i.e. the two functions $G$ and $F_1(G)$ cross in $G=0$. It can be also noted that $F_1(G) \rightarrow \sum_{t=1}^\TMPR (1- P_{\text{E}|t})$ as $G \rightarrow \infty$.
Let us now investigate the derivative of $F_1(G)$ given by 
\begin{align}
 F'_1(G) = \sum_{t=1}^\TMPR (1- P_{\text{E}|t}) \frac{1}{\eta} \exp \left( - \frac{G}{\eta} \right)  \frac{1}{(t-1)!} \left( \frac{G}{\eta} \right)^{(t-1) } . 
\end{align}
It is always positive for $G\geq 0$ and $F'_1(0)=(1/\eta)(1-P_{\text{E}|1})$ (note that the term $(G/\eta)^{t-1}$ has to be replaced with $1$ when $t=1$). If $F'_1(0)>[1 -\Lambda_1 P_{\text{E}|1}(1+e)- (1-\Lambda_1)P_{\text{E}|1}^2(1+e)^2]$, i.e., $\eta<(1-P_{\text{E}|1})/[1 - \Lambda_1 P_{\text{E}|1}(1+e)- (1-\Lambda_1)P_{\text{E}|1}^2(1+e)^2]$, there must be at least one positive solution of \eqref{eq:G_threshold_bound_new} with equality, which is also \eqref{eq:G_threshold_bound2_new}, between $0$ and $F_1(\infty)$.
Since $F''_1(G)$ is found to be negative, this solution is unique.
Let us now denote with $F_2(G)$ the left-hand side of \eqref{eq:G_threshold_bound_2} with $G$ replacing $\avgG^{\star}$. We can rewrite \eqref{eq:G_threshold_bound_2} as $ F_2(\avgG)\leq P_{\text{E}|1}(1+e)$. The function $F_2(.)$ has the same structure of function $f_s(.)$, which is monotonically increasing with its argument, starting from $F_2(0)=P_{\text{E}|1}$ and ending at $F_2(\infty)=1$. Thus, there is only one positive solution of \eqref{eq:G_threshold_bound_2} with equality, which is also \eqref{eq:G_threshold_bound2_new_2}, if $P_{\text{E}|1}(1+e)<1$. Finally, since the two conditions of Theorem~\ref{th:threshold_bound_new} must be valid, the load threshold $\avgG^*$ is upper bounded as in \eqref{eq:G_threshold_bound3_new}.
\end{IEEEproof}
The corollary applies to the two classes of \ac{IRSA} protocols, i.e., the one with $\Lambda_1=0$ for which \eqref{eq:G_threshold_bound_2} is always satisfied and $\eta$ can not be larger than $0.5$, and the one with $\Lambda_1\neq 0$ for which $0<\eta\leq 1$.

\begin{figure}[t]
\centering
\includegraphics[width=0.9\columnwidth]{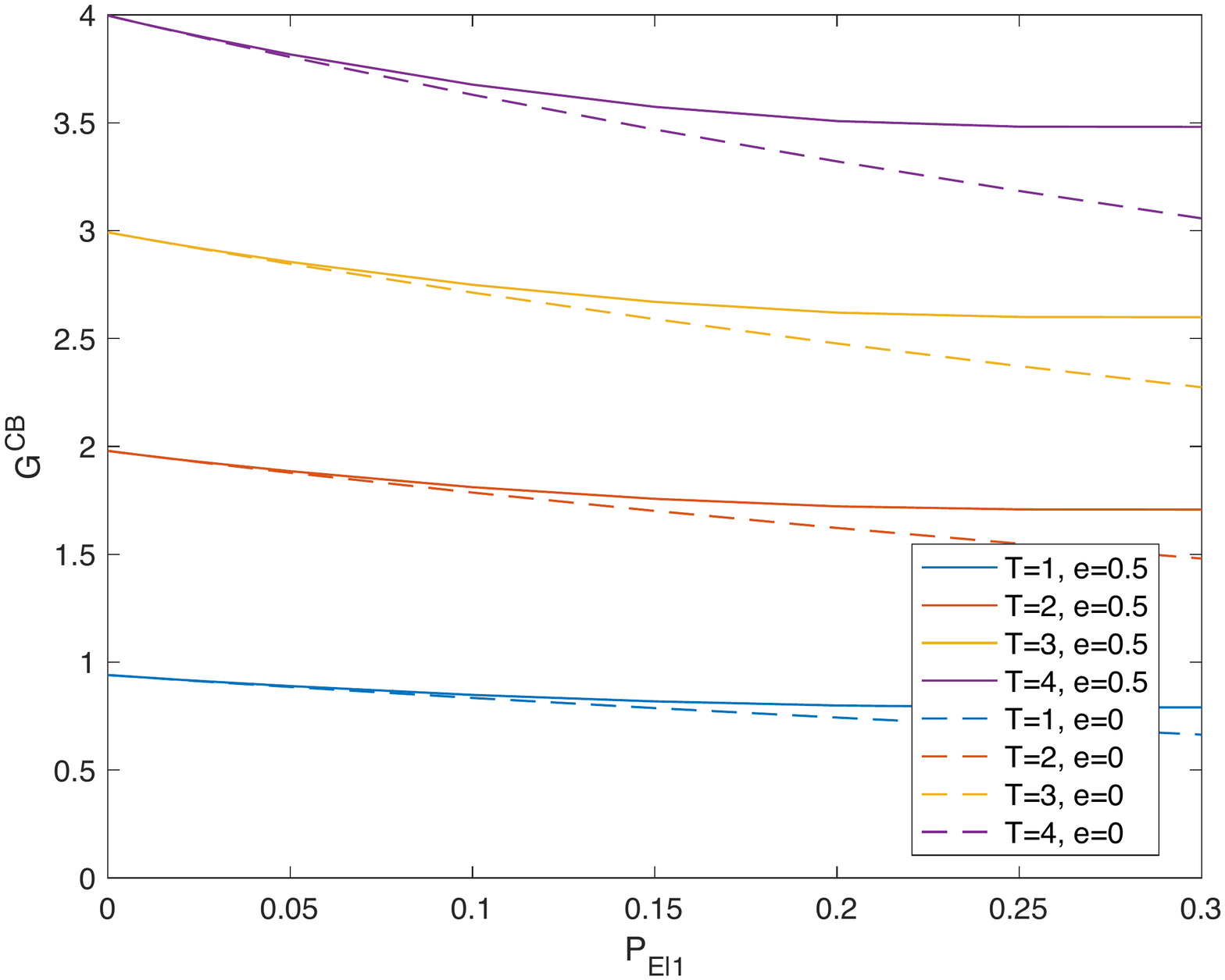}
      \caption{Convergence boundary $\avgG^{\mathrm{CB}}=\mathbb{G}(\eta,\Lambda_1,\TMPR,\mathbf{P}_{\text{E}};e)$ of \ac{IRSA} protocols with $\Lambda_1=0$ and $\eta=1/3$ as function of  $P_{\text{E}|1}$ ($=P_{\text{E}|t}$ if $\TMPR>1$).}
    \label{fig:Boundary_pe}
\end{figure}

\begin{figure}[t]
\centering
\includegraphics[width=\columnwidth]{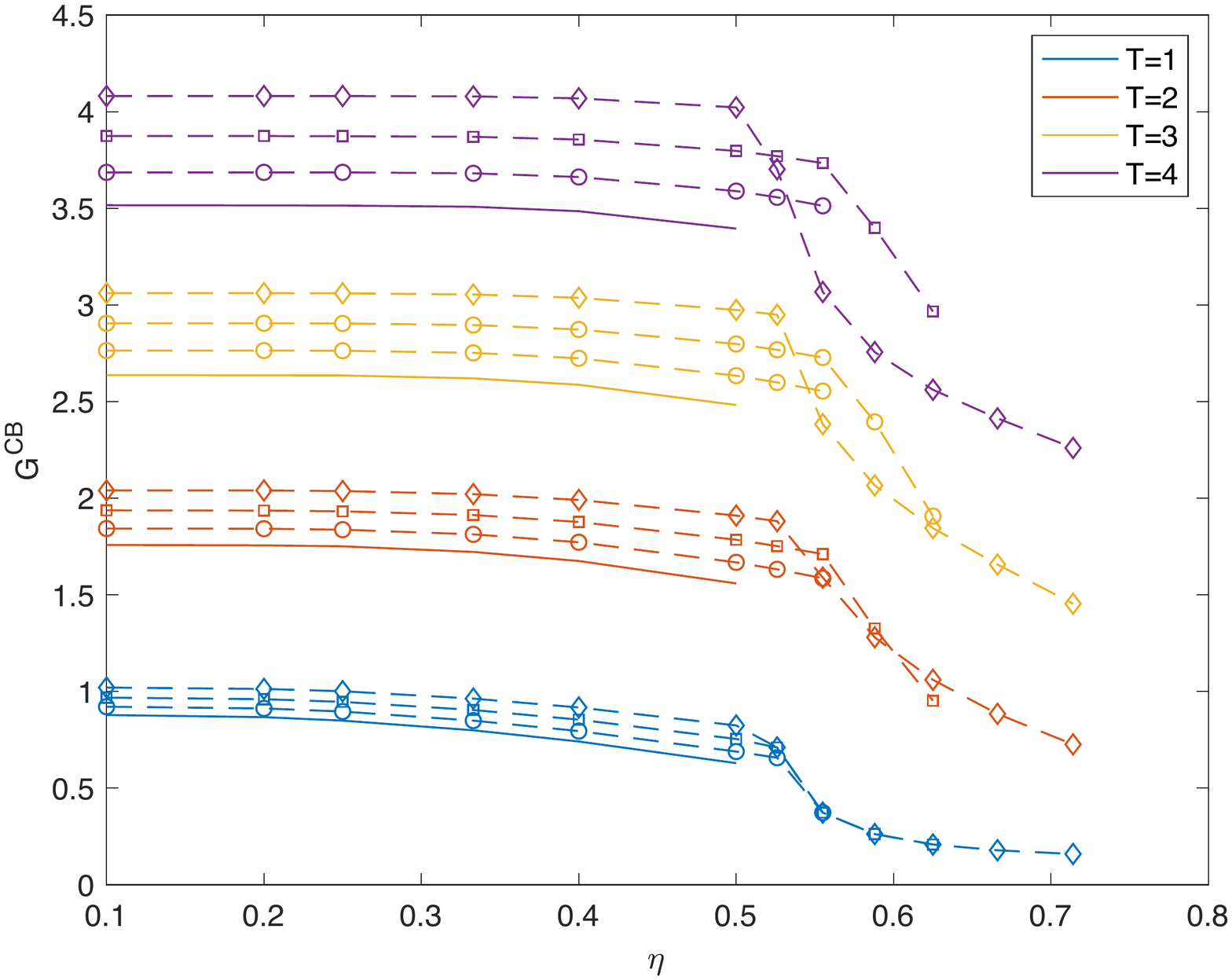}
      \caption{Convergence boundary $\avgG^{\mathrm{CB}}=\mathbb{G}(\eta,\Lambda_1,\TMPR,\mathbf{P}_{\text{E}};e)$ of \ac{IRSA} protocols versus  $\eta$ for $P_{\text{E}|t}=0.2$, $\forall t$,  $e=0.5$ and different values of $\Lambda1$: 0 (no marker), 0.2 (circle), 0.4 (square), 0.6 (diamond).}
    \label{fig:Boundary_eta}
\end{figure}

\begin{remark}
As observed in the previous subsection, \eqref{eq:G_threshold_bound3_new} is a \emph{converse bound} which defines a region for the values of $\avgG$ where the \ac{SIC} process  does not converge to $p_\infty \leq P_{\text{E}|1}(1+e)$ as $\ell \rightarrow \infty$.
The boundary of this region is $\mathbb{G}(\eta,\Lambda_1,\TMPR,\mathbf{P}_{\text{E}};e)$
and is referred to  as the \emph{convergence boundary} of an \ac{IRSA}-based random access scheme with $\TMPR$-\ac{MPR} capability over a channel with finite slot decoding error probabilities $\mathbf{P}_{\text{E}}$. Note also that in this region where $p_\infty > P_{\text{E}|1}(1+e)$ the packet loss probability is lower-bounded by
\begin{align}\label{eq:PLfloor}
P_{L\infty} > \sum_d \Lambda_d P_{\text{E}|1}^d (1+e)^d \geq P_{L\min} (1+e)^{d_{\min}} .
\end{align}
\end{remark}

As for the collision channel, the achievability of this boundary for any rate $\eta$ has never been proved to date.
However, we have to note that this boundary is obtained through Lemma~\ref{lemma_conv_pe} (Appendix C) which includes an upper bound on the area $A_\cap$ (see the proof). This bound is tight when $P_{\text{E}|1}(1+e)$ is small, but becomes loose as $P_{\text{E}|1}$ increases beyond $0.1$-$0.2$.  A tighter approximation would be obtained by removing $e$ from \eqref{eq:area_tunnel}, and, consequently, from \eqref{eq:G_threshold_bound_new}, but it is not guaranteed that this approximation is always a bound. This approximated boundary is simply $\mathbb{G}(\eta,\Lambda_1,\TMPR,\mathbf{P}_{\text{E}};0)$ and behaves well in all the cases where the load threshold appears clearly and almost independent  of parameter $e$, i.e., when the two conditions in \eqref{eq:pmin_conditions} hold.
Plots of the convergence boundary as function of $P_{\text{E}|1}$ and $\eta$, for some values of $\TMPR$, are shown in the Figs. \ref{fig:Boundary_pe} and \ref{fig:Boundary_eta}. 
In Fig.~\ref{fig:Boundary_pe}, \ac{IRSA} protocols with $\Lambda_1=0$ and $\eta =1/3$ are considered and convergence boundaries, versus $P_{\text{E}|1}$, for  $e=0$ and $e=0.5$ are compared. In Fig.~\ref{fig:Boundary_eta} convergence boundaries versus $\eta$ are plotted for four classes of \ac{IRSA} protocols, i.e., with $\Lambda_1= 0, 0.2, 0.4, 0.6$, and for $P_{\text{E}|1}=0.2$. Note that, from the definition of $\eta$, for a given $\Lambda_1$ the maximum admitted value of $\eta$ is $1/(2-\Lambda_1)$, obtained with $d_{\max}=2$. The figure shows that convergence boundary increases with $\Lambda_1$ when $\eta<0.5$. When $\eta>0.5$ there is a tradeoff between the maximum value of $\eta$ and maximum value of the boundary.


\section{PHY Layer Encoding and Decoding Schemes } \label{Sec:PHY_performance}

In this section we discuss two PHY layer encoding and decoding schemes that can be used in \ac{IRSA}-based random access with \ac{MPR}. They can also be used in the simple frame-based \ac{SA} random access that will be taken as a reference case for comparison. We also discuss models and methods for evaluating the performance of PHY layer slot decoding in terms of  probability of message error, $P_{\text{M}| t}$, and probability of  decoding error,  $P_{\text{E}| t}$, both conditioned to $t$ transmissions  in the slot.
These models will be exploited in Section~\ref{Sec:achievable_EbNo}  to evaluate the achievable $E_b/N_0$  and the tradeoff between spectrum and energy efficiency for \ac{IRSA}-based random access schemes.

To define the slot decoding procedure let us assume that in slot $i$ there are $t_i$ transmitted codewords not yet detected and cancelled by the \ac{SIC} process. Under the assumption of perfect estimation of $t_i$, in each slot
 $i$ the slot decoder tries to estimate the list of $t_i$ transmitted messages by using the decoding function
$$
g_P: \;\; Y^n=(y_{i1},\ldots,y_{in})  \longmapsto g_P(Y^n)=\{m_1,\ldots, m_{t_i}\} 
$$
with $m_j \in [M] $. 
The output is an empty list when $t_i>T$. An error flag $f_i$ marks the slot $i$ as unresolved if the list is empty or if one or more elements of the list are not correct.
This kind of decoder is an incomplete decoder that relies on both the perfect estimation of $t_i$ and the error detection capability of the PHY layer code.
The two coding options are described in the following subsections, starting from the most general scheme.

\subsection{Option 1: Optimum Coding for MPR with Binary Error Detection Coding}

Each message $m_{s,i}\in [M]$, represented by a $k$-bit block with $M\leq 2^k$, is first encoded into a $k'$-bit block with a binary code for error detection. The number of parity check bits  $r=k'-k$ should be large enough to keep the undetected error probability $P_{ue}$ below a given fixed value. It is known that there exist $(k+r,k)$ block codes with $P_{ue}$ bounded by $2^{-r}$. We have to keep in mind that  an undetected error in the PHY layer code at the slot level may results in a failure in MAC layer decoding with some additional packet losses due to interference cancellation errors. The effects of a non zero probability of cancellation error are investigated in \cite{dumas2021:canc_err}. By looking at the results of this work and considering cancellation errors due to undetected errors at PHY layer only, we can design $r$ to keep  $P_{ue}$ at least one order of magnitude below the target packet loss probability.
The $k'$-bit block is then encoded into an $n$-symbol block by using an extended codebook of $M'= 2^{k'}=M2^r$ elements and encoding function $f_{P1}(m')$, where $m'\in [M']$ is one of the possible $k'$-bit messages.

The symbol sequence received from the GMAC in a given slot is first decoded with the following minimum distance decoder:
$$
g_{P1}(Y^n)=  \underset{\{m'_1,\ldots, m'_{t_i}\}, m'_j\in [M']}{\arg\min} \Big\| \sum_{j=1}^{t_i} f_{P1}(m'_j) - Y^n \Big\|^2 .
$$
The decoder is activated only if $1\leq t_i\leq T$. The messages in the output list are then checked with binary error detection. If at least one message of the list fails error detection, the output flag is set to $0$ (unresolved slot).

The random coding bound from \cite{Polyanskiy2017:Perspective} (reviewed right before Section~\ref{subsec:random_access_codes}) provides an upper-bound  for the probability of message error when $t$ active transmitters use a codebook of size $M'$ to send their message over a slot with finite size of $n$ channel-uses. It is given by
\begin{align}
P_{\text{M}| t} \leq F_1(n,M',P/\sigma^2,t)
\end{align}
where 
$P/\sigma^2=(2\eta\log_2M/n)E_b/N_0$, according to \eqref{eq:ebno}, sets an energy constraint in the transmission which is  related to $E_b/N_0$.
This result can be also exploited to derive a bound for the decoding error probability $P_{\text{E}| t}$, by considering, in the most general way, the union-bound 
\begin{align} \label{eq:option1_bound}
P_{\text{E}| t} \leq t P_{\text{M}| t} \leq t F_1(n,M',P/\sigma^2,t).
\end{align}
However,  the expression of the bound in \cite{Polyanskiy2017:Perspective} may be suitably manipulated to obtain directly $P_{\text{E}| t}$ as the sum of  the $t$ probabilities that $i$ transmitted messages, with $1\leq i\leq t$, are not in the output list of the decoder.
As pointed out in \cite{glebov2019}, the set of probabilities $P_{\text{E}| t}$ for $t=1,\ldots,T$ are the results of a random coding bound and all of them have to be achieved  by the same code. There exist codes with probability 1 in the random set able to achieve the set of probabilities $\{ P_{\text{M}| t} / \alpha_t, \; t=1,\ldots,T \}$ if the coefficients $\alpha_t$ satisfy $\sum_t \alpha_t <1$. A code that achieves the set of random coding bounds is denoted as an optimum code for \ac{MPR}.

\subsection{Option 2: BPR and Binary Linear Concatenated Coding with Error Detection Coding}

We consider the \ac{BPR} codebook construction proposed  in \cite{Bar-David1993:Forward}, with  modifications as in \cite{Ordentlich2017:low_complexity}  to adapt it to the unsourced setting considered here. 
This coding technique is able to achieve near-zero 
error probability over a  binary adder channel (BAC) with $1\leq t\leq T$ active transmitters, where $T$ is a code design parameter.
This channel is noiseless, has $t$ inputs $X_1$, $\dots$, $X_{t}$, where $X_i \in \{0,1\} \subset \mathbb{R}$, and one output $Y \in \{0,\dots,t\} \subset \mathbb{R}$, given by 
$   Y = \sum_{i=1}^{t} X_i $, the sum being over the reals.
 In the simple case of $T=1$, i.e., no \ac{MPR} capability, there is no need of BPR coding and the inner coding is not constrained to be binary or linear.
 
The \ac{BPR} construction starts from the parity-check matrix of a $\T$-error-correcting binary \ac{BCH} code of length $2^k-1$, 
which has $k\T$ binary rows and $2^k-1$ columns.
The set of all columns is the common codebook with codewords of length $k\T$. Its size is $M=2^k-1$.
Note that the codebook is nonlinear and does not contain the all-$0$ codeword.
As mentioned above, the \ac{BPR} construction achieves near-zero error probability over the \ac{BAC} with $1 \leq t\leq T$ inputs.
In fact, since by construction any $2\T$ columns of the parity-check matrix of a $\T$-error-correcting linear block code are linearly independent, any two different $\T$-tuples of codewords cannot have the same sum.
Hence, a set of up to $\T$ messages, if they are all different, can be decoded with no error at the receiver with standard low complexity algorithms for \ac{BCH} codes.  An error can occur only if some messages are equal. The probability of this event is usually negligible when the codebook size $M$ is large.

The slot encoding function is described as follows. A  concatenated coding scheme with an outer \ac{BPR}  code and an inner binary linear code, as proposed by \cite{Ordentlich2017:low_complexity}, is considered.
Each message $m_{s,i}\in [M]$, represented by a $k$-bit block, is first encoded  into a codeword of length $k\T$ belonging to the \ac{BPR} codebook. 
The codeword is further encoded into a block of $k'=k\T+r$ bits using a binary linear code for error detection. The redundancy  $r$ should be large enough to keep the undetected error probability below a given fixed value. The $k'$-bit block is then encoded into an $n$-symbol block by using a binary linear codebook of $M'=2^{k'}$ elements and encoding function $f_{P2}(m')$ where $m'\in [M']$ is one of the possible $k'$-bit messages. The symbols in the final codeword $
\mathbf{x}_i$ are binary antipodal symbols with values $\pm \sqrt{P}$ obtained from the encoded bits $\mathbf{b}_i$,  as $\mathbf{x}_i=\sqrt{P}(2\mathbf{b}_i-1)$.

The symbol sequence $\mathbf{y}_i$ received from the GMAC with $t_i$ transmissions in the generic slot $i$ is first processed to map it onto a sequence of symbols belonging to interval $[0,2)$, by using modulo-2 operation, as follows:
\begingroup
\allowdisplaybreaks
\begin{align}\label{eq:mod2-awgn}
\hat{\mathbf{y}}_i &= \left[ \frac{\mathbf{y}_i}{2\sqrt{P}}+\frac{t_i}{2} \right] \; \mathrm{mod} \;2 \nonumber \\
& = \left[ \sum_{j=1}^{t_i} \mathbf{b}_j+\frac{\mathbf{z}_i}{2\sqrt{P}} \right] \; \mathrm{mod} \;2 \nonumber \\
& = \left[ \mathbf{B}_i+{\mathbf{z'}}_i\right] \; \mathrm{mod} \;2 .
\end{align}
\endgroup
Here, $\mathbf{B}_i=[ \sum_{j=1}^{t_i} \mathbf{b}_j ]  \; \mathrm{mod}\;2$ is the bit-wise modulo-2 sum of the transmitted codewords and ${\mathbf{z'}}_i={\mathbf{z}_i}/(2\sqrt{P})$ is the noise term. 
Since the transmitted codewords are obtained from the same linear code, used to encode each  \ac{BPR} codeword, $\mathbf{B}_i$ is necessarily a codeword of the linear code obtained by encoding the modulo-2 sum of the transmitted \ac{BPR} codewords.

Therefore, in the receiver there is a first inner decoder that tries to estimate the codeword $\mathbf{B}_i$ 
 from $\hat{\mathbf{y}}_i$ through a maximum likelihood search, as 
$$
\hat{\mathbf{B}}_i=g_{P2}(\hat{\mathbf{y}}_i)= \underset{\mathbf{B}_i }{\mathrm{argmin}} \| \mathbf{B}_i - \hat{\mathbf{y}}_i \|^2 .
$$
By exploiting the error detection code embedded as outer component of the linear code, the decoder is also able to check the validity of the codeword with negligible decoding failure probability. If decoding is successful, i.e. $\hat{\mathbf{B}}_i=\mathbf{B}_i$, then the modulo-2 sum of the transmitted \ac{BPR} codewords can be extracted from $\hat{\mathbf{B}}_i$ and  sent to the \ac{BPR} decoder. If decoding is not successful, the output list is left empty 
and the error flag $f_i$ marks the slot $i$ as unresolved, under the assumption that decoding is activated when $t_i\leq T$.

The \ac{BPR} decoder  uses a low-complexity decoding procedure derived from standard BCH decoding algorithm, as illustrated in \cite{Ordentlich2017:low_complexity,Bar-David1993:Forward}. If $t_i\leq T$ and all the $t_i$ messages are distinct, then  the \ac{BPR} decoder is able to decode the list $\{m_1,\ldots, m_{t_i}\} $ of transmitted messages with zero error probability. An error event  occurs only if two or more transmitted messages are the same. This kind of error can be detected in presence of perfect knowledge of $t_i$, due to the mismatch between list length and $t_i$, and the slot is marked as unresolved. 

In this scheme the error events are generated by the first part of decoding, related to inner ideal linear coding. If a correct inner codeword is decoded, BPR decoding produces the correct list of messages transmitted in a slot, provided that their number is not greater than $T$. If an error event is detected at the inner decoder a decoding error is declared and an empty list is provided at the output of BPR decoder. 
Thus, there is no difference between message error and decoding error for any number of transmissions $t\leq T$ in the slot, and we use notation $P_{\text{E}| t} = P_{\text{M}| t} = P_{\text{E}}$.

An expression for the decoding error probability is derived in the Appendix E as function of $P/\sigma^2$, coding rate and other system parameters.
Regarding the inner binary linear coding we are considering 
ideal coding able to achieve coding rate limits of a binary-input memoryless channel with AWGN in the finite-blocklength regime.

\section{Achievable $E_b/N_0$ }\label{Sec:achievable_EbNo}

The scope of this section is to set up the model for the evaluation of packet loss probability for a slotted random access code over a \ac{GMAC}, as a function of the system parameters, by integrating the main methods and results developed in Section~\ref{Sec:MAC_performance} and Section~\ref{Sec:PHY_performance}. 
With this model we will be able to derive the achievable $E_b/N_0$ values, indicating the energy efficiency of the random access code, able to guarantee a predefined (per-user) packet loss probability.

In a general framework, the packet loss probability derived in Section \ref{Sec:MAC_performance}, for a given number $\Ka$, or  $\mathbb{E}[\Ka]$, of active devices, is obtained as
\begin{align}
P_L=L(\TMPR,\mathbf{P}_{\text{E}},\mathbb{E}[\Ka],\nslot)
\end{align}
where the function $L(\cdot)$ can be specified for each MAC layer coding scheme, 
for both \ac{SA} and \ac{IRSA}.
We remind that the analytic expression of $L(\cdot)$ is available for \ac{SA}, whereas for  \ac{IRSA}  this relationship can be generally obtained  through simulation. When $\mathbb{E}[\Ka]$ and $\nslot$ are sufficiently large and with $\avgG=\mathbb{E}[\Ka]/\nslot$, the packet loss probability can be approximated by its asymptotic value $P_{L\infty}$ which is evaluated through recursive equations summarized in the  general form
\begin{align}
P_{L\infty}=L_{\infty}(\TMPR,\mathbf{P}_{\text{E}},\avgG) \, .
\end{align}

One input of the functions $L(\cdot)$ and $L_{\infty}(\cdot)$  is the set of slot decoding error probabilities, related to the PHY layer coding options analyzed in Section~\ref{Sec:PHY_performance}.
The conditional  error probabilities are obtained as
\begin{align}
{P}_{\text{E}|t}=F(n,M,P/\sigma^2,t)
\end{align}
where the function $F(\cdot)$ can be specified, for each PHY layer coding scheme, as function of $M$, $n=N/\nslot$ and $P/\sigma^2=(2\eta\log_2M/n)E_b/N_0$ depending  on $E_b/N_0$.
The codebook size $M$ is strictly related to the message length $k$ in bits and some expressions shown in Section~\ref{Sec:PHY_performance} (and Appendix~\ref{appendix:decoding_option2}) are functions of $k$, directly or through other coding parameters.
It is important to note that for a given fixed amount of channel uses $N=nN_\mathrm{s}\,$, and a fixed average number of
active devices $\mathbb{E}[\Ka]$, it is possible to play with parameters $n$ and $N_\mathrm{s}$. A large value of $n$ (low-rate PHY layer code) leads to  small values of $P_{\text{E}|t}$, whereas a large value of $N_\mathrm{s}$ leads to a small value of $\mathsf{G}$ in the MAC layer code, both improving packet loss probability. A tradeoff can therefore be found to optimize this probability.

\begin{definition}
For given fixed values of codebook size $M$, frame length $N$, number of active transmitters $\Ka$ or $\mathbb{E}[\Ka]$, and a given coded random access scheme, the \emph{achievable} $E_b/N_0$ with given (per-user) packet loss probability $\epsilon$ is defined as
\begin{align}
[E_b/N_0]^{\star}= \inf \{ E_b/N_0> 0 : P_L<\epsilon, n>0\} 
\end{align}
where the slot size $n$ 
is taken as a free configuration parameter of the coded random access scheme to capture the best tradeoff between $n$ and $N_\text{s}$.

\end{definition}
When $\mathbb{E}[\Ka]$ and $\nslot$ are sufficiently large, and the packet loss probability is approximated by its asymptotic expression $P_{L\infty}$, we obtain the asymptotic approximation of $[E_b/N_0]^{\star}$, denoted as $[E_b/N_0]^{\star}_{\infty}$ and given by
%
%
\begin{align}
[E_b/N_0]^{\star}_{\infty} 
&= \inf \Big\{ E_b/N_0> 0 : 
\avgG<\avgG^{\star}( \bm{\Lambda},T,\mathbf{P}_{\text{E}};e), \nonumber\\
&\quad\quad\quad\,\, \sum_d \Lambda_d [P_{\text{E}|1}(1+e)]^d<\epsilon , n>0, e>0 \Big\} \label{eq:ebno_star} 
\end{align}
according to Remark 2.

\subsection{Achievable $E_b/N_0$ for IRSA on the Convergence Boundary}

When the performance of an \ac{IRSA}-based random access code is evaluated through the asymptotic approximation, which is tight when $\mathbb{E}[\Ka]$ and $\nslot$ are sufficiently large, it is possible to define an energy efficiency limit that holds for any \ac{IRSA} scheme with given $\TMPR$ and PHY layer coding option. It is based on the convergence boundary  of \ac{IRSA} over a channel with slot decoding error probabilities $\mathbf{P}_{\text{E}}$. Although it can not be considered as a strictly achievable limit, it provides a useful benchmark for the achievable energy efficiency of any \ac{IRSA} scheme, considering that some specific optimized schemes in the more general class of \ac{CSA} protocols \cite{paolini2015:csa} have been found that approach such limit quite closely.

To define this energy efficiency limit, let us first remind that Corollary \ref{cor:boundary} holds for \ac{IRSA} in the asymptotic conditions.
Moreover, we also consider the inequality $P_L=\sum_d \Lambda_d p_{\infty}^d = \Lambda_1 p_{\infty} +\sum_{d>1} \Lambda_d p_{\infty}^d \leq \Lambda_1 p_{\infty} +(1-\Lambda_1) p_{\infty}^2$ that holds for \ac{IRSA} protocols with any choice of $\Lambda_1$.
Therefore, for given fixed values of codebook size $M$, frame length $N$, and number of active users $\Ka$ or $\mathbb{E}[\Ka]$, we define the achievable value of $E_b/N_0$ \textit{on the convergence boundary} for any \ac{IRSA}-based scheme with given $T$-MPR capability, probability of 1 replica $\Lambda_1$ and  (per-user) packet loss probability $\epsilon$, as
\begin{align}
[E_b/N_0]^{\mathrm{CB}} 
 &=\inf \Big\{ E_b/N_0> 0 :  \; \avgG <\mathbb{G}(\eta,\Lambda_1,\TMPR,\mathbf{P}_{\text{E}},e), \nonumber \\ 
&\quad \Lambda_1 P_{\text{E}|1}(1+e)+ (1-\Lambda_1)P_{\text{E}|1}^2(1+e)^2 < \epsilon ,e>0, n>0, \nonumber \\ 
&\quad \eta \leq \frac{1- P_{\text{E}|1}}{[1 - \Lambda_1 P_{\text{E}|1}(1+e)- (1-\Lambda_1)P_{\text{E}|1}^2(1+e)^2]} \Big\}\label{eq:ebno_CB}
\end{align}
where the slot size, $n$, and $\eta$  are taken as free configuration parameters of the coded random access schemes.
Due to Corollary \ref{cor:boundary}, we also have, for any \ac{IRSA} scheme with given $\TMPR$ and PHY layer coding option,
\begin{align}
[E_b/N_0]^{\star}\approx [E_b/N_0]^{\star}_{\infty}\geq [E_b/N_0]^{\mathrm{CB}} .
\end{align}
We finally note that also the parameter $e$, relating the load threshold to error floor, is a free variable in the evaluation. However, in all the cases where the load threshold appears clearly and almost independent  of parameter $e$, i.e., when the two conditions of Theorem \ref{cor:suf_conditions} hold or when the channel has zero slot decoding error probability, a value of $[E_b/N_0]^{\mathrm{CB}}$ can be obtained by exploiting the approximated boundary $\mathbb{G}(\eta,\Lambda_1,\TMPR,\mathbf{P}_{\text{E}};0)$ and evaluating \eqref{eq:ebno_CB} with $e=0$ in a simplified way.

\subsection{Tradeoff between Energy Efficiency and Spectrum Efficiency}\label{Sec:tradeoff}

In this subsection we finally  try to capture the fundamental tradeoff between energy efficiency and spectrum efficiency for slotted random access schemes when the working condition approches the asymptotic regime. We define the asymptotic scenario with the following setting: 
 $\mathbb{E}[\Ka]\rightarrow \infty$, $N\rightarrow \infty$, $\nslot\rightarrow \infty$, with $\mathbb{E}[\Ka]/\nslot=\avgG$, $\avgG$ finite real number,  $N/\nslot=n$, $n$ finite integer number. 
Note that this scenario is in principle not compatible with the unsourced access where all users share the same finite-size codebook and  with the slotted random access codes that use the finite-size codebook to encode frames with a diverging number of slots. Therefore, the asymptotic setting is not a  feasible working condition, but is just a limit condition that can be approached when $K$ and $\nslot$ are very large, but still compatible with a finite-size common codebook.

\begin{definition}
In the asymptotic scenario the spectrum efficiency is defined as
\begin{align}
S=\frac{\mathbb{E}[\Ka] \log_2M}{N}=\avgG \frac{\log_2M}{n}.
\end{align}
For a slotted coded random access scheme, $\avgG$ and $\log_2M/n$ quantify the portions of the efficiency due to the MAC layer component and to the PHY layer coding component, respectively. 
\end{definition}
The energy efficiency is given by the achievable value of $E_b/N_0$ that guarantees a per user packet loss probability $P_{L\infty} <\epsilon$ and can be evaluated as a function of $S$ with the same methods previously presented, 
 by still considering  $n$ as a free optimization parameter.
 It is sufficient to consider \eqref{eq:ebno_star} and \eqref{eq:ebno_CB} and replace $\avgG$ with $nS/(\log_2M)$.
We obtain two expressions that define, for each random access scheme, a relationship between energy efficiency parameter $[E_b/N_0]^{\star}_{\infty}$, or $[E_b/N_0]^{\mathrm{CB}}$, and spectrum efficiency $S$.

The evaluation on the convergence boundary provides a fundamental limit in the tradeoff between energy and spectrum efficiency, for  any \ac{IRSA} scheme with $T$-MPR capability, in terms of a converse bound, although not yet proved as achievable bound. This tradeoff, as well as its limit, will be explored in the next section for the \ac{IRSA}-based schemes presented in the paper.

\section{Numerical results}\label{Sec:results}

\begin{figure}[t]
\centering
\includegraphics[width=\columnwidth]{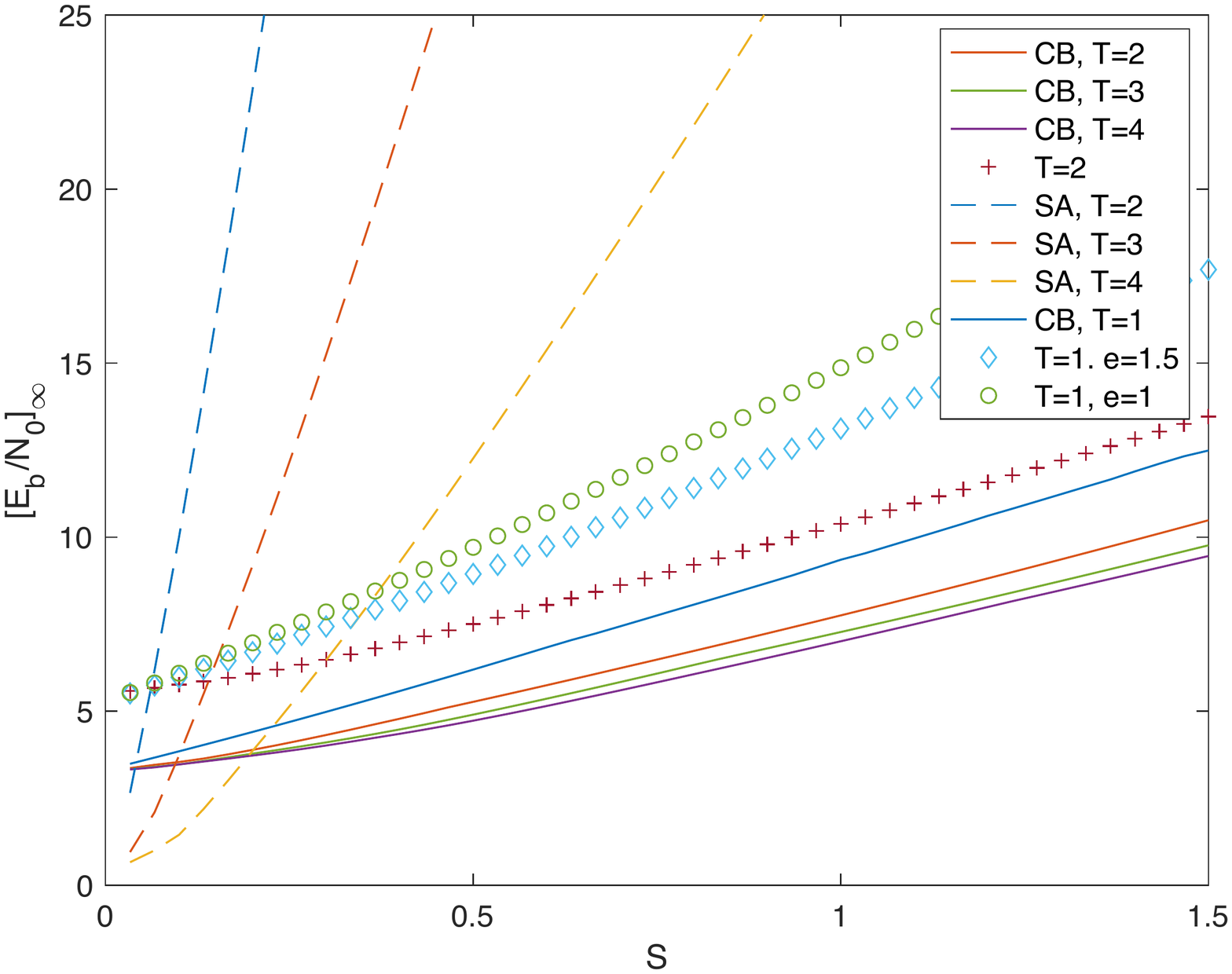}
      \caption{Achievable $E_b/N_0$ vs spectrum efficiency $S$ for \ac{IRSA} protocols with $\Lambda_1=0$ and option 1 PHY layer coding, $\log_2M=100$ and target packet loss probability $\epsilon=0.005$. CB: on convergence boundary. SA: slotted ALOHA.}
    \label{fig:Toff_ideal}
\end{figure}

\begin{figure}[t]
\centering
\includegraphics[width=\columnwidth]{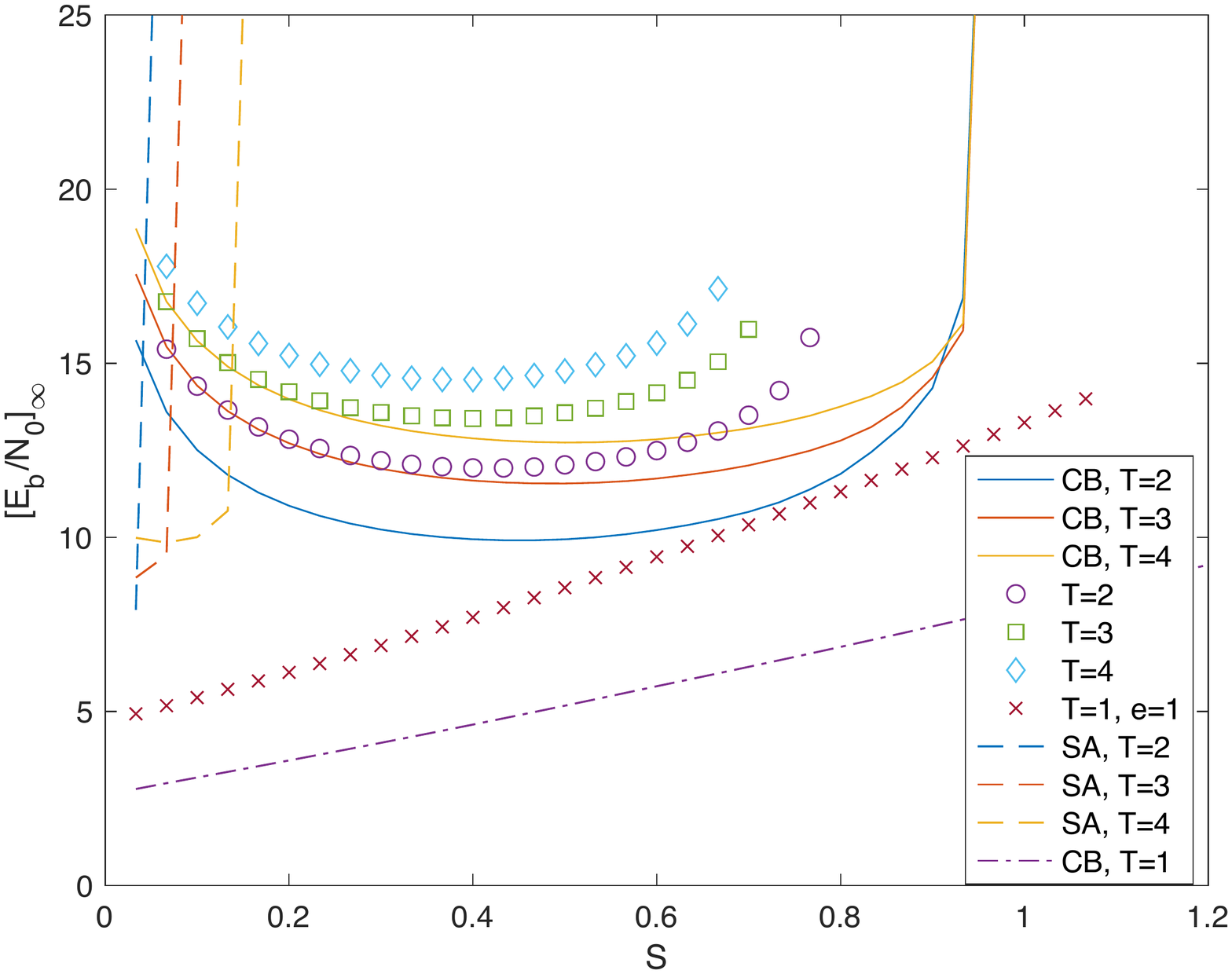}
      \caption{Achievable $E_b/N_0$ vs spectrum efficiency $S$ for \ac{IRSA} protocols with $\Lambda_1=0$ and option 2 PHY layer coding, $\log_2M=100$ and target packet loss probability $\epsilon=0.005$. CB: on convergence boundary. SA: slotted ALOHA. }
    \label{fig:Toff_BPR}
\end{figure}

In this section we illustrate the tradeoff  between energy efficiency and spectrum efficiency achievable by \ac{IRSA} based random access codes in the asymptotic setting. 
As far as energy efficiency is concerned, we always compare the achievable $E_b/N_0$ for some specific \ac{IRSA} protocols  with the achievable  $E_b/N_0$ on the convergence boundary for the class of IRSA schemes, with specific $\TMPR$ and PHY layer coding option, which the protocol belongs to. 
Since the latter is based on the convergence boundary of the IRSA-based schemes, which is a converse bound, and is evaluated for a channel with an achievable error probabilities $\mathbf{P}_{\text{E}}$, it is important to remind that it can not be considered as a strictly achievable limit, although it provides a useful benchmark.
We first consider a scenario where the packet size is $\log_2M=100$ and target packet loss probability is $\epsilon=0.005$.\footnote{This value of packet loss probability, lower than the ones often targeted in the unsourced random access literature (usually spanning from $0.05$ to $0.1$), may well represent the reliability requirement of a next-generation massive machine-type communications application.} 

The results in Fig.~\ref{fig:Toff_ideal} refer to \ac{IRSA} protocols with $\Lambda_1=0$ and to option 1 for PHY layer slot encoding with an optimum code that achieves the random coding bound at slot level with perfect knowledge of the number of transmissions per slot. The results on the convergence boundary provide a limit beyond which the target error probability can never be achieved. The energy efficiency limit in terms of $E_b/N_0$ for \ac{IRSA} schemes with $\Lambda_1=0$ and $T>1$ is between $7$ and $8$ dB at a spectrum efficiency of $1 \, \mathrm{bit/channel \,\, use}$. $E_b/N_0$ also decreases with $T$, i.e., the \ac{MPR} capability.
A very large  improvement over the reference case of \ac{SA} with \ac{MPR}  is obtained. \ac{SA} is a convenient choice only when $S$ is small and the network is working with a small load. The figure also reports with markers the results obtained for a specific \ac{IRSA} protocol
with parameters $d_{\min}=2$, $d_{\max}=4$, $\mathbf{\Lambda}=\{0, 0.5102, 0, 0.4898\}$ and $\eta=1/\bar{d}=0.3356$ (also considered in Example 1). We can note that for $T>1$, the case where the load threshold is well defined, the energy efficiency values stay within 2 dB from the convergence boundary. This gap increases, as expected, for $T=1$ (random access code without \ac{MPR}) when the dependency on parameter $e$ becomes significant. We report for this case two plots with different fixed values of $e$. The best result in this specific case is obtained with $e=1.5$.

The results in Fig.~\ref{fig:Toff_BPR} refer to \ac{IRSA} protocols with $\Lambda_1=0$ and to option~2  PHY layer slot encoding, where BPR coding is used to support \ac{MPR} capability. 
We can note from the figure that the results on the convergence boundary for $T>1$ get far from the limits set by the optimum case (option 1) for PHY layer slot encoding. 
Spectrum efficiency beyond $0.9 \, \mathrm{bit/channel \,\, use}$ is not achievable due to the code rate limitations coming from BPR encoding. 
Also energy efficiency suffers the low code rate of BPR encoding, which decreases as $T$ increases. 
The results with the markers, obtained for the same \ac{IRSA} distribution $\bm{\Lambda}$ considered in Fig.~\ref{fig:Toff_ideal}, show that the best tradeoff between energy and spectrum efficiency is obtained for the scheme without \ac{MPR}\footnote{We should also note that for $T=1$, options 1 and 2 refer to the same scheme, but the evaluation of decoding error probability $P_{\text{E}|1}$ through the approximation presented in Appendix \ref{appendix:decoding_option2} leads to  values smaller than those obtained from the bound in \eqref{eq:option1_bound}.}  ($T=1$) which also outperforms the \ac{SA} schemes in the region of small $S$.
This is essentially due to the rate loss introduced by the \ac{BPR} code for the \ac{BAC}, having a code rate of approximately $1/\TMPR$, whose effect is not compensated by the enhanced \ac{MPR} capability.

\begin{figure}[t]
\centering
\includegraphics[width=\columnwidth]{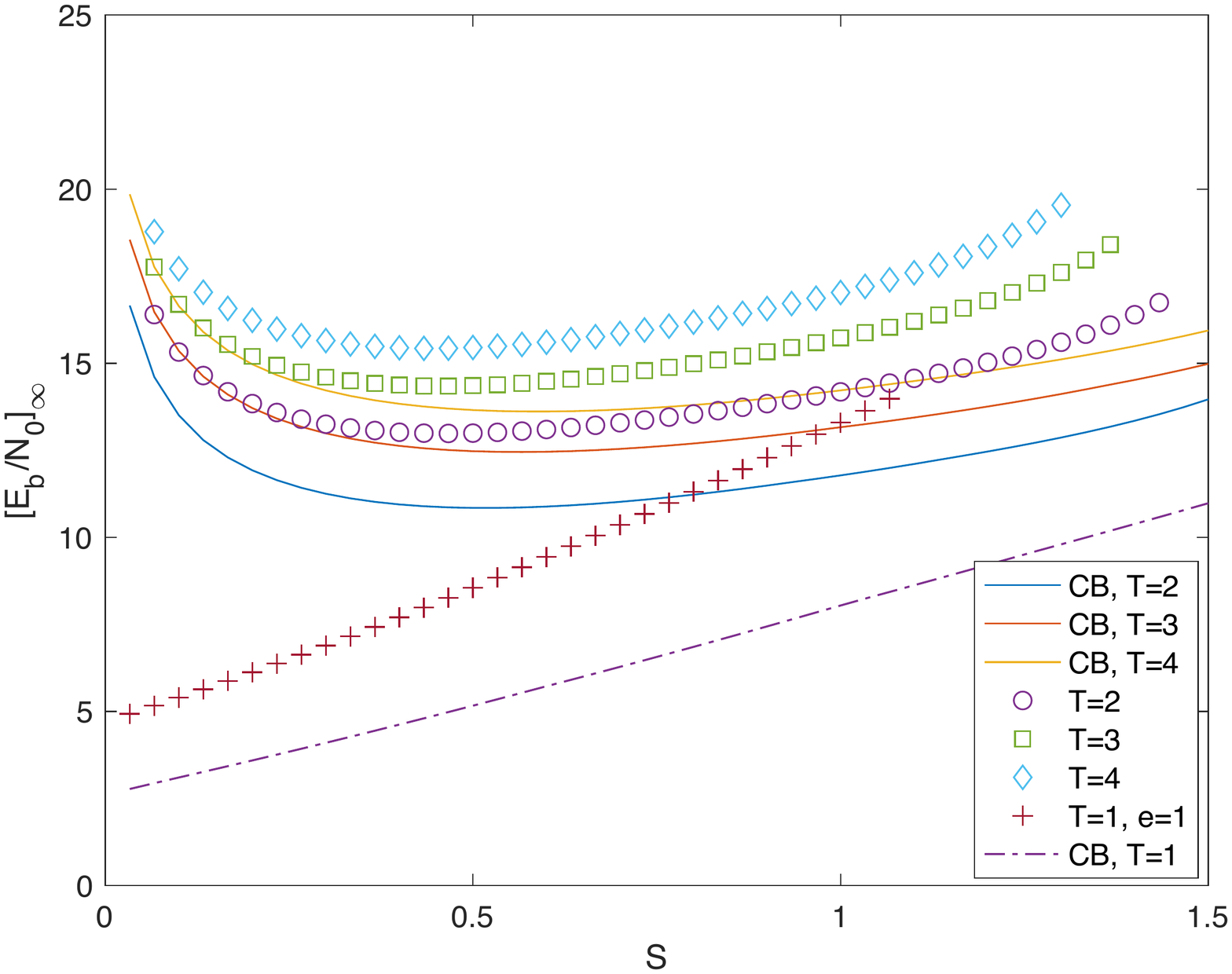}
      \caption{Achievable $E_b/N_0$ vs spectrum efficiency $S$ for \ac{IRSA} protocols with $\Lambda_1=0$ and option 2 PHY layer coding (2-layer extension, as in \cite{Ordentlich2017:low_complexity}), $\log_2M=100$ and target packet loss probability $\epsilon=0.005$. CB: on convergence boundary. }
    \label{fig:Toff_BPR_R2}
\end{figure}

Fig.~\ref{fig:Toff_BPR} has shown the limitations coming from BPR slot encoding to support \ac{MPR}. To overcome these limitations,  the work in \cite{Ordentlich2017:low_complexity} proposed a multilayer extension of  the basic binary scheme to  increase spectrum efficiency. We explored this solution by deriving the results presented in Fig.~\ref{fig:Toff_BPR_R2}. Here, a 2-layer BPR scheme is considered. We can note that this scheme is effective to improve spectrum efficiency for both convergence boundary and specific \ac{IRSA} schemes. No significant improvement is obtained in terms of achievable $E_b/N_0$. However, we can see from the figure that the scheme with $T=2$ allows working in the region with $S>1.1$ with efficiency comparable to that of the scheme with $T=1$.


\begin{figure}[t]
\centering
\includegraphics[width=\columnwidth]{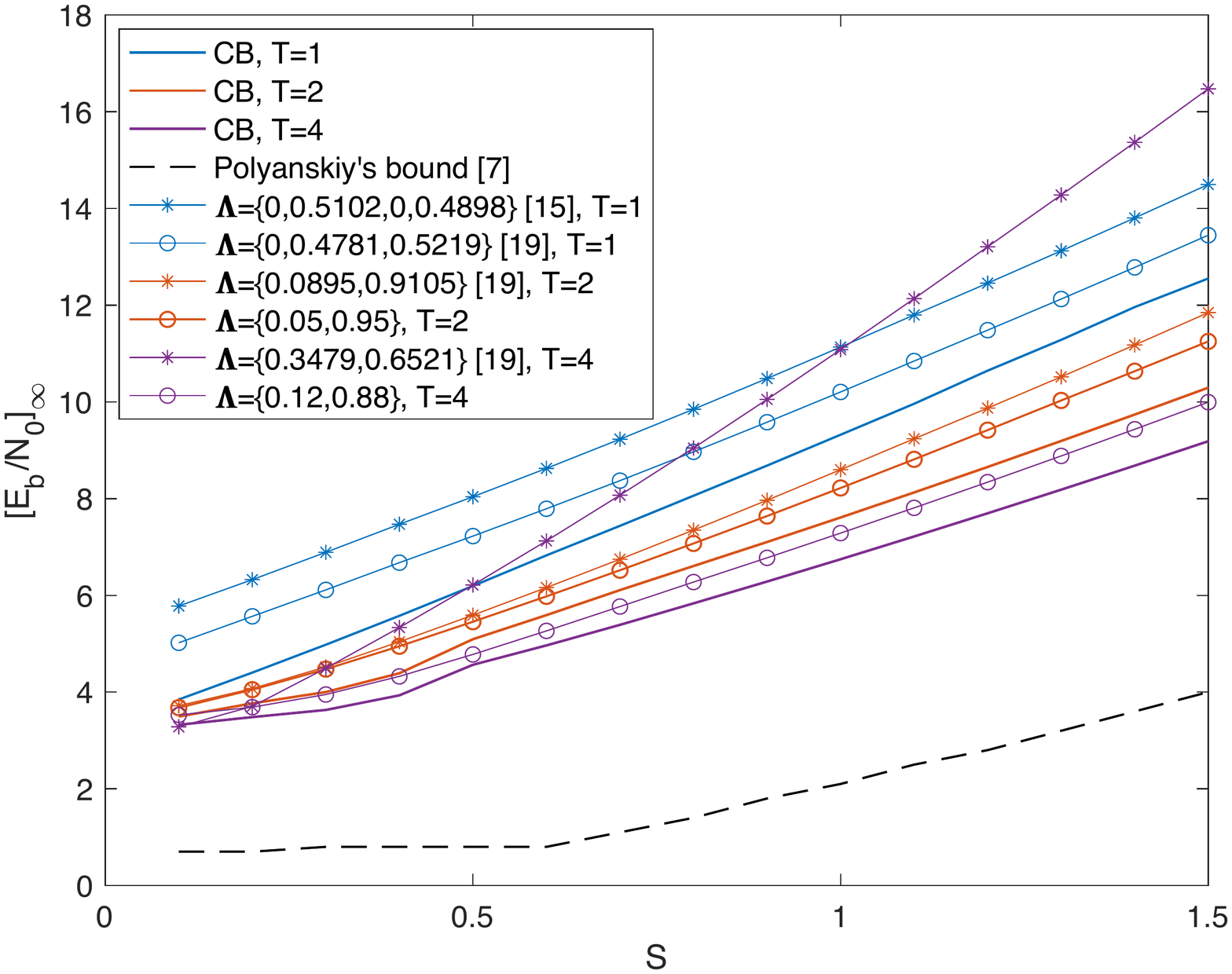}
      \caption{Achievable $E_b/N_0$ vs spectrum efficiency $S$ for the class of \ac{IRSA} protocols including those with $\Lambda_1=0$ and those with $\mathbf{\Lambda}=\{\Lambda_1,(1-\Lambda_1)\}$, $\Lambda_1>0$. Option 1 PHY layer coding, $\log_2M=100$ and target packet loss probability $\epsilon=0.005$. CB: on convergence boundary.}
    \label{fig:Toff_ideal_comp}
\end{figure}

\begin{figure}[t]
\centering
\includegraphics[width=\columnwidth]{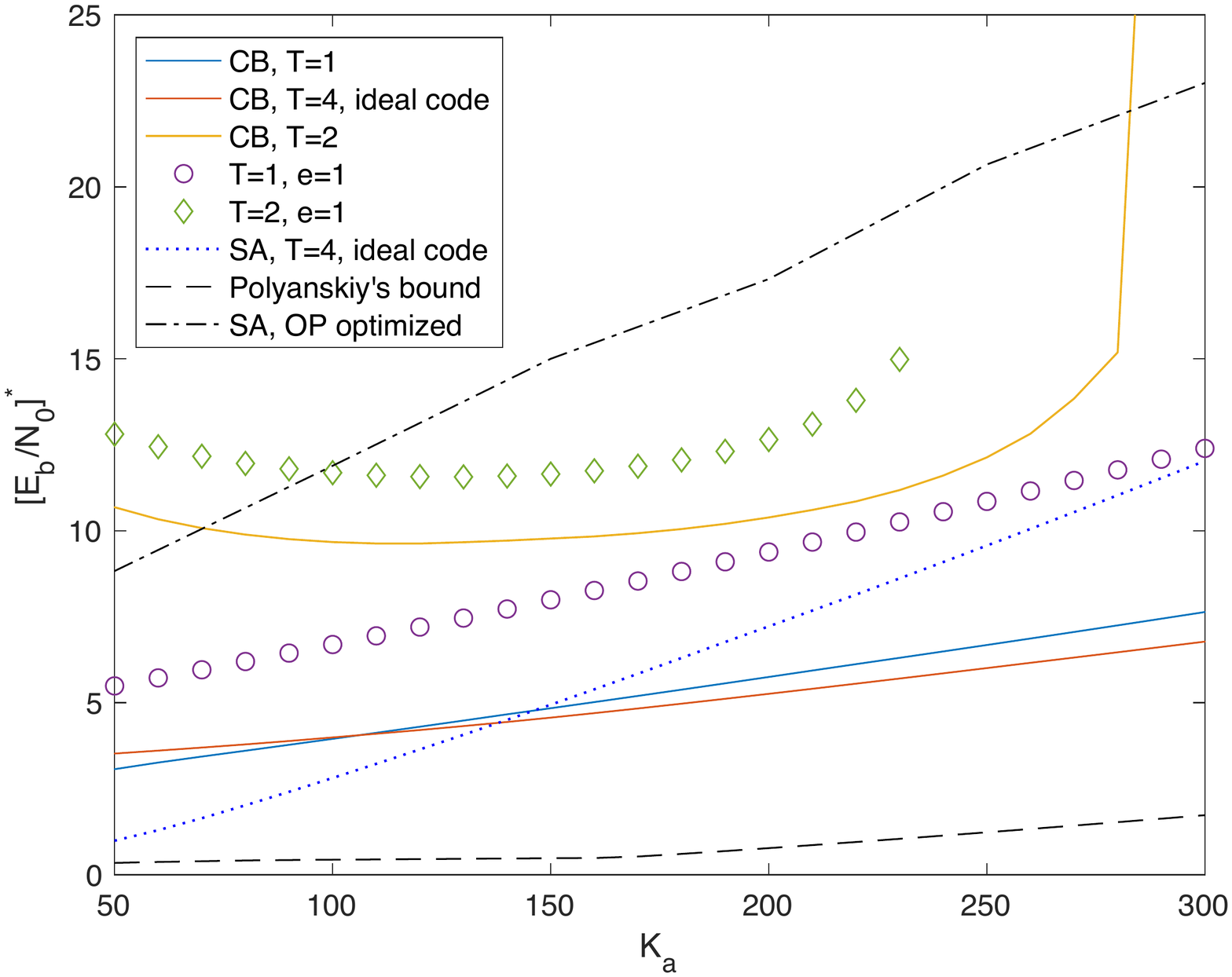}
      \caption{Achievable $E_b/N_0$ vs number of active transmitter $\Ka$ for \ac{IRSA}-based random access codes \ac{IRSA}  with $\Lambda_1=0$, using option 2 PHY layer coding (results for option 1 PHY layer coding are also reported denoted to as ideal code), with $\log_2M=100$, $N=30000$ and target packet loss probability $\epsilon=0.05$. CB: on convergence boundary. SA: slotted ALOHA. OP: results from \cite{Ordentlich2017:low_complexity}.}
    \label{fig:CompPoly}
\end{figure}

\begin{figure}[t]
\centering
\includegraphics[width=\columnwidth]{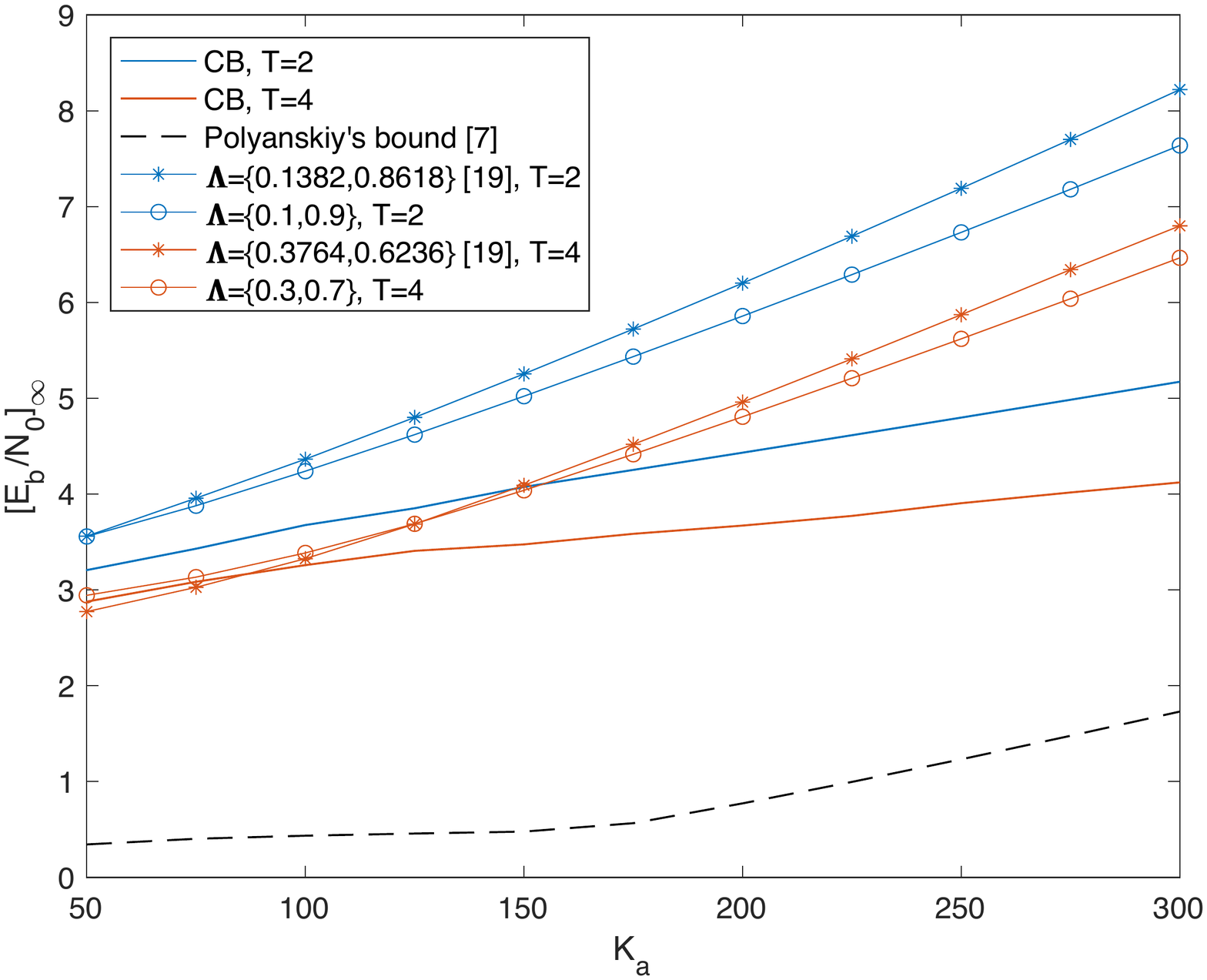}
      \caption{Achievable $E_b/N_0$ vs number of active transmitter $\Ka$ for the class of \ac{IRSA} protocols including those with $\Lambda_1=0$ and those with $\mathbf{\Lambda}=\{\Lambda_1,(1-\Lambda_1)\}$, $\Lambda_1>0$. Options 1  PHY layer coding, $\log_2M=100$, $N=30000$ and target packet loss probability $\epsilon=0.05$. CB: on convergence boundary. }
    \label{fig:CompPoly_comp}
\end{figure}

The next figure,  Fig.~\ref{fig:Toff_ideal_comp}, refers to schemes with option 1  PHY layer slot encoding and has the aim of comparing specific \ac{IRSA} protocols with different \acp{PGF} and different values of $\TMPR$. The figure shows the performance of six protocols with $\mathbf{\Lambda}$ and $\TMPR$ reported in the legend. Among them, there is the protocol mentioned in the Example 1, originally designed in \cite{liva2011:irsa} for a noiseless channel, and three protocols indicated in Tables I, II, and III of reference \cite{glebov2019} originally designed for random access schemes with $\epsilon=0.05$. The other protocols have been empirically designed for the system configuration investigated here, in the class of protocols with $\mathbf{\Lambda}=\{\Lambda_1,(1-\Lambda_1)\}$ having efficiency $\eta\geq0.5$. Although these protocols do not have a well defined threshold, according to Theorem~\ref{cor:suf_conditions}, they can achieve good energy efficiencies, as noted in \cite{vem2019:unsourced,glebov2019}, due to the small average number of packet replicas transmitted. The figure also shows the benchmarks on the convergence boundary and the random coding achievability bound given in \cite{Polyanskiy2017:Perspective} for the asymptotic scenario.
We can note from the figure that suitably designed \ac{IRSA} schemes with $\Lambda_1>0$ are able to get close to the convergence boundary for $\TMPR>1$. Also the schemes designed in \cite{glebov2019} have a good behavior in this setting, for $\TMPR=1, 2$.
We finally note, in comparison to Fig.~\ref{fig:Toff_ideal}, that the energy efficiency limit in terms of $E_b/N_0$ for \ac{IRSA} schemes with \ac{MPR} and without constraints on $\Lambda_1$ slightly improves with respect to schemes with $\Lambda_1=0$, approaching $6.7$ dB at the spectrum efficiency of $1 \, \mathrm{bit/channel \,\, use}$ for $T=4$.

In the last two figures, Fig.~\ref{fig:CompPoly} and Fig.~\ref{fig:CompPoly_comp}  we move to a finite length scenario with a fixed frame size $N=30000$ and target packet loss probability $\epsilon=0.05$. The aim of Fig.~\ref{fig:CompPoly} is to compare  the energy efficiencies of \ac{IRSA}-based  and \ac{SA}-based schemes, including also the optimized scheme reported in \cite{Ordentlich2017:low_complexity}. The figure shows the achievable $E_b/N_0$   as a function of the number of active transmitter $\Ka$. For \ac{IRSA}-based scheme the asymptotic approximation is evaluated, which is expected to be optimistic for small values of $\Ka$. The result obtained from the Polyanskiy's random coding bound in \cite{Polyanskiy2017:Perspective} on per-user message error probability is also plotted in the figure. The plots with the markers refers to specific \ac{IRSA} protocol considered in Example 1. 
The figure shows the positioning of \ac{IRSA}-based schemes, with $\Lambda_1=0$ in this specific case, with respect to other schemes and the gap with respect to random coding bound. 

We can see that \ac{IRSA}-based random access schemes with \ac{MPR} and option 1 PHY layer coding have a convergence boundary limit within no more than 5 dB from random coding bound and perform better than \ac{SA} schemes when the number of active transmitters is not too small. The use of option 2 PHY layer coding as a practical coding scheme to realize \ac{MPR} does not appear effective to achieve the potential gain of \ac{MPR} in \ac{IRSA}-based schemes. In fact, the simple \ac{IRSA}-based scheme with $T=1$ performs better than the schemes with \ac{MPR} and  coding option 2, as shown for the \ac{IRSA} protocol of Example~1 (see the plot with circles in the figure). As seen in Fig.~\ref{fig:Toff_BPR_R2}, the multilayer extension of option 2 PHY layer coding is needed to make it effective for \ac{MPR} at large values of required spectrum efficiencies, i.e. values of $\Ka$ larger than those  in Fig.~\ref{fig:CompPoly}.  However, \ac{IRSA}-based schemes with coding option 2 perform better than \ac{SA} schemes, even when the latter have  optimized configuration as in \cite{Ordentlich2017:low_complexity}.
A more effective, not optimum, PHY layer coding option  that could better capture the potential gain of \ac{MPR} in \ac{IRSA}-based schemes seems to be the coding solution proposed in \cite{vem2019:unsourced}  where the message is split into two parts and the first portion, transmitted through a compressive sensing scheme, selects the interleaver for an LDPC code, while the second part is encoded using the LDPC code. We report here the single simulation results given in \cite{vem2019:unsourced} for a random access code with $T=2$ that achieves nearly 7.5 dB of $E_b/N_0$ for $\Ka=100$ active users.

The last figure,  Fig.~\ref{fig:CompPoly_comp}, only refers to schemes with option 1  PHY layer slot encoding and has the aim of comparing specific \ac{IRSA} protocols with different \acp{PGF} and different values of $\TMPR$. The figure shows the performance of four protocols with $\mathbf{\Lambda}$ and $\TMPR$ reported in the legend. Two protocols are taken from Tables  II, and III in \cite{glebov2019}, while  the other two protocols have been empirically designed in the class of protocols with $\mathbf{\Lambda}=\{\Lambda_1,(1-\Lambda_1)\}$ having efficiency $\eta\geq0.5$. The figure also shows the results on the convergence boundary and the random coding achievability bound given in \cite{Polyanskiy2017:Perspective}.
We can first note that for the class of protocols enlarged to include schemes with $\Lambda_1>0$ the results on the convergence boundary appear quite optimistic with respect to the results of specific protocols, in this setting with target packet loss probability $\epsilon=0.05$, because the accuracy of bounding technique in Theorem 3 decreases as error probabilities increase. However,  suitably designed \ac{IRSA} schemes with $\Lambda_1>0$ are able to get close or even better than  the schemes designed in \cite{glebov2019} for this setting. The results still show $3$ to $5\,\mathrm{dB}$ gap from random coding bound and can be also compared to the results achieved by other state-of-the-art techniques for unsourced random access reported in Fig.~3 of \cite{pradhan2021:icc}. In this figure, it is shown that using random spreading to control multiple access interference and a robust coding scheme allows to fill the gap to random coding bound, when $\Ka$ is smaller than 125, and to stay within 2 dB from the bound for larger $\Ka$. The comparison with these techniques, which appear more efficient to counteract random interference,  allows us to position \ac{IRSA}-based schemes within the framework of state-of-the-art unsourced random access schemes.




\section{Conclusions}\label{Sec:conclusions}

In this paper,   \ac{IRSA}-based  grant-free random access  schemes operating over a \ac{GMAC} have been investigated, presented as a special instance of the class of slotted random access codes in the unsourced setting, which are built by concatenation of a MAC layer code extended over the frame and a PHY layer code for transmission in each slot of the frame. 
A framework for performance analysis in asymptotic conditions of \ac{IRSA} MAC layer codes has been first provided, to allow the evaluation of 
minimum packet loss probability and average load threshold, in presence of both collisions and slot decoding errors. A converse bound  for threshold values, 
which defines a region for the average load  where packet loss probability can never converge to values below a suitable minimum, has been also derived.
Then, by exploiting this analysis the performance of \ac{IRSA}-based random access schemes has been evaluated with two basic options for PHY layer codes: i) optimum coding for random access with up to $T$ active transmitters, ii) \ac{BPR} coding for $T$-user binary adder channel, concatenated with a binary linear inner code as in \cite{Ordentlich2017:low_complexity}.  

The results have shown the performance limits  of \ac{IRSA}-based random access schemes in terms of achievable $E_b/N_0$ and tradeoff between energy and spectrum efficiency. More specifically, it is found that the energy efficiency limit in terms of $E_b/N_0$ is between $7$ and $8\,\mathrm{dB}$ at a spectrum efficiency of $1\,\mathrm{bit/channel\,\,use}$, with a very large  improvement over the reference case of \ac{SA} with \ac{MPR}  when the network load is high. 
In the scenario with finite frame size, the convergence boundary limit is within no more than $5\,\mathrm{dB}$ from the random coding bound of \cite{Polyanskiy2017:Perspective}. 
It is also found that the use of BPR coding option for  PHY layer coding  as a practical coding scheme to realize MPR does not appear effective to achieve the potential gain of MPR in IRSA-based schemes. The gap may be partially reduced by using a multilayer extension or by considering other PHY layer coding schemes as for example the one proposed in \cite{vem2019:unsourced}.

We close the paper by mentioning some possible extensions to different \ac{MAC} layer schemes and \ac{PHY} channel models.
Regarding the first point, we mention the possibility to extend the results to \acf{CSA} \cite{paolini2015:csa}, of which \ac{IRSA} represents a special case.
In \ac{CSA} the user message is split into fragments, that are encoded by a fragment-oriented encoder for erasure correction; encoded fragments are then individually encoded by the PHY layer code and the obtained data units are sent in random sub-slots of the frame.
Extension to this setting of the density evolution analysis developed in Section~\ref{Sec:MAC_performance} (and, consequently, of the asymptotic threshold) is immediate: $\bar{d}$ in \eqref{eq:density_evolution_mpr} should be replaced by $1/\eta$, being $\eta$ the \ac{CSA} efficiency \cite{paolini2015:csa}, the probabilities $P_{\text{E}|t}$ should be interpreted at sub-slot level, and  \eqref{eq:density_evolution_mprq} should be replaced by equation (5) in \cite{paolini2015:csa}.
The extension of other results, including Theorem~2, Theorem~3 and its corollary, and the study of the $E_b/N_0$ limit on the convergence boundary, is instead less direct and would deserve an ad-hoc investigation.
Regarding \ac{PHY} layer extensions, generalization of the results to fading channel models would be particularly worthwhile.
In this respect, we point out that aspects to be carefully handled towards such an extension concern non-coherent estimation of the number of transmissions per slot, channel state information to be acquired directly from the signal samples received in a slot (since the access scheme is grant-free) according to some pilot design and assignment scheme, imperfect subtraction of interference due to possible contamination of the pilot of the packet replica to be subtracted.
Finally, we mention that the convergence speed of the iterative \ac{SIC} procedure, related to the latency of the scheme, has not been addressed in this paper. 
An extension of the proposed analysis to incorporate, for example, the effect of a finite number of \ac{SIC} iterations would certainly represent a worthwhile direction of investigation.


\appendices

\section{Proof of Theorem~\ref{th:density_evolution_mpr}}\label{appendix:proof_th1}

Let us start with a preliminary Lemma \cite{liva2011:irsa,paolini2015:csa} which will be exploited in the proof.
\begin{lemma}
In the asymptotic conditions, when $\nslot \rightarrow \infty$ and $\mathbb{E}[\Ka] \rightarrow \infty$ for constant $\avgG=\mathbb{E}[\Ka]/\nslot=\pi \alpha$,
the probability that the generic edge is connected to a slot-node of degree $h$ is obtained as
\begin{align}\label{rho_h}
\rho_h=\frac{({\avgG}\bar{d})^{(h-1)}}{(h-1)!} \exp ({-{ \avgG}\bar{d}})
\end{align}
which is the probability that there are other $h-1$ transmissions colliding in the slot.
\label{lemma_prelim}
\end{lemma}
We will first prove the theorem in the special case with $T=1$, for better understanding of proof method. 
We derive density evolution equations by considering the effects of a slot decoding  error event in the iterative decoding procedure. If $T=1$, we can say that at the iteration $\ell$ the probability $1-p_{\ell}$, that a generic edge is connected to  a slot-node that  becomes resolved after the perfect cancellation of interference coming from detected burst-nodes is given by the probability that $h-1$ packets are canceled out of $h$ colliding packets, and slot-decoding is successful, i.e.,
\begin{align}\label{main_prob}
1-p_{\ell}=(1-P_{\text{E}|1})\sum_{h\geq1} \rho_{h} (1-q_\ell)^{h-1} 
\end{align}
where $\rho_h$ is the probability that the generic edge is connected to a slot-node of degree $h$ and $1-q_\ell$ is the probability that each one of the other  $h-1$ edges is connected to a detected burst-node (that can be used to cancel interference in the slot).
The result of the theorem with $T=1$ is obtained by exploiting the results of Lemma~\ref{lemma_prelim}. 
By using \eqref{rho_h} into \eqref{main_prob}, after some manipulations we get 
the statement.
Moreover, $q_\ell$ can be evaluated from $p_{\ell-1}$ as from Lemma~\ref{th:density_evolution_mpr_ch}.

We can now follow the same method for the case $T>1$.  We can say that at the iteration $\ell$ the probability, $1-p_{\ell}$, that a generic edge is connected to  a slot-node that  becomes resolved after the perfect cancellation of interference coming from detected burst-nodes is given by the probability that any combination of $h-t$ packets, with $t\in [1,T]$,  is canceled out of $h$ colliding packets, and slot-decoding is successful with $t$ remaining packets in the slot, i.e.,
\begin{align}\label{main_prob_T}
1-p_{\ell}=\sum_{h\geq1} \rho_{h} \! \sum_{t=1}^{\min (T,h-1)} \!\! (1-P_{\text{E}|t}) \binom{h-1}{t-1}q_\ell^{t-1}(1-q_\ell)^{h-t} 
\end{align}
where $\rho_h$ is the probability that the generic edge is connected to a slot-node of degree $h$ and $1-q_\ell$ is the probability that each one of the edges is connected to a detected burst-node (that can be used to cancel interference in the slot).
By 
reordering the pair of 
indexes $(t,h)$, we get
\begingroup
\allowdisplaybreaks
\begin{align}\label{main_prob_T2}
1-p_{\ell} &=\sum_{t=1}^{T} (1-P_{\text{E}|t}) q_\ell^{t-1}\sum_{h\geq t} \rho_{h}  \binom{h-1}{t-1}(1-q_\ell)^{h-t} \nonumber \\
&=\sum_{t=1}^{T} (1-P_{\text{E}|t}) q_\ell^{t-1}\sum_{h'\geq 0} \rho_{h'+t}  \binom{h'+t-1}{t-1}(1-q_\ell)^{h'} \, .
\end{align}
\endgroup
The final result of the theorem is obtained by exploiting the results in Lemma~\ref{lemma_prelim}, as follows
\begin{align}\label{rho_h_T}
\rho_{h'+t}=\frac{({ \avgG}\bar{d})^{(h'+t-1)}}{(h'+t-1)!} \exp ({-{ \avgG}\bar{d}}) \, .
\end{align}
By using \eqref{rho_h_T} into \eqref{main_prob_T2}, after some manipulations we get the result in the statement.
The starting point of the recursion is obtained as $p_0=f_s(1)$.
%


\section{Proof of Theorem~\ref{cor:suf_conditions}}\label{appendix:proof_th2}

Let us consider \ac{IRSA} schemes with $d_{\min}>1$. We first prove that a value $\avgG_0$ always exists such that for $\avgG\leq \avgG_0$ there is only one intersection $(p_0,q_0)$ in the $(p,q)$ plane between the functions $q=f_b(p)$ and $p=f_s(q)$,  and $p_0=O(P_{\text{E}|1})$, $q_0=O(P_{\text{E}|1})$ when $P_{\text{E}|1}$ is close to zero.
In fact, by observing that $f'_b(p)$ 
is positive and monotonically increasing starting from $f'_b(0)<1$, there is a value $A$ such that $f'_b(A)=1$ and $f'_b(p)\leq1$ for $p\leq A$. Since $f'_s(q)$ 
is always positive and dependent on $\avgG$, and $f_s(q)$ starts from $f_s(0)=P_{\text{E}|1}$ and reaches $f_s(1)\leq P_{\text{E}|1}+[1-\exp(-{\avgG} \bar{d})]$ there is always a value $\avgG_0$ such that $f_s(q)<A$ (there is at least one intersection) and  $f'_s(q)<1$ (the intersection is single) for $q\in [0,1]$ and $\avgG\leq \avgG_0$. \\
Moreover, since from equation \eqref{eq:density_evolution_mpr} we can easily obtain that $f_s(q) \leq P_{\text{E}| 1} + (1 - P_{\text{E}| 1}) [ 1-\exp (-{\avgG} \bar{d} q)] \leq P_{\text{E}| 1} +  \avgG \bar{d} q $, and since $f_b(p)\leq p$ holds, the point $(p_0,q_0)$ will be always located in the region bounded by the two lines $q=p$ and $p=P_{\text{E}| 1} + \avgG \bar{d} q $. These two lines have intersection in the point with $p=q=P_{\text{E}| 1} /(1-\avgG \bar{d} )$, if $\avgG \bar{d} <1$. Hence, $p_0=O(P_{\text{E}|1})$ and $q_0=O(P_{\text{E}|1})$, i.e.,  $p_0, q_0$ are close to zero when $P_{\text{E}|1}$ is close to zero.

As a final step, by leveraging the fact that the point $(p_0,q_0)$ is close to $(0,0)$, we try to evaluate it as the intersection in the $(p,q)$ plane of the two functions $q= \tilde{f}_b(p)+ O(p^{d_{\min}})$ and $p=\tilde{f}_s(q)+O(q^2)$ that include the lowest order approximations of $q=f_b(p)$ and $p=f_s(q)$, respectively, for $p, q$ close to 0.
The intersection  is obtained as the solution  of the equation $p=f_s(f_b(p))$, which can be rewritten as 
\begin{align}
p=\tilde{f}_s(\tilde{f}_b(p))+O(p^{d_{\min}}) \, .
\end{align}
Using \eqref{eq:ftilde_b} and \eqref{eq:ftilde_s}, it becomes
\begin{align}
p-\Lambda_{d_{\min}} d_{\min} \avgG (P_2 - P_{\text{E}|1}) p^{d_{\min}-1}+O(p^{d_{\min}})= P_{\text{E}|1}
\end{align}
where $P_2=1$ if $T=1$ and $P_2=P_{\text{E}|2}$ if $T>1$. 
The left-hand side of the equation for $p$ close to 0, as $P_{\text{E}|1}$, is approximated by the first-order term. Therefore, we can conclude as follows.
If $d_{\min} \geq 3$, the solution becomes $p_0\approx P_{\text{E}|1}$. If $d_{\min} =2$, the solution becomes $p_0 \approx P_{\text{E}|1}/ [1-2\Lambda_{2} \avgG (P_2 - P_{\text{E}|1}) ]$, for all $\avgG\leq\avgG_0$. In particular,  when $T>1$, it reduces to $p_0\approx P_{\text{E}|1}$ if $P_2=P_{\text{E}|2}$ satisfies the condition $P_{\text{E}|2}-P_{\text{E}|1}\ll 1/(2\Lambda_{2} \avgG)$, which can be restricted to $P_{\text{E}|2}-P_{\text{E}|1}\ll 1/(2\Lambda_{2} \avgG_0)$.


\section{Proof of Theorem~\ref{th:threshold_bound_new}}\label{appendix:proof_th3}

Let us first consider the following lemma \cite{paolini2015:csa}.
\begin{lemma}\label{lemma_conv}
A necessary (but not sufficient) condition for the convergence to 0 as $\ell \rightarrow \infty$ of the recursion $p_\ell=f(p_{\ell-1})$ describing the evolution of the \ac{SIC} process in the decoding of \ac{IRSA} protocols with $\Lambda_1=0$, when $P_{\text{E}|t}=0$ $\forall t$, is 
\begin{align}
\int_0^1 f_s(q) \text{d}q + \eta \leq1 .
\end{align}
\end{lemma}
\noindent The condition of the lemma describes the existence of an ``open tunnel" in the EXIT chart of the the \ac{IRSA} protocol. This condition can be reformulated for the cases with non zero slot decoding error probability, as follows.
\begin{lemma}\label{lemma_conv_pe}
A necessary (but not sufficient) condition for convergence to $p_\infty \leq P_{\text{E}|1}(1+e)$ as $\ell \rightarrow \infty$ of the recursion $p_\ell=f(p_{\ell-1})$ describing the evolution of the \ac{SIC} process in the decoding of \ac{IRSA} protocol, in presence of slot decoding errors with probabilities in the set $\mathbf{P}_{\text{E}}$, is 
\begin{align}\label{eq:area_tunnel}
&\int_0^1 f_s(q) \text{d}q + \eta - \eta [\Lambda_1P_{\text{E}|1}(1+e)+(1-\Lambda_1)P_{\text{E}|1}^2(1+e)^2] \notag \\ 
&\leq 1
\end{align}
that holds if $P_{\text{E}|1}(1+e)<0.5$. 
\end{lemma}
\begin{IEEEproof}
The result of the lemma can be proved with the same method used for Lemma \ref{lemma_conv}. Let us take Fig. \ref{fig:exit} as a reference example. There is an open tunnel to the intersection $(p_{\infty},q_{\infty})$ of the two functions $q=f_b(p)$ and $p=f_s(q)$ if the sum of the areas of the region below $f_b(p)$ and the region on the left of $f_s(q)$, minus the area of the intersection of these two regions, is smaller than 1. The area of the region below $q=f_b(p)$ is always $\eta$. The area of the region on the left of $f_s(q)$ is evaluated with the integral in the first term of \eqref{eq:area_tunnel}. The area of the intersection of the two regions is denoted with $A_{\cap}$ and depends on the two functions $f_b(p)$ and $f_s(q)$. 
We find now an upper bound for $A_{\cap}$ that only depends on the \ac{IRSA} efficiency $\eta$ and the probability $\Lambda_1$. 
Let us consider the function $\hat{f}_b(p)=\eta [\Lambda_1+(1-\Lambda_1)2p]$. 
It is a line and the area of the region below this line is  $\eta$, as for $f_b(p)$. Since $f_b(p)$ is convex, we have $\hat{f}_b(p)\geq f_b(p)$ in a range $[0,p_x]$, where $p_x$ must be greater than $0.5$. Since $p_{\infty}\leq P_{\text{E}|1}(1+e)$ and $q_{\infty}=f_b(p_{\infty})\leq \hat{f}_b(p_{\infty})$, we obtain $A_{\cap} \leq \eta [\Lambda_1P_{\text{E}|1}(1+e)+(1-\Lambda_1)P_{\text{E}|1}^2(1+e)^2]$.
\end{IEEEproof}

We can prove the first inequality of Theorem~\ref{th:threshold_bound_new} by exploiting Lemma~\ref{lemma_conv_pe}.
Let us consider an \ac{IRSA} scheme with average load $\avgG \leq \avgG^{\star}$ that operates with successful decoding in asymptotic conditions, i.e., $\nslot \rightarrow \infty$ and $\mathbb{E}[\Ka] \rightarrow \infty$ for constant $\avgG=\mathbb{E}[\Ka]/\nslot$. The \ac{SIC} process is governed by the recursion given in Theorem~\ref{th:density_evolution_mpr}. We first evaluate the integral of the slot-node function $f_s(q)$ in \eqref{eq:density_evolution_mpr}, by setting $\eta=1/\bar{d}$. Application of integration by parts $t$ times yields
\begingroup
\allowdisplaybreaks
\begin{align}
\int_0^1 f_s(q) \text{d}q 
&=1- \sum_{t = 1}^{\TMPR } (1 - P_{\text{E}| t}) \int_0^1 \exp \left( -\avgG q/\eta   \right)\frac{ \left( \avgG q/\eta  \right)^{t-1}}{(t-1)!} \text{d}q  \nonumber \\
&=1- \sum_{t = 1}^{\TMPR } (1 - P_{\text{E}| t}) \left[ \frac{\eta}{\avgG} - \frac{\eta}{\avgG} \exp \left( -\frac{\avgG}{\eta } \right) \sum_{k=0}^{t-1} \frac{1}{k!} \left( \frac{\avgG}{\eta}\right)^k \right] \, .
\end{align}
\endgroup
Next, we use this expression to evaluate the necessary condition in Lemma \ref{lemma_conv_pe} for the convergence of IRSA decoding process, obtaining
\begingroup
\allowdisplaybreaks
\begin{align}
&- \frac{\eta}{\avgG} \sum_{t = 1}^{\TMPR } (1 - P_{\text{E}| t}) \left[ 1 - \exp \left( -\frac{\avgG}{\eta}  \right) \sum_{k=0}^{t-1} \frac{1}{k!} \left( \frac{\avgG}{\eta}\right)^k \right] \nonumber\\
&+ \eta [1 -\Lambda_1P_{\text{E}|1}(1+e)-(1-\Lambda_1)P_{\text{E}|1}^2(1+e)^2] \leq 0
\end{align}\endgroup
if $P_{\text{E}|1}(1+e)<0.5$, which leads directly to the result \eqref{eq:G_threshold_bound_new} of the theorem.

The second inequality of Theorem~\ref{th:threshold_bound_new} can be proved by observing that the convergence point $(p_\infty,q_\infty)$ is a point of both functions $q=f_b(p)$ and $p=f_s(q)$. Since $f_b(p)\geq f_b(0)=\Lambda_1$, we have $q_\infty\geq\Lambda_1$. Since $f_s(q)$ is  monotonically increasing with $q$, we have $p_\infty=f_s(q_\infty)\geq f_s(\Lambda_1)$. Finally, the condition $p_\infty\leq P_{\text{E}|1}(1+e)$ leads to $f_s(\Lambda_1)\leq P_{\text{E}|1}(1+e)$.


\section{IRSA-based Random Access over \ac{GMAC} with imperfect \ac{SIC}}\label{appendix:imperfect_sic}

In this appendix we explore the effects of an imperfect \ac{SIC} process, with the aim to understand how it affects the asymptotic analysis in Section~\ref{Sec:MAC_performance} and what is its impact on performance results.
We consider the model proposed in \cite{dumas2021:canc_err} to describe the effectiveness of interference cancellation. According to this model, in absence of additive noise the probability of successfully decoding the signal in a slot with $h$ colliding packets, after the subtraction of $h-t$ known packets (whose replicas have been decoded in other slots), is given by $\gamma^{h-t}$, where the parameter $\gamma \leq 1$ is called the ``\ac{SIC} efficiency'' ($\gamma=1$ is the ideal case). 
Note that this model only accounts for \ac{SIC} process imperfections, as the packets to be cancelled are assumed to be perfectly decoded, which is fully consistent with our assumption of ideal error detection.
The following analysis is carried out by extending this model to the \ac{GMAC},
where the probability of successful decoding for a slot with $h$ colliding packets and $h-t$ cancelled packets can be evaluated as
\begin{align}
(1-P_{\text{E}|t})\gamma^{h-t} \, .
\end{align}
The first outcome of this analysis is an extension of Theorem~\ref{th:density_evolution_mpr}, as follows.

\begin{corollary}\label{th:density_evolution_mpr_impsic}
Let $\nslot \rightarrow \infty$ and $\mathbb{E}[\Ka] \rightarrow \infty$ for constant $\avgG=\mathbb{E}[\Ka]/\nslot=\pi \alpha$. 
Let $\csaefficiency=1/\bar{d}$ be the efficiency of the \ac{IRSA} protocol as defined in \eqref{eq:csa_efficiency}. 
Then, at the $\ell$-th iteration of the \ac{SIC} process we have
\begingroup
\allowdisplaybreaks
\begin{align}\label{eq:density_evolution_gamma}
p_{\ell} &=1-\exp [-{ \avgG} \bar{d}(1-\gamma+\gamma q_{\ell})] \sum_{t = 1}^{\TMPR } (1 - P_{\text{E}| t}) \frac{\left({\avgG} \bar{d}q_{\ell}\right)^{t-1}}{(t-1)!} 
=f_s(q_\ell) \\
q_{\ell} &=(1/\bar{d}) \sum_{d} \Lambda_d \, d \, p_{\ell-1}^{d-1} = \Lambda'(p_{\ell-1})/\bar{d}=f_b(p_{\ell-1})
\end{align}
\endgroup
where the starting point of the recursion is $p_0=1-\exp ( - \avgG \bar{d} ) \sum_{t = 1}^{\TMPR } (1 - P_{\text{E}| t})\left(  \avgG \bar{d} \right)^{t-1} / (t-1)!$.   
\end{corollary}
\begin{IEEEproof}
We can follow the same proof method of Theorem~\ref{th:density_evolution_mpr}, with the following variation. At \ac{SIC} iteration $\ell$ the probability that a generic edge is connected to a slot node that becomes resolved after the cancellation of interference coming from detected burst nodes, $1-p_{\ell}$, is now given by
\begin{align}\label{main_prob_gamma}
1-p_{\ell} 
&=\sum_{h\geq1} \rho_{h} \sum_{t=1}^{\min (T,h-1)} (1-P_{\text{E}|t}) \binom{h-1}{t-1}q_\ell^{t-1}[(1-q_\ell)\gamma]^{h-t} \, .
\end{align}
The last term in \eqref{main_prob_gamma} includes the model for the probability of successful slot-decoding with $h-t$ canceled packets, out of $h$ colliding packets, and  $t$ remaining packets in the slot.
\end{IEEEproof}
We first note that when $\gamma=1$ we obtain the result of Theorem~\ref{th:density_evolution_mpr}, as expected. 
We can now compare the results with $\gamma<1$ with those of Theorem~\ref{th:density_evolution_mpr}. 
By looking at $f_s(q)$ we easily see that
\begin{align}
f_s(0) &= 1- \exp [-{ \avgG} \bar{d}(1-\gamma)] (1 - P_{\text{E}| 1}) \notag \\ 
&= P_{\text{E}| 1}+ (1 - P_{\text{E}| 1}) (1- \exp [-{ \avgG} \bar{d}(1-\gamma)])
\end{align}
which is the lower limit for the asymptotic value $p_{\infty}$, that is related to the floor of the packet loss probability $P_L$. 
It is given by $P_{\text{E}|1}$ when $\gamma=1$, but an extra term, dependent on $\avgG$, arises when $\gamma<1$. 
This extra term is approximated by $(1 - P_{\text{E}| 1}){ \avgG} \bar{d}(1-\gamma)$ when $1-\gamma$ is sufficiently small, and captures the effects of imperfect \ac{SIC} on the the floor of $P_L$.
Moreover, since the derivative $f'_s(p)$ is still positive, considering that $p_{\infty}\geq P_{\text{E}| 1}+ (1 - P_{\text{E}| 1}) (1- \exp [-{ \avgG} \bar{d}(1-\gamma)])\geq P_{\text{E}| 1}$, Corollary~\ref{cor:pl_limit} remains valid.

The numerical validation is presented in Fig.~\ref{fig:PLev_gamma} which compares the results reported in Fig.~\ref{fig:PLev} ($\gamma=1$) with those obtained with imperfect \ac{SIC} with $\gamma=0.97$. 
It is noted that the floor of $P_L$ is affected by the additional contribution dependent on $\avgG$, while the load threshold slightly decreases.
To make imperfect \ac{SIC} effects negligible, $\gamma$ should approach~$1$ in a way such that ${ \avgG} \bar{d}(1-\gamma) \ll P_{\text{E}| 1}$.

\begin{figure}[t]
\centering
\includegraphics[width=\columnwidth]{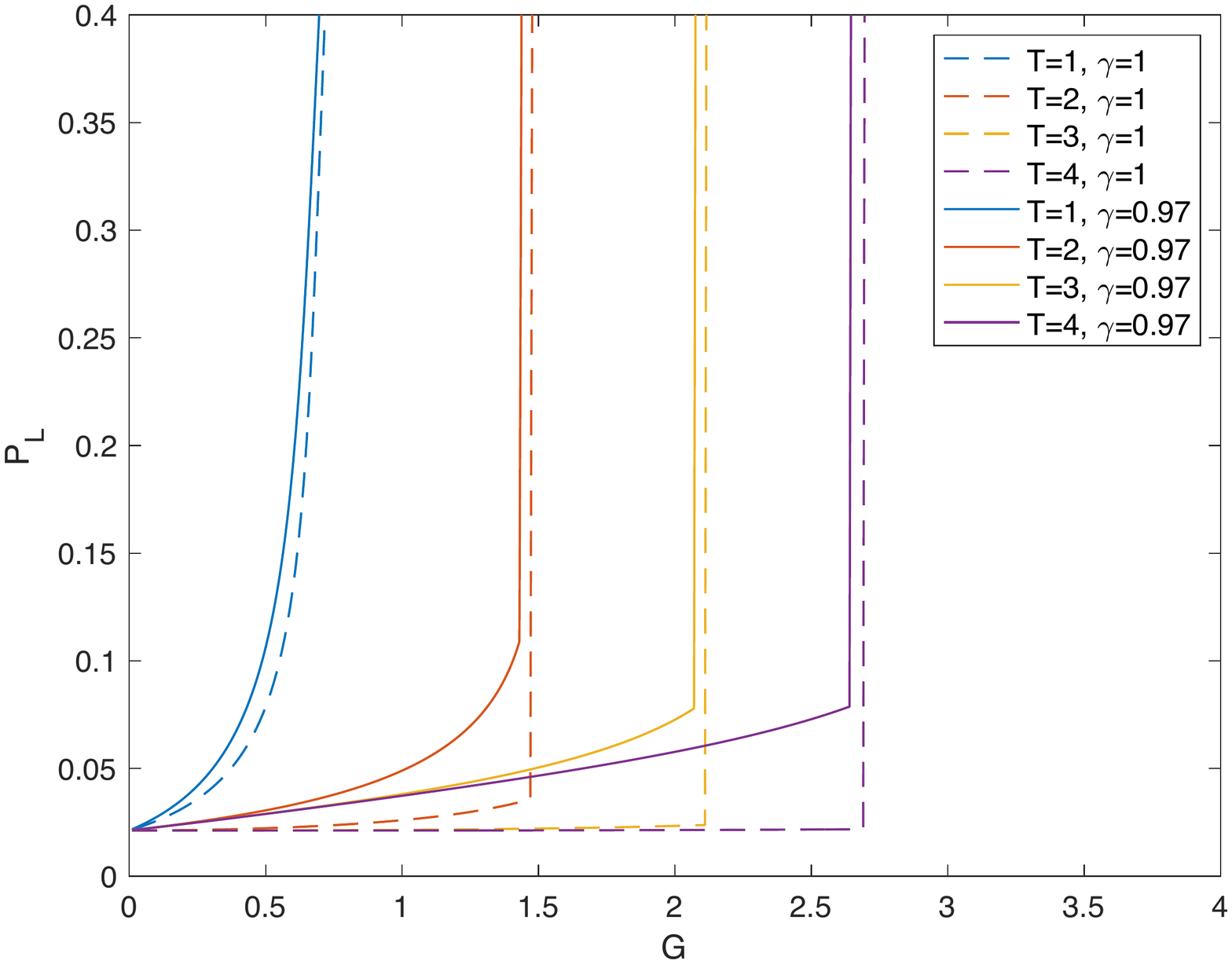}
      \caption{Packet loss probability as function of average load  $\avgG$ for \ac{IRSA} protocol obtained with $d_{\max}=4$, $\mathbf{\Lambda}=\{0,0.5102, 0, 0.4898\}$, $P_{\text{E}|t}=0.2, \;\forall t$. Solid line is for imperfect \ac{SIC} with $\gamma=0.97$. Dashed line is for perfect \ac{SIC}.}
    \label{fig:PLev_gamma}
\end{figure}


\section{Decoding error probability for option-2 PHY layer coding}\label{appendix:decoding_option2}

Let us first consider the case of ideal coding achieving coding rate limits in the finite blocklength regime.
As for coding option 1, we need a relationship 
that connects the coding rate with the error probability. We consider here the Gaussian approximation for the maximal coding rate achievable with error probability $P_{\text{E}}<\epsilon$ and finite codeword length, proposed  in \cite{Polyanskiy2010}, given by 
\begin{align}
\frac{\log_2 M'}{n}=C-\sqrt{\frac{V}{n}}Q^{-1}(\epsilon)+O\bigg(\frac{\log_2 n}{n}\bigg)
\end{align}
where $C$ is the channel capacity, $V$ is the channel dispersion, $\log_2M'=kT+r$, and $Q(x)=(1/\sqrt{2\pi})\int_x^\infty e^{-z^2/2}dz$. This leads to:
\begin{align}\label{eq:PPV}
P_{\text{E}}<Q\left[ \sqrt{\frac{n}{V}} \left( C-\frac{\log_2 M'}{n}+O\bigg(\frac{\log_2 n}{n}\bigg) \right)\right] .
\end{align}

For the simple case of $T=1$, i.e., without \ac{MPR} capability, there is no need of BPR coding and the inner coding is not constrained to be binary or linear. In this case, the coding rate limits are those of a simple AWGN channel. Therefore, to evaluate error probability, \eqref{eq:PPV} can be specialized for AWGN channel \cite{Polyanskiy2010} with
\begin{align}
C=\frac{1}{2}\log_2 \left( 1+P/\sigma^2 \right)
\end{align}
\begin{align}
V=\frac{1}{2}\log_2 e \frac{(P/\sigma^2+2)P/\sigma^2}{(P/\sigma^2+1)^2}
\end{align}
and by replacing $O(\log_2n/n)$ with $\log_2n/(2n)$. As for coding option 1, $P/\sigma^2=(2\eta\log_2M/n)E_b/N_0$ sets an energy constraint in the transmission which is, according to \eqref{eq:ebno},  related to $E_b/N_0$.

In the most general case with $T>1$,  binary random coding over a binary-input modulo-2 AWGN channel has to be considered, according to \eqref{eq:mod2-awgn}, to evaluate channel capacity $C$ and dispersion $V$ in  \eqref{eq:PPV}. Starting from the mutual information of the channel with equiprobable binary input, the two parameters are derived in \cite{Ordentlich2017:low_complexity} as
\begin{align}
C=\mathbb{E}[i(\tilde{{z}})], \qquad V=\text{Var}[i(\tilde{{z}})]
\end{align}
where
\begin{align}
i(x) =   \log_2\frac{2f_{\tilde{{z}}}(x)}{f_{\tilde{{z}}}(x)+f_{\tilde{{z}}_1}(x)} ,\;\; 0\leq x<2
\end{align}
and $\tilde{{z}}=[{{z'}}] \;\text{mod}\; 2$, $\tilde{{z}}_1=[{{z'}}+1] \;\text{mod} \; 2$ are random variables related to the noise term in \eqref{eq:mod2-awgn}, which is $z'\in \mathcal{N}(0,\sigma'^2)$ with $\sigma'^2=\sigma^2/4P$. The probability density function 
of the modulo-2 random variables $\tilde{z}$ and $\tilde{z}_1$ can be easily derived, for $0\leq x<2$, as
\begingroup
\allowdisplaybreaks
\begin{align}
f_{\tilde{{z}}}(x)  &= \frac{1}{\sqrt{2\pi\sigma'^2}} \sum_{i=-\infty}^\infty \exp \left( \frac{-(x-2i)^2}{2\sigma'^2}\right) \\
 {f_{\tilde{{z}}_1}(x)} &= f_{\tilde{{z}}}(x-1) .
\end{align}
\endgroup
Note that $\sigma'^2=\sigma^2/4P$ includes the energy constraint on transmission which is related to $E_b/N_0$.

\section{Estimation of the number $t_i$ of transmissions in slot $i$ and effects of estimation errors on PHY-layer decoding}\label{appendix:t_estimation} 

In Section II.B the model of the slot decoder has been defined by assuming that it includes a block able to provide perfect estimation of the number $t_i$ of transmissions in slot $i$. However, in practical schemes the estimator may fail to  estimate $t_i$ leading to a possible increase of packet losses. In fact, the estimator has to be carefully designed to keep the probability of estimation error negligible. 

A simple estimation method, which does not require additional power or spectrum resources, is based on the evaluation of the received signal energy in the slot. The energy-based estimator of $t_i$ is defined as
\begin{align}
\hat{t}_i= \underset{t}{\mathrm{argmin}} \big| \, \|Y^n\|^2 - n(\sigma^2 +tP) \, \big|
\end{align}
where $Y^n=\sum_{j=1}^{t_i} X_j^n + Z^n$ is the sequence at the output of the channel in slot $i$ and $X_j^n$ is the symbol sequence transmitted by user $j$ in slot $i$.
Under the assumption of i.i.d. zero-mean Gaussian input symbols with power $P$ it is possible to analytically evaluate the probability of estimation failure, defined as $P_{\mathrm{F|t_i}}=\mathbb{P}\{\hat{t}_i\neq t_i | t_i\}$, as function of $t_i$, $P/\sigma^2$ and slot size $n$. 
In fact, in this case $\|Y^n\|^2/(\sigma^2 +t_iP)$ becomes a chi-squared random variable with $n$ degrees of freedom. Hence, the probability of estimation failure can be evaluated as
\begin{align}
P_{\mathrm{F|t_i}} &= \mathbb{P}\{A>n(\sigma^2 +t_iP)+nP/2)\} \notag \\
&+\mathbb{P}\{A<n(\sigma^2 +t_iP)-nP/2)\}
\end{align}
where $A\sim(\sigma^2 +t_iP)\chi^2(n)$.
This probability significantly increases as $t_i$ gets large. 
As an example, by using $n=200$ and $P/\sigma^2=5$, we obtain $P_{\mathrm{F|1}}<0.0001$, $P_{\mathrm{F|2}}=0.023$, $P_{\mathrm{F|3}}=0.117$, suggesting that this estimator appears useful only for random access schemes with $T$ up to $1,\, 2$.

An alternative estimation method, whose behavior does not depend on $t_i$, is based on the use of additional pilot symbols transmitted in each slot with the encoded message. Pilot symbols are fixed symbols with amplitude $\sqrt{P}$. 
In case of $t_i$ transmissions in  slot $i$, the receiver sees at the output of the channel, in the position of the pilot, a  noisy sample with the sum of $t_i$ pilot symbols, which can be used to estimate $t_i$. 
By considering the use of $n_\mathrm{p}$ pilot symbols per slots, the ML estimator can be defined as
\begin{align}
\hat{t}_i= \underset{t}{\mathrm{argmin}} \sum_{j=1}^{n_\mathrm{p}} \big( y_j -t \sqrt{P}  \big) ^2
\end{align}
where $y_j$, $j=1,\ldots, n_\mathrm{p}$  is the set of received samples in the $n_\mathrm{p}$ positions of the transmitted pilots. Since this is a minimum distance estimator, the probability of estimation failure can be easily derived  as 
\begin{align} 
P_{\mathrm{F|t_i}}=\mathrm{erfc}\left(\sqrt{\frac{n_\mathrm{p}P}{8\sigma^2}}\right)=P_{\mathrm{F}}(n_\mathrm{p})
\end{align}
which is independent of $t_i$ and dependent on the number of pilot symbols. 
As an example, by using $P/\sigma^2=5$, we obtain $P_{\mathrm{F}}(1)=0.26$, $P_{\mathrm{F}}(6)<0.01$, $P_{\mathrm{F}}(12)<0.0001$, highlighting the tradedoff between reliability of the estimation and additional resources required to implement pilot symbols. In this case a reliable estimation of $t_i$ can be done, for any value of $T$, at the expense of a suitable number of pilot symbols. 

\begin{figure}[t]
\centering
\includegraphics[width=\columnwidth]{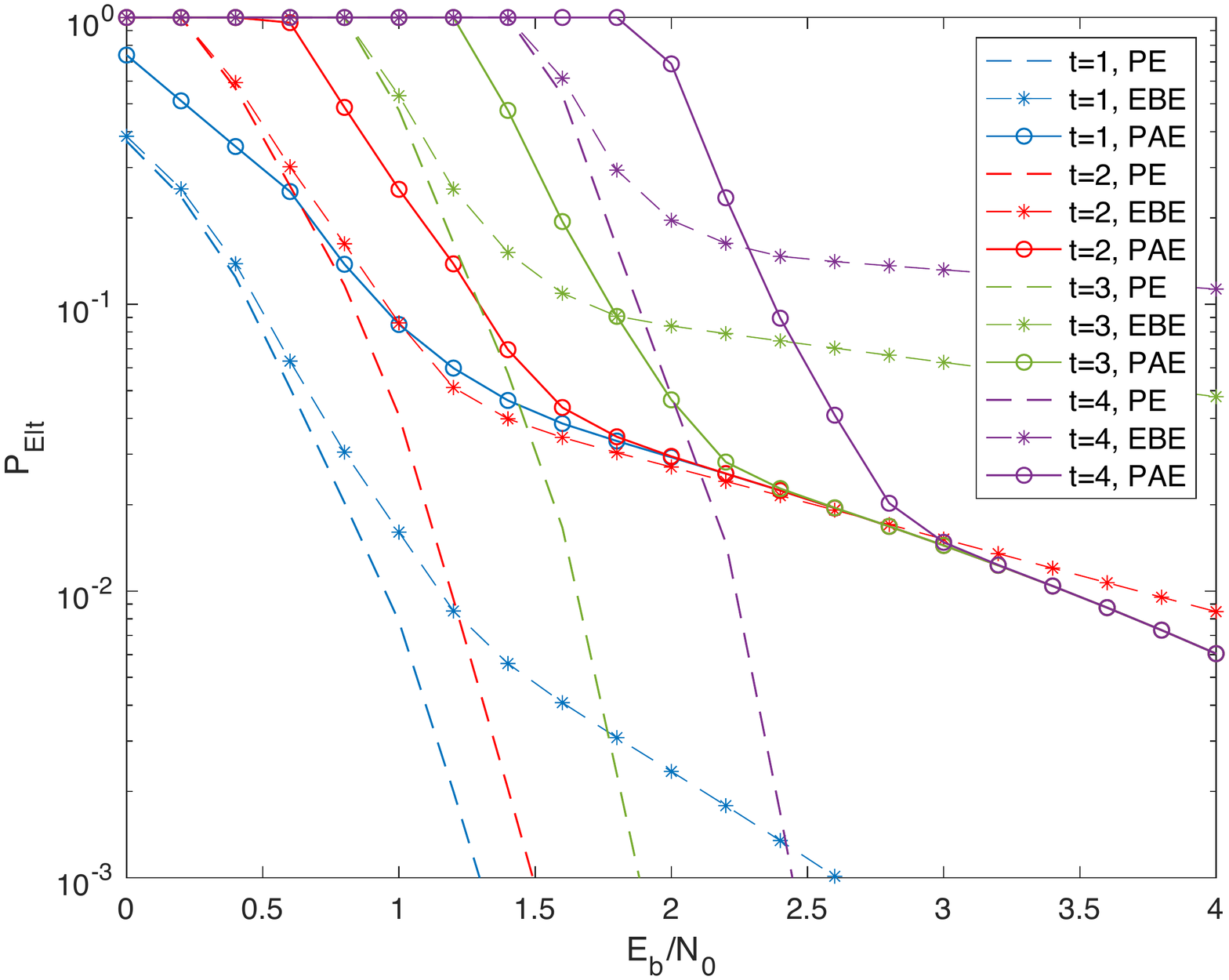}
      \caption{Conditional slot decoding error probability $P_{\text{E}|t}$, $t=1,\;2,\;3,\;4$ as function of $E_b/N_0$ for coding option 1 with perfect estimation (PE) of $t$, energy based estimation (EBE) and pilot assisted estimation (PAE).  Other parameters: $n=500$, $n_\mathrm{p}=30$, $2\eta=1$, $\log_2M=100$.}
    \label{fig:PEvsEbNo_1}
\end{figure}
\begin{figure}[t]
\centering
\includegraphics[width=\columnwidth]{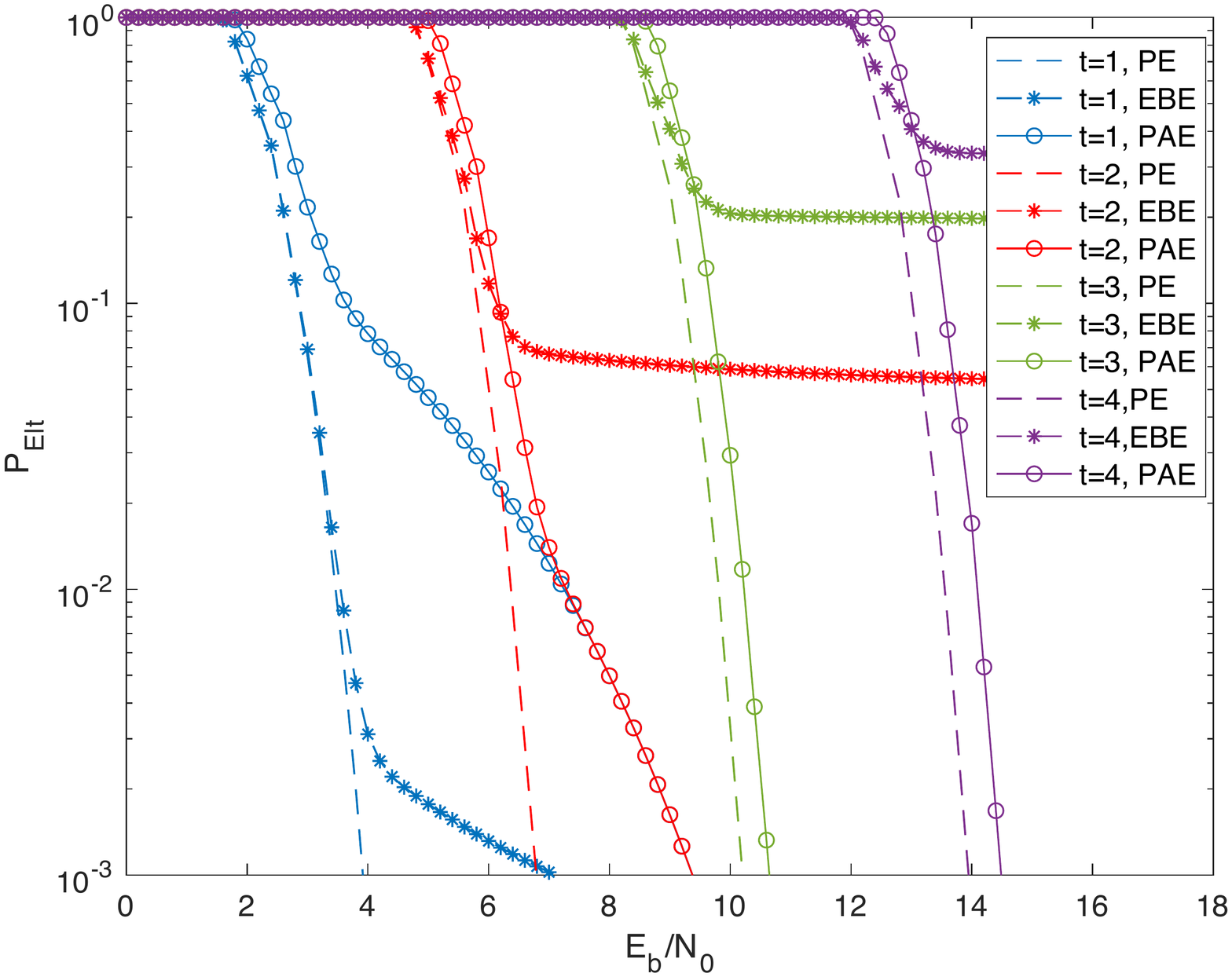}
      \caption{Conditional slot decoding error probability $P_{\text{E}|t}$, $t=1,\;2,\;3,\;4$ as function of $E_b/N_0$ for coding option 1 with perfect estimation (PE) of $t$, energy based estimation (EBE) and pilot assisted estimation (PAE).  Other parameters: $n=120$, $n_\mathrm{p}=3$, $2\eta=1$, $\log_2M=100$.}
    \label{fig:PEvsEbNo_2}
\end{figure}

To evaluate the effects of imperfect $t_i$ estimation on the packet loss probability, we first remind that it depends on the set of slot decoding error probabilities $P_{\text{E}|t}$, being $t\geq1$ the number of colliding packets not yet subtracted by the \ac{SIC} process in slot under analysis.
In absence of estimation errors the conditional  error probabilities are expressed, according to the PHY layer coding options presented in Section~\ref{Sec:PHY_performance}, as
$
{P}_{\text{E}|t}=F(n,M,P/\sigma^2,t)
$
where the function $F(\cdot)$  is specific of each PHY layer coding scheme, 
and $P/\sigma^2=(2\eta\log_2M/n)E_b/N_0$ depends on $E_b/N_0$.
Under the assumption, stated in the system model of this paper, of ideal error detection, when the estimate of the number of colliding packets $\hat{t}$ is not correct, i.e.,  $\hat{t}\neq t$, the decoder will react with high probability with an error that will be detected, and the slot will be marked as unresolved. 
Therefore, in presence of imperfect estimation of $t$, the slot decoding error probability $P_{\text{E}|t}$, $t\geq 1$ should be replaced by 
\begin{align}
\widetilde{P}_{\text{E}|t} &=P_{\mathrm{F|t}}+(1-P_{\mathrm{F|t}}) F(n-n_\mathrm{p},M,P/\sigma^2,t) \nonumber \\
&=  F(n-n_\mathrm{p},M,P/\sigma^2,t) 
+P_{\mathrm{F|t}} [1-F(n-n_\mathrm{p},M,P/\sigma^2,t)].
\end{align}
We can first note that an additional term arises, function of the estimation failure probability previously derived for the two estimation schemes. 
In practice, an imperfect estimation of $t$ behaves as an additional source of decoding errors which increases the value of $P_{\text{E}|t}$. 
We also note that only a subset of the channel uses, of size $n-n_\mathrm{p}$, remains available in the slot for the data symbols, when pilot symbols are transmitted ($n_\mathrm{p}=0$ for the energy-based estimator). This penalty is taken into account into the function $F(\cdot)$.
The new decoding error probability $\widetilde{P}_{\text{E}|t}$ depends on the set of parameters $(n,M,P/\sigma^2,t,n_\mathrm{p})$ and can be easily compared to ${P}_{\text{E}|t}$. 
The results are shown in Fig.~\ref{fig:PEvsEbNo_1} and Fig.~\ref{fig:PEvsEbNo_2} as function of $E_b/N_0$,  considering PHY layer coding option 1 (Section~\ref{Sec:PHY_performance}-A) and $2\eta=1$, $\log_2M=100$. Fig.~\ref{fig:PEvsEbNo_1} refers to the case with $n=500$  (low-rate PHY layer code), while Fig.~\ref{fig:PEvsEbNo_2} refers to the case with $n=120$ (high-rate PHY layer code). 
For pilot assisted estimation the value of $n_\mathrm{p}$ has been chosen to minimize $E_b/N_0$ at the target error probability $P_{\text{E}|1}=0.1$.
We can note from the figures that the $E_b/N_0$ gap of pilot assisted estimation with respect to perfect estimation of $t$ is always limited to $0.5\,\mathrm{dB}$ (with the exception of high-rate code for $t=1$), whereas the gap of energy based estimation is negligible for $t=1$, but rapidly diverges for $t\geq 3$.

\vskip 5mm
\section*{Acknowledgements}
The authors would like to thank Marco Chiani for useful discussions. 
They would also like to thank the anonymous Reviewers for their insightful and very detailed comments, that helped to improve the paper considerably.

Supported in part by the CNIT National Laboratory WiLab and the WiLab-Huawei Joint Innovation Center and in part by 
the European Union under the Italian National Recovery and Resilience Plan  of NextGenerationEU, partnership on ``Telecommunications of the Future'' (PE00000001 - ``RESTART'').
 

\end{document}